\documentclass[a4paper,fleqn,usenatbib]{mnras}

\usepackage{newtxtext,newtxmath}

\usepackage[T1]{fontenc}
\usepackage{ae,aecompl}

\usepackage{graphicx}	
\usepackage{amsmath}	
\usepackage{amssymb}	
\usepackage{enumitem}
\usepackage{bm}

\title[Clustering of \ensuremath{\rm{H}\alpha} emitters]{The  \ensuremath{\rm{H}\alpha} luminosity-dependent clustering of star-forming galaxies from $z\sim0.8$ to $z\sim2.2$ with HiZELS}

\author[R.K. Cochrane et al.]{
R. K. Cochrane,$^{1}$\thanks{E-mail: rcoch@roe.ac.uk}
P. N. Best,$^{1}$
D. Sobral,$^{2,3}$
I. Smail,$^{4}$
D. A. Wake,$^{5}$
J. P. Stott,$^{2,6}$
\newauthor
J. E. Geach$^{7}$
\\
$^{1}$SUPA, Institute for Astronomy, Royal Observatory Edinburgh, EH9 3HJ, UK\\
$^{2}$Department of Physics, Lancaster University, Lancaster, LA1 4YB \\
$^{3}$Leiden Observatory, Leiden University, P.O. Box 9513, NL-2300 RA Leiden, The Netherlands \\
$^{4}$Centre for Extragalactic Astronomy, Department of Physics, Durham University, South Road, Durham DH1 3LE, UK  \\
$^{5}$Department of Physical Sciences, The Open University, Milton Keynes MK7 6AA, UK \\
$^{6}$Sub-department of Astrophysics, Department of Physics, University of Oxford, Denys Wilkinson Building, Keble Road, Oxford OX1 3RH, UK \\
$^{7}$Centre for Astrophysics Research, Science \& Technology Research Institute, University of Hertfordshire, Hatfield, AL10 9AB, UK \\
}

\date{Accepted XXX. Received YYY; in original form ZZZ}

\pubyear{2017}

\begin{document}
\label{firstpage}
\pagerange{\pageref{firstpage}--\pageref{lastpage}}
\maketitle

\begin{abstract}\\
We present clustering analyses of identically-selected star-forming galaxies in 3 narrow redshift slices (at $z=0.8$, $z=1.47$ and $z=2.23$), from HiZELS, a deep, near-infrared narrow-band survey. The HiZELS samples span the peak in the cosmic star-formation rate density, identifying typical star-forming galaxies at each epoch. Narrow-band samples have well-defined redshift distributions and are therefore ideal for clustering analyses. We quantify the clustering of the three samples, and of $\rm{H}\alpha$ luminosity-selected subsamples, initially using simple power law fits to the two-point correlation function. We extend this work to link the evolution of star-forming galaxies and their host dark matter halos over cosmic time using sophisticated dark matter halo models. We find that the clustering strength, $r_{0}$, and the bias of galaxy populations relative to the clustering of dark matter increase linearly with $\rm{H}\alpha$ luminosity (and, by implication, star-formation rate) at all three redshifts, as do the host dark matter halo masses of the HiZELS galaxies. The typical galaxies in our samples are star-forming centrals, residing in halos of mass $M_{\rm{halo}}\sim$ a few times $10^{12}M_{\odot}$. We find a remarkably tight redshift-independent relation between the $\rm{H}\alpha$ luminosity scaled by the characteristic luminosity, $L_{\rm{H}\alpha}/L_{\rm{H}\alpha}^{*}(z)$, and the minimum host dark matter halo mass of central galaxies. This reveals that the dark matter halo environment is a strong driver of galaxy star-formation rate and therefore of the evolution of the star-formation rate density in the Universe.\\
\end{abstract}




\section{Introduction}

The galaxies we observe exist in a wide range of environments, from rich clusters to underdense void regions. They are thought to trace an underlying distribution of dark matter, with more highly clustered galaxies occupying massive dark matter overdensities \citep{Zwicky1933,Peebles1982}. This is commonly explained via the paradigm of hierarchical growth: weak density fluctuations in an expanding, homogeneous Universe are amplified by gravitational instabilities, with smaller structures forming first. Galaxies form due to the collapse of baryonic matter under the gravity of dark matter halos \citep{White1991}, with the progenitors of the most massive clusters starting to form earliest. Dark matter halos assemble via successive mergers and accretion of small halos, which naturally leads to the formation of galaxy groups and clusters, with a single dark matter halo capable of hosting many galaxies. \\
\indent While the observed `cosmic web' spatial distribution of dark matter in the Lambda Cold Dark Matter ($\Lambda$CDM) paradigm can be successfully modelled using N-body simulations \citep{Davis1985} as advanced as the Millennium Simulation \citep{Springel2005}, resolution is limited and the evolution of galaxies within this web is harder to model. This complexity reflects the additional baryonic processes present: we must consider not only the underlying distribution of dark matter but also the non-linear physics of galaxy formation and evolution. Key processes such as gas cooling, star-formation, and the physics of feedback due to star-formation and black hole accretion all act on different timescales with different galaxy mass and environment dependencies. The latest generation of hydrodynamical simulations such as Illustris \citep{Vogelsberger2014} and EAGLE \citep{Crain2015,Schaye2015} and semi-analytic models \citep[e.g.][]{Baugh2006} currently do fairly well in modelling such processes, broadly reproducing key observed relations such as galaxy luminosity and stellar mass functions, and the bimodal galaxy colour distribution, but a wealth of observational data is required to fine-tune parameters. \\
\indent Many details of the environmental drivers of galaxy evolution, and how they relate to galaxy mass, remain poorly understood. It has long been known that at low redshifts, galaxies in rich clusters are preferentially passive ellipticals \citep{Oemler1977,Dressler1980}, whereas field galaxies tend to be star-forming and disk-like, with increasing star-formation rates and star-forming fractions further from cluster centres \citep{Lewis2002,Gomez2003}. High mass galaxies are also far less likely to be star-forming than their low mass counterparts \citep{Baldry2006}. Despite these well-established observational trends, the effects of mass and environment have remained hard to distinguish, given the inter-dependence of the two quantities (galaxies of higher masses tend to reside in higher density environments).\\
\indent The latest observational data at both low and high redshifts has provoked a flurry of recent work aiming to understand the relationships between stellar mass, star-formation rate and environment \citep[e.g.][]{Peng2010,Sobral2011,Scoville2013a,Darvish2015}. Both mass and environment are associated with transformations in colour, star-formation rate and morphology, popularly known as `quenching'. Supplementing low redshift data from the SDSS \citep{York2000} with higher redshift data from the zCOSMOS survey \citep{Lilly2007}, \cite{Peng2010} proposed that two primary quenching mechanisms, `mass quenching' and `environment quenching', act independently and dominate at different epochs and galaxy masses. `Environment quenching', which primarily affects satellite galaxies \citep{Peng2012}, is attributed to some combination of gas stripping (due to ram pressure \citep{Boselli2006} or tidal effects) and `strangulation' \citep*{Larson1980,Peng2015}, whereby gas is prevented from cooling onto the galaxy from its hot halo, perhaps upon accretion onto a massive halo. Mass quenching, which dominates the cessation of star-formation for massive galaxies, is also attributed to a shut-down of cold gas accretion, via shock heating by the hot halo \citep{Dekel2006}, possibly in combination with AGN heating \citep{Croton2006,Best2006}.\\
\indent There is evidence that the trends observed at low redshift hold to at least $z\sim1$. At $z\sim1$, \cite{Sobral2011} and \cite{Muzzin2012} both find that the fraction of galaxies that are star-forming decreases once we reach group densities and at high galaxy masses. However, things become less clear at even higher redshifts. \cite{Scoville2013a} find a flattening in the relationship between environmental overdensity and both star-forming fraction and star-formation rate above $z\sim1.2$ for galaxies in the COSMOS field, and note that this flattening holds out to their highest redshift galaxies at $z\sim3$. Other studies have found an apparent reversal of the low-z star-formation rate (or morphology)-density relation at higher redshifts \citep{Melorose1978}. Both \cite{Sobral2011} and \cite{Elbaz2007} found that at $z\sim1$, median galaxy star-formation rates increase with overdensity until cluster densities are reached, at which point star-formation rates decrease with overdensity, as in the local universe. Attempting to explain these opposing trends, \cite{McGee2009} propose that the pressure of the intra-cluster medium on infalling galaxies in the outskirts of galaxy clusters actually compresses gas and enhances star-formation prior to stripping in the denser environment of the cluster core. Increased galaxy-galaxy interactions may also trigger intense star-formation via the disruption of gas disks.  At high redshifts, high gas fractions \citep[e.g.][]{Tacconi2010} permit more efficient starburst responses. Thus at high redshifts, the richest environments may provide the combination of large gas reservoirs and ICM pressures which fuel high star-formation rates and later lead to quenching via gas exhaustion and stripping \citep{Smail2014}. \\
\indent Quantifying the environmental dependence of star-formation activity at high redshift directly is inherently challenging. An alternative approach to studying this is through auto-correlation functions of star-forming galaxies. The dark matter correlation function is the inverse fourier transform of the dark matter power spectrum. Observing the projected real-space galaxy correlation function, which is a linear scaling of the dark matter correlation function, provides a natural connection between galaxies and the underlying matter distribution which determines their large-scale environments. Modelling these correlation functions using Halo Occupation Distribution model frameworks \citep{Peacock2000} can yield more information about galaxy host halos, in particular their masses. It also provides a powerful technique for exploring the central/satellite dichotomy in galaxy populations. The `one-halo' term represents clustering on small scales, within a single dark matter halo, and is determined by the spatial separation of central galaxies and their satellites. The `two-halo' term, in contrast, is controlled by the larger-scale clustering of galaxies in different dark matter halos (driven primarily by the halo mass), and incorporates central-central pairs as well as satellite-satellite and central-satellite correlations. A consistent picture has emerged in which more luminous and more massive star-forming galaxies tend to be more strongly clustered, as a result of lying preferentially in high mass dark matter halos. This holds at both low redshifts \citep[e.g.][]{Norberg2001,Zehavi2011} and at high redshifts \citep[e.g.][]{Sobral2010,Wake2011,Geach2012,Hatfield2015}.\\
\begin{figure} 
	\centering
	\includegraphics[scale=0.7]{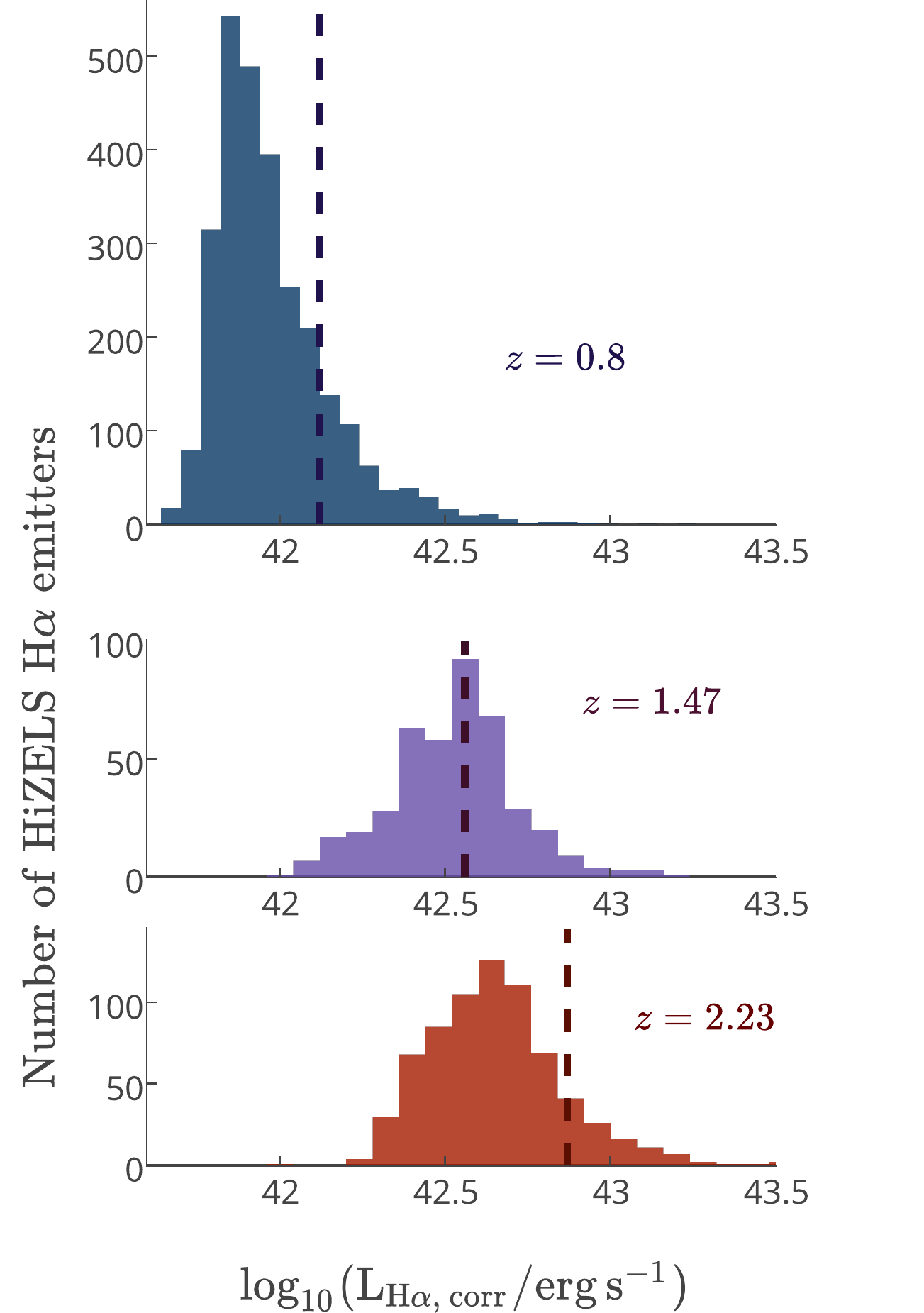}	
	\caption{Distribution of dust-corrected $\rm{H}\alpha$ luminosities of HiZELS emission line-selected galaxies in our samples at the three epochs. Vertical dashed lines show the characteristic luminosity, $L_{\rm{H}\alpha}^{*}$, at each redshift. HiZELS galaxies span a large luminosity range at each epoch, probing well below $L_{\rm{H}\alpha}^{*}$.}
    \label{fig:nbj_fields}
\end{figure}
\indent In this paper we build upon the work presented in \cite{Sobral2010}, which studied the clustering of $\sim700$ $\rm{H}\alpha$ emitters at $z=0.84$ from the High-Redshift(Z) Emission Line Survey (HiZELS, see Section \ref{sec:sample_selection}). Narrow band $\rm{H}\alpha$ surveys such as HiZELS select only those galaxies with emission lines within a very narrow redshift range ($\Delta z \sim0.02$), and with a well-defined redshift distribution. For clustering measurements, these types of survey are therefore superior to photometric ones, which are often hampered by systematic uncertainties and require a more complex treatment of the spatial distribution in the clustering analysis. Furthermore, unlike many spectroscopic surveys, the narrow band approach provides a clean selection function down to a known flux (star-formation rate) limit. \cite{Sobral2010} found evidence for a strong $\rm{H}\alpha$ luminosity dependence of the clustering strength of $\rm{H}\alpha$ emitters at $z=0.84$, along with evidence for a single relation with $L_{\rm{H}\alpha}/L_{\rm{H}\alpha}^{*}$ from $z\sim0.2$ to $z\sim2.2$. \cite{Geach2008,Geach2012} supplemented this work with the first analyses of the clustering of HiZELS galaxies at $z=2.23$, though the sample was not sufficiently large to permit binning by luminosity. \\
\indent Here we analyse a larger sample of $\sim3000$ emitters at $z=0.8$ spanning three fields: COSMOS, UDS and SA22. Crucially, we also use larger samples of $\rm{H}\alpha$ emitters at $z=2.23$, and include new data at $z=1.47$ \citep{Sobral2012a,Sobral2012}. Our samples, which span large ranges in $\rm{H}\alpha$ luminosity and redshift, provide optimal data for revealing the drivers of galaxy evolution over cosmic time. We provide details of the HiZELS sample selection in Section \ref{sec:sample_selection}. In Section \ref{sec:clustering_one}, we lay out our approach to quantifying the clustering of these sources via two-point correlation functions, and in Section \ref{sec:r0_results} we present the results of simple power-law fits to these. Given the high quality of the correlation functions obtained, we extend these analyses to incorporate a sophisticated Halo Occupation Distribution (HOD) modelling treatment. In Section \ref{sec:halo_fitting} we set up the HOD framework and present derived halo properties for our HiZELS galaxies, in particular typical halo masses and galaxy central/satellite fractions. We discuss the implications of these results in Section \ref{sec:discussion}.\\
\indent We use a $H_{0} = 70\rm{kms}^{-1}\rm{Mpc}^{-1}$, $\Omega_{M} = 0.3$ and $\Omega_{\Lambda} = 0.7$ cosmology throughout this paper.
\begin{table*}
\begin{center}
\begin{tabular}{l|c|c|c}
Field & ${\overline{z}}_{\rm{H}\alpha\,\rm{emitters}}$ & No. emitters & Area ($\rm{deg}^{2}$)\\
\hline
NBJ COSMOS \& UDS & $0.845\:\ensuremath{\pm}\:0.011$ & $503$ & $1.6$ \\
LOW0H2 SA22 & $0.81\:\ensuremath{\pm}\:0.011$ & $2332$ & $7.6$ \\
NBH COSMOS \& UDS & $1.47\:\ensuremath{\pm}\:0.016$ & $451$ & $2.3$ \\
NBK COSMOS \& UDS & $2.23\:\ensuremath{\pm}\:0.016$ & $727$ & $2.3$ \\
\end{tabular}
\caption{Numbers and mean redshifts of H\ensuremath{\alpha} emitters identified by the HiZELS survey and selected for this analysis. \textcolor{black}{HiZELS uses standard and custom-made narrow-band filters, complemented by broad-band imaging, over well-studied fields.} Only emitters which exceed the limiting flux, $f_{50}$, of their frames are included.}
\label{table:total_numbers}
\end{center}
\end{table*}
 \begin{figure*} 
	\centering
	\includegraphics[scale=0.45]{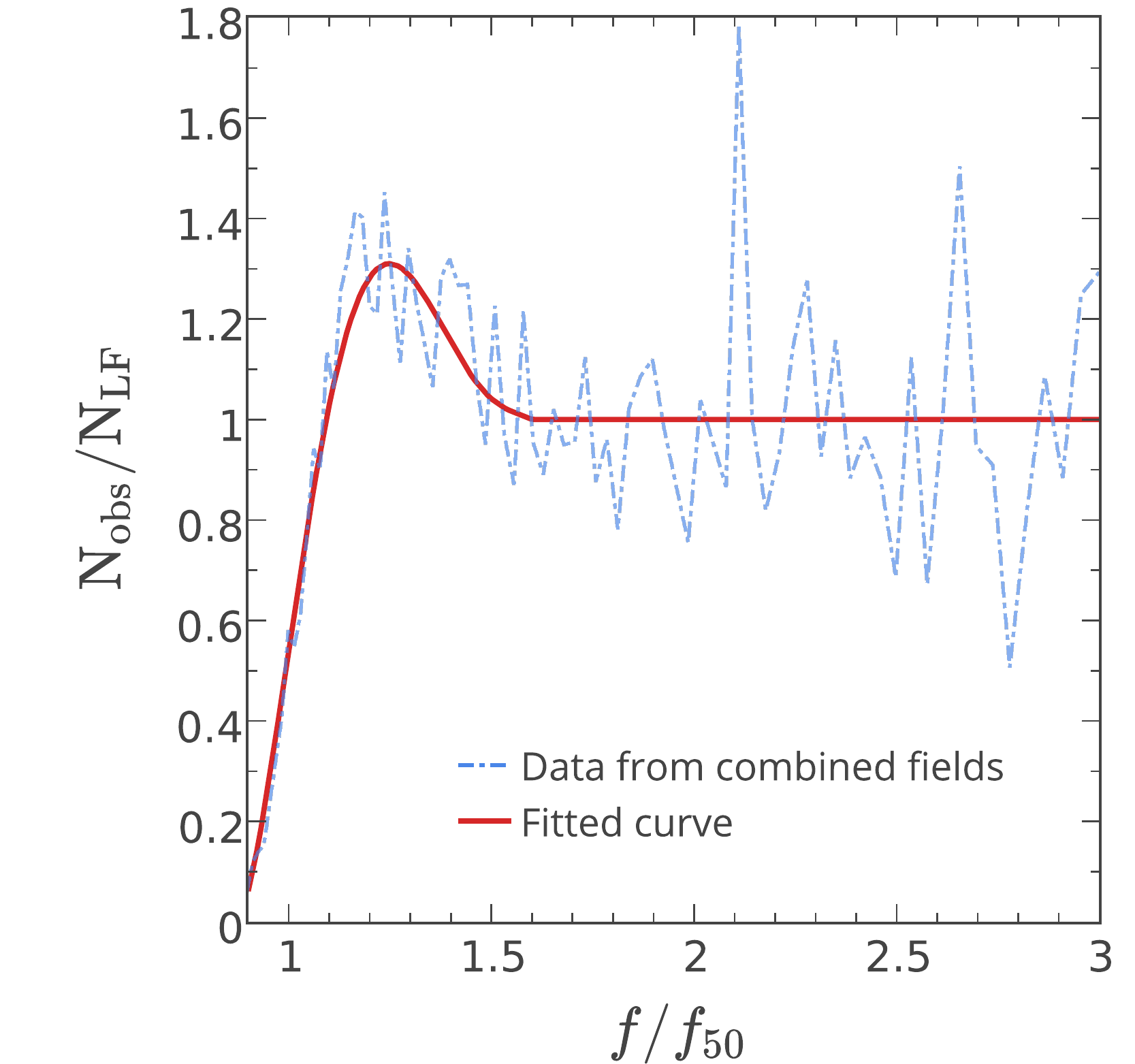}
	\includegraphics[scale=0.45]{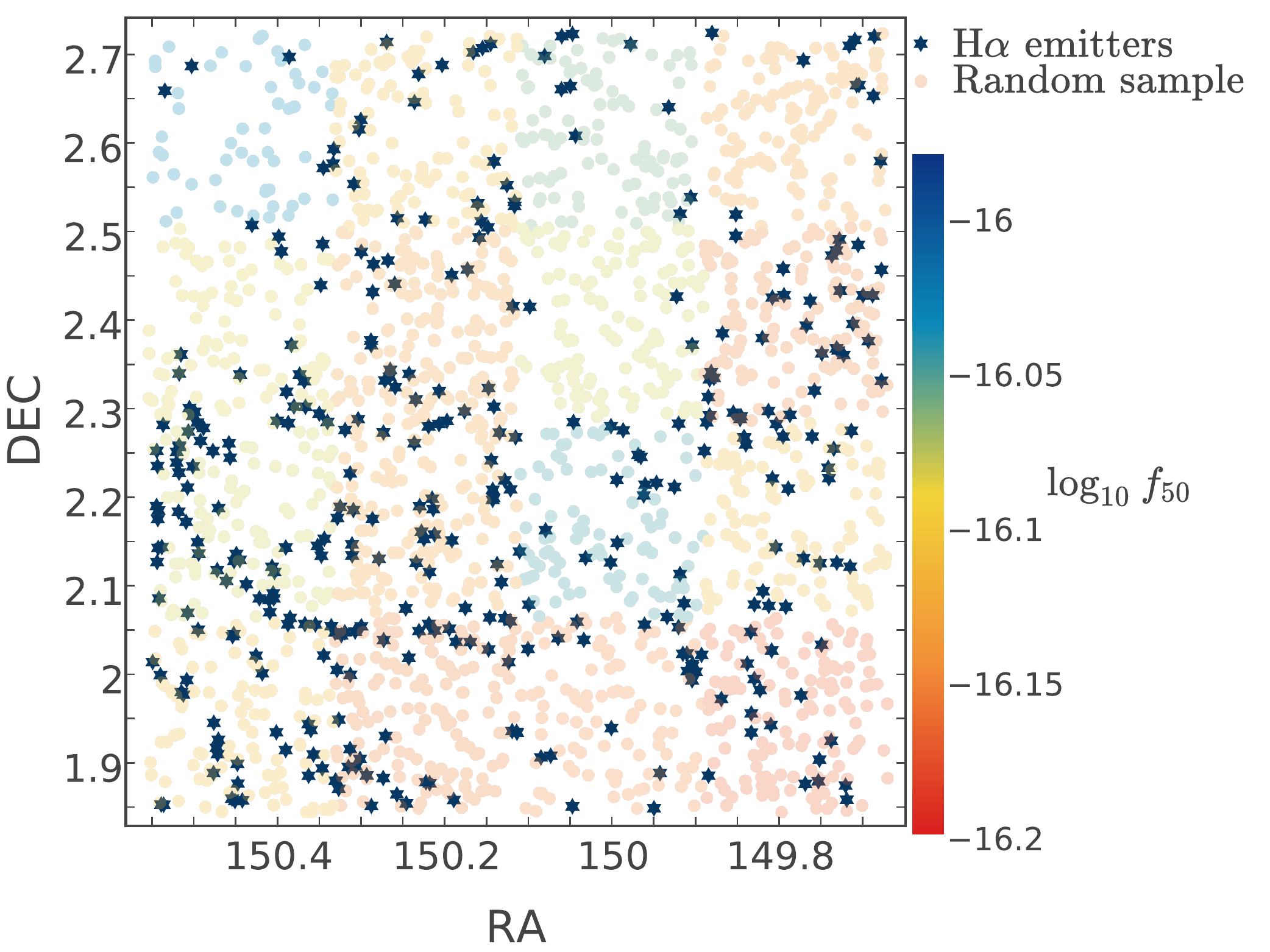} 
    \caption{Left: the completeness curve used to place sources in frames with flux limit $f_{50}$. We account for a small number of excess sources due to flux boosting around the detection limit. Right: example of random sources in the COSMOS field, colour coded by the limiting flux of their frame, with real sources shown by stars overlaid. Fluxes are given in units of $\rm{erg\,s^{-1}cm^{-2}}$.}
    \label{fig:completeness_curve}
\end{figure*}
\section{The HiZELS survey and sample selection}\label{sec:sample_selection}
\subsection{Sample of $\rm{H}\alpha$ emitters}\label{sec:sample_details}
HiZELS (\citealt{Geach2008,Sobral2009,Sobral2012a,Sobral2012}) used the United Kingdom Infra-Red Telescope (UKIRT)'s Wide Field CAMera (WFCAM), the Subaru Telescope's Suprime-Cam with the NB921 filter, the Very Large Telescope (VLT)'s HAWK-I camera and the Canada France Hawaii Telescope (CFHT) with MegaCam (CFHiZELS; \citealt{Sobral2015}) to detect line emitters over large areas within well-studied fields. We present only a brief overview of the survey here, referring the curious reader to \cite{Sobral2012} for a full description of the HiZELS COSMOS and UDS data, and to \cite{Sobral2015} for details of the SA22 CFHiZELS campaign. \\
\indent HiZELS uses standard and custom-made narrow-band (NB) filters, complemented by broad-band (BB) imaging. Sources identified by the narrow-band filters are matched to those in the broad-band images by using the same aperture size and a search radius of $0.9"$. True emitters are selected based on their NB-BB colour excess, with a signal-to-noise cut of $S/N>3$ and an equivalent width selection corresponding to ${\rm{EW}>25\AA}$ for $\rm{H}\alpha$. High quality photometric redshifts derived from data spanning from optical to mid-IR wavelengths (e.g. \citealt{Cirasuolo2010,Ilbert2009,Lawrence2007}) were used to identify which emission line is being selected for each emitter, and thus select a clean sample of H\ensuremath{\alpha} emitters. This technique enables the identical selection of H$\alpha$ emitting galaxies at $z=0.81,0.84$ (NBJ: COSMOS, UDS, SA22), $z=1.47$ (NBH: COSMOS, UDS) and $z=2.23$ (NBK: COSMOS, UDS); see Table \ref{table:total_numbers} for details. Spectroscopic redshifts confirmed that the large sample of galaxies we obtain lies within well-defined redshift ranges (see also \citealt{Sobral2016a,Stott2016}). \\
\indent $\rm{H}\alpha$ fluxes are corrected for contamination by the adjacent $\rm{[NII]\lambda6548,6584}$ lines within the NB filter using the relationship between $\log(\rm{[NII]}/\rm{H}\alpha)$ and $\rm{EW}_{0}(\rm{[NII]}+\rm{H}\alpha)$ derived by \cite{Sobral2012} and confirmed spectroscopically in \cite{Sobral2015}. They are also corrected for dust attenuation assuming $\rm{A}_{\rm{H}\alpha}=1.0\,\rm{mag}$ \citep{Garn2010a,Ibar2013}. The median combined correction is $0.307\,\rm{dex}$ at $z=0.8$, $0.325\,\rm{dex}$ at $z=1.47$ and $0.335\,\rm{dex}$ at $z=2.23$.
\subsection{Generating random samples}\label{sec:random_sample}
We generated unclustered random samples in order to quantify the clustering of the observed $\rm{H}\alpha$ emitters. Variations in coverage and observing conditions have resulted in individual HiZELS frames having different depths, meaning that robustly-constructed random samples are essential to differentiate between true clustering and that introduced by the observing strategy. In this section we describe the construction of random samples which reflect these depths. \\
\indent Most simply, random sources may be generated by calculating a limiting flux at which each frame is essentially $100\%$ complete, drawing sources from the luminosity function down to this flux, and distributing these randomly across the frame. For this analysis we aim to push further in flux, so as to include as many sources as possible. We include sources down to luminosities corresponding to the $50\%$ completeness flux, $f_{50}$, as calculated by \cite{Sobral2012,Sobral2015} for each frame using Monte Carlo simulations. To study source detection as a function of the limiting flux (taking account of both incompleteness and flux boosting biases), we have calculated the ratio of the number of sources recovered, $N_{\rm{obs}}$, to the number of sources expected from the luminosity function, $N_{\rm{LF}}$, as a function of $f_{50}$ in each frame. We found a small boost in the number of sources with recovered fluxes around the flux limit, suggesting that flux-boosting effects dominate over incompleteness. We tested different filters, and both deep and shallow fields separately, and found that all show the same general form. We have therefore fitted a single empirically-derived effective completeness curve (Figure \ref{fig:completeness_curve}, left) and taken this into account when generating the random catalogues. Numerous tests have confirmed that our results are qualitatively unchanged if the random sources are simply drawn from the luminosity function down to $f_{50}$ or constructed using a slightly different completeness curve. \\
\indent In this paper, we use luminosity functions of the form:
 \begin{equation}
 	\phi(L)dL = \phi^{*}\Bigg(\frac{L}{L^{*}}\Bigg)^{\alpha}e^{-(L/L^{*})}d\Bigg(\frac{L}{L^{*}}\Bigg).
 \end{equation}
Here, $L^{*}$ represents the characteristic luminosity `break' of the LF, $\phi^{*}$ is the corresponding characteristic comoving space density, and $\alpha$ is the `faint-end' slope of the power law, dominant at low luminosities. The parameters we adopt, given in Table \ref{table:lf_params}, were derived using the samples of $\rm{H}\alpha$ emitters from \cite{Sobral2012,Sobral2015}.
\begin{table}
\begin{center}
\begin{tabular}{l|c|c|c}
z & $L^{*}_{\rm{H}\alpha}\rm{(erg\,s^{-1})}$ & $\phi^{*} \rm{(Mpc^{-3})}$ &  $\alpha$ \\
\hline
$0.810$ \& $0.845$ & $42.12^{+0.03}_{-0.02}$ & $-2.31^{+0.04}_{-0.05}$ & $-1.6^{+0.2}_{-0.2}$\\
$1.466$ & $42.56^{+0.06}_{-0.05}$ & $-2.61^{+0.08}_{-0.09}$ & $-1.62^{+0.25}_{-0.29}$\\
$2.231$  & $42.87^{+0.08}_{-0.06}$ & $-2.78^{+0.08}_{-0.09}$ & $-1.59^{+0.12}_{-0.13}$\\
\end{tabular}
\caption{LF parameters used in this paper, derived in \protect\cite{Sobral2012,Sobral2015}. At $z\sim0.8$, we use the Schechter function fit to the much larger $z=0.81$ sample by \protect\cite{Sobral2015}, which is more accurate than that presented by \protect\cite{Sobral2012} and is also a good fit for the $z=0.84$ data.}
\label{table:lf_params}
\end{center}
\end{table}
We generated a random position for each random source, carefully taking into account the boundaries of each frame and the masked regions due to bright stars and artefacts. The final number of sources generated within a frame depends on both its unmasked chip area and its depth. All random samples are substantially larger (e.g. $1000\times$) than the real samples. When constructing correlation functions for samples binned by flux, we also require knowledge of the fluxes of the random sources, to account for faint sources being preferentially detected in the deepest frames. The fluxes of random sources are drawn from the luminosity functions given in Table \ref{table:lf_params}, scaled by the fitted completeness curve (Figure \ref{fig:completeness_curve}) for a given $f_{50}$. We have also incorporated average corrections for dust and [NII] emission line contamination. We did not include any real or random sources with flux $f < f_{50}$ in this analysis.  
\subsection{Effects of potential contaminants}
Here we discuss three classes of possible contaminants: sources that are not true emitters; true emitters which are different lines misclassified as $\rm{H}\alpha$; and AGN interlopers. As discussed in Section \ref{sec:sample_details}, HiZELS emitters are selected based on their NB-BB colour excess, with a signal-to-noise cut of $S/N>3$. To check the possibility of including false emitters, we have repeated the clustering measurements using a more conservative cut of $S/N>4$ for various luminosity bins. We find no significant differences in the clustering strengths. We also note that the exclusion of sources with fluxes below their frame's $f_{50}$ serves to remove some potential low-flux contaminants. Contamination from misclassified lines is also estimated to be small, at $\sim5\%$, as estimated by \cite{Sobral2012}. Such contaminants will generally have the effect of a small decrease in $w(\theta)$, with much smaller effects than our observed trends.\\
\indent Our sample could suffer from contamination from AGN, for which $\rm{H}\alpha$ emission is not a reliable tracer of star-formation rate. Using extensive multi-wavelength data to identify AGN candidates within HiZELS samples in the COSMOS and UDS fields, \cite{Garn2010} estimate an AGN fraction of $\sim10\%$, but \cite{Sobral2016} find that this can be much higher at very high $\rm{H}\alpha$ luminosities. We expect that the effect of AGN contamination may only be very important in the highest luminosity bins. However, these bins show no evidence of deviation from the linear trend of the low-luminosity regime (see Section \ref{sec:r0_lum_dependence}). Given that it is difficult to exclude these individual sources from our analyses, we present all results using $\rm{H}\alpha$ luminosity rather than converting to star-formation rate explicitly. We invoke star-formation rate only in our gas-regulator model interpretation in Section \ref{sec:gas_reg}. 
\section{Quantifying galaxy clustering using the two-point correlation function}\label{sec:clustering_one}
Broadly, the two-point correlation function compares the clustering of an observed sample to a uniformly distributed random sample with the same areal coverage. It quantifies overdensities on a large range of scales; unlike nearest-neighbour estimators, it can yield insights into both the local environment within halos and the large scale environment. When quantifying galaxy clustering, we construct correlation functions based on angular or projected distances between pairs of galaxies on the sky.
\subsection{Angular two-point clustering statistics}\label{sec:two_point}
The angular two-point correlation function, $w(\theta)$, is a popular estimator of the clustering strength of galaxies. It is defined as the excess probability of finding a pair of galaxies separated by a given angular distance, relative to that probability for a uniform (unclustered) distribution. The probability $dP(\theta)$ of finding objects in solid angles $d\Omega_{1}$ and $d\Omega_{2}$ separated by angular distance $\theta$ is:
\begin{equation}
dP(\theta) = N^{2}(1+w(\theta))\,d\Omega_{1}d\Omega_{2},
\end{equation}
where N is the surface density of objects.\\\\
Many estimators of $w(\theta)$ have been proposed. We use the minimum variance estimator proposed by \cite{Landy1993}, which was shown to be less susceptible to bias from small sample sizes and fields:
\begin{equation} \label{eq:w_theta}
w(\theta) = 1 + \left(\frac{N_R}{N_D}\right)^{2}\frac{DD(\theta)}{RR(\theta)} - 2\frac{N_{R}}{N_{D}}\frac{DR(\theta)}{RR(\theta)},
\end{equation}
where $N_{R}$ and $N_{D}$ are the total number of random and data galaxies in the sample, and $RR(\theta)$, $DD(\theta)$ and $DR(\theta)$ correspond to the number of random-random, data-data, and data-random pairs separated by angle $\theta$. $w(\theta)$ is normally fitted with a power law, $w(\theta) = A\theta^{\beta}$, where $\beta=-0.8$.
Traditionally, Poissonian errors are used:
\begin{equation}
\Delta w(\theta) = \frac{1+w(\theta)}{\sqrt{DD(\theta)}}.
\end{equation}
However, these errors are underestimates \citep[e.g. see][]{Norberg2009}, since they do not account for cosmic variance or correlations between adjacent $\theta$ bins. Using these errors also gives unjustifiably large weightings to the largest angular separations, where large DD pair counts result in very low $\Delta w(\theta)$. \\
\indent \cite{Norberg2009} conclude that while no internal estimator reproduces the error of external estimators faithfully, jackknife and bootstrap resampling methods perform reasonably well, although both overestimate the errors. They note that jackknife resampling estimates the large-scale variance accurately but struggles on smaller scales ($\sim 2-3\rm{h}^{-1}\rm{Mpc}$), with the resulting bias strongly dependent on the number of sub volumes. Bootstrap resampling, meanwhile, overestimates the variance by approximately 50\% on \it{all} \normalfont{scales}, which may be minimised by oversampling the sub-volumes. In this paper, we use the bootstrap resampling method with each correlation function constructed from 1000 bootstraps, taking the error on each $w(\theta)$ bin as the diagonal element of the bootstrap covariance matrix.\\
\indent We also implement the integral constraint, IC, \citep{Groth1977}, a small correction to account for the underestimation of clustering strength due to the finite area surveyed. 
\begin{equation}
\rm{IC} = \frac{\sum\nolimits_{\theta} A\theta^{\beta}RR(\theta)}{\sum\nolimits_{\theta}RR(\theta)}
\end{equation}
IC is small where fields are large. HiZELS fields reach square-degree scales, and so IC corrections are largely negligible.
\subsection{Obtaining a real-space correlation length}\label{sec:correlation_length}
In order to compare the clustering strengths of populations of star-forming galaxies at different redshifts quantitatively, we convert the angular correlation function to a spatial one. This conversion is often performed using Limber's approximation \citep{Limber1953}, which assumes that spatial correlations which follow $\xi = (r/r_{0})^{\gamma}$ are projected as angular correlation functions with slopes ${\beta}=\gamma+1$. This results in the approximate relation between $\xi$ and $w(\theta)$:
\begin{equation}
w(\theta) = \int_{0}^{+\infty}p_{1}(r)p_{2}(r)dr \int_{-\infty}^{+\infty}r\xi(R,r)d\Delta,
\end{equation}
where $R = \sqrt{r^{2}\theta^{2}+\Delta r^{2}}$, and $p_{1}(r),p_{2}(r)$ are the filter profiles for projected fields 1\&2.
Substituting  ${\xi = (r/r_{0})^{\gamma}}$ yields:
\begin{equation}
\begin{split}
w(\theta) = r_{0}^{\gamma} \theta({\rm{rad}})^{1-\gamma} \times \frac{\Gamma(\gamma/2 - 1/2)\Gamma(1/2)}{\Gamma(\gamma/2)} \times \\ \int_{0}^{+\infty} p_{1}(r)p_{2}(r) r^{1-\gamma}dr,
\end{split}
\end{equation}\\
\textcolor{black}{where $\Gamma(x)$ is the gamma function.} This is a good approximation for small angular scales, where $\theta \ll \sigma/\mu$, and can thus be used to evaluate $r_{0}$ from the fitted $w(\theta)$ profile. However, the integral diverges for narrow filters. \cite{Simon2007} shows that in the limiting case of a delta function filter, the observed $w(\theta)$ is no longer a projection, but simply a rescaled $\xi_{\rm{gal}}(r)_{0}$ (thus $\beta = \gamma$ at large separations). Since Limber's approximation is not reliable for our samples of galaxies, which span fields with separations of degrees and use very narrow filters, we perform a numerical integration of the exact equation: 
\begin{equation}\label{eq:filter_integration}
w_{\rm{model}}(\theta) = \psi^{-1}\int_{0}^{+\infty}\int_{s\sqrt{2\phi}}^{2s}\frac{2f_{s}(s-\Delta)f_{s}(s+\Delta)}{R^{-\gamma-1}r_{0}^{\gamma}\Delta}dRds.
\end{equation}\\
Here, $\psi=1+\cos\theta$, $\phi=1-\cos\theta$, $\Delta=\sqrt{(R^{2}-2s^{2}\phi)/2\psi}$, and $f_{s}$ is the profile of the filter, fitted as a Gaussian profile with $\mu$ and $\sigma$ that depend on the filter being considered (see Table \ref{table:gauss_params} for the parameters of our filters). We assume the standard value of $\gamma = -1.8$. $\chi^{2}$ fitting of observed against modelled $w(\theta)$, generated using different $r_{0}$, allows us to estimate $r_{0}$ and its error \citep[following][]{Sobral2010}.
\subsubsection{Projected-space two-point clustering statistics}\label{sec:two_point_proj}
The clustering statistic required as input for the halo fitting routine we use in Section \ref{sec:fitting_halomod} is the projected-space ($r_{p}$) two-point correlation function, $w_{p}(r_{p})$. We therefore transform our measured $w(\theta)$ to $w_{p}(r_{p})$. $w_{p}(r_{p})$ is defined by first considering the spatial two-point correlation function along the line of sight ($r_{l}$) and perpendicular to the line of sight ($r_{p}$):
\begin{equation}\label{eq:xi_rp}
\xi(r_{p},r_{l}) = 1 + \left(\frac{N_R}{N_D}\right)^{2}\frac{DD(r_{p},r_{l})}{RR(r_{p},r_{l})} - 2\frac{N_{R}}{N_{D}}\frac{DR(r_{p},r_{l})}{RR(r_{p},r_{l})}.
\end{equation}\\
$\xi(r_{p},r_{l})$ is then integrated over $r_{l}$ to obtain $w_{p}(r_{p})$:
\begin{equation}\label{eq:wp_rp}
w_{p}(r_{p}) = 2\int_{0}^{r_{l,\rm{max}}}\xi(r_{p},r_{l})dr_{l}.
\end{equation}\\
This is related to the real-space correlation function by:
\begin{equation}\label{eq:pi_int}
w_{p}(r_{p}) = 2\int_{r_{p}}^{+\infty} \frac{r\xi(r)}{(r^{2}-r_{p}^{2})^{1/2}}dr
\end{equation}
in the limit of a wide filter, and the solution tends to:
\begin{equation}
w_{p}(r_{p}) = r_{p}\Bigg(\frac{r_{p}}{r_{0}}\Bigg)^{-\gamma}\frac{\Gamma(\gamma/2 - 1/2)\Gamma(1/2)}{\Gamma(\gamma/2)}.
\end{equation}
In the case of a narrower top-hat filter, we integrate over a finite range of $r_{l}$, using $(r_{p}^{2}+r_{l,\rm{max}}^{2})^{1/2}$ as the upper limit to the integral in Equation \ref{eq:pi_int}. \\\\
In this paper, we calculate $w_{p}(r_{p})$ from our observed $w(\theta)$. However, our filter profiles are not top-hat (as assumed for the integral in Equation \ref{eq:pi_int}) but are better approximated by Gaussian profiles (see Table \ref{table:gauss_params} for parameters). To account for this difference, we perform numerical integrations to determine the factor by which $w(\theta)$ differs (for a given $\xi(r)$) if observed over a top-hat of width $2\sigma$ as opposed to a Gaussian of width $\sigma$ (changing $f_{s}$ in Equation \ref{eq:filter_integration}); we find a required correction of $\sqrt{\pi}$. Using this, and combining Equations~\ref{eq:w_theta},~\ref{eq:xi_rp}, ~\ref{eq:wp_rp}, with $r_{p} = \rm{D}_{ang}\theta(\rm{rad})$, we then obtain: 
\begin{equation}
w_{p}(r_{p}) \sim 2\sigma\sqrt{\pi}\,w\Bigg(\theta =  \frac{r_{p}}{\rm{D}_{ang}}\Bigg)\,(1+z)^{0.8}.
\end{equation}
\begin{table}
\begin{center}
\begin{tabular}{l|c|c|c|c}
Redshift & $\mu(h^{-1}\rm{Mpc})$ & $\sigma(h^{-1}\rm{Mpc})$\\
\hline
$0.81\:\ensuremath{\pm}\:$0.011  & 1970 & 14 \\
$1.47\:\ensuremath{\pm}\:$0.016  & 3010 & 18 \\
$2.23\:\ensuremath{\pm}\:$0.016  & 3847 & 18 \\
\end{tabular}
\caption{Parameters of \textcolor{black}{gaussian} filter profile fits for the three HiZELS redshifts studied.}
\label{table:gauss_params}
\end{center}
\end{table}
\vspace{-0.5cm}
\section{Results from power-law fits to the angular correlation function}\label{sec:r0_results}
\begin{figure*} 
	\centering
	\includegraphics[scale=0.5]{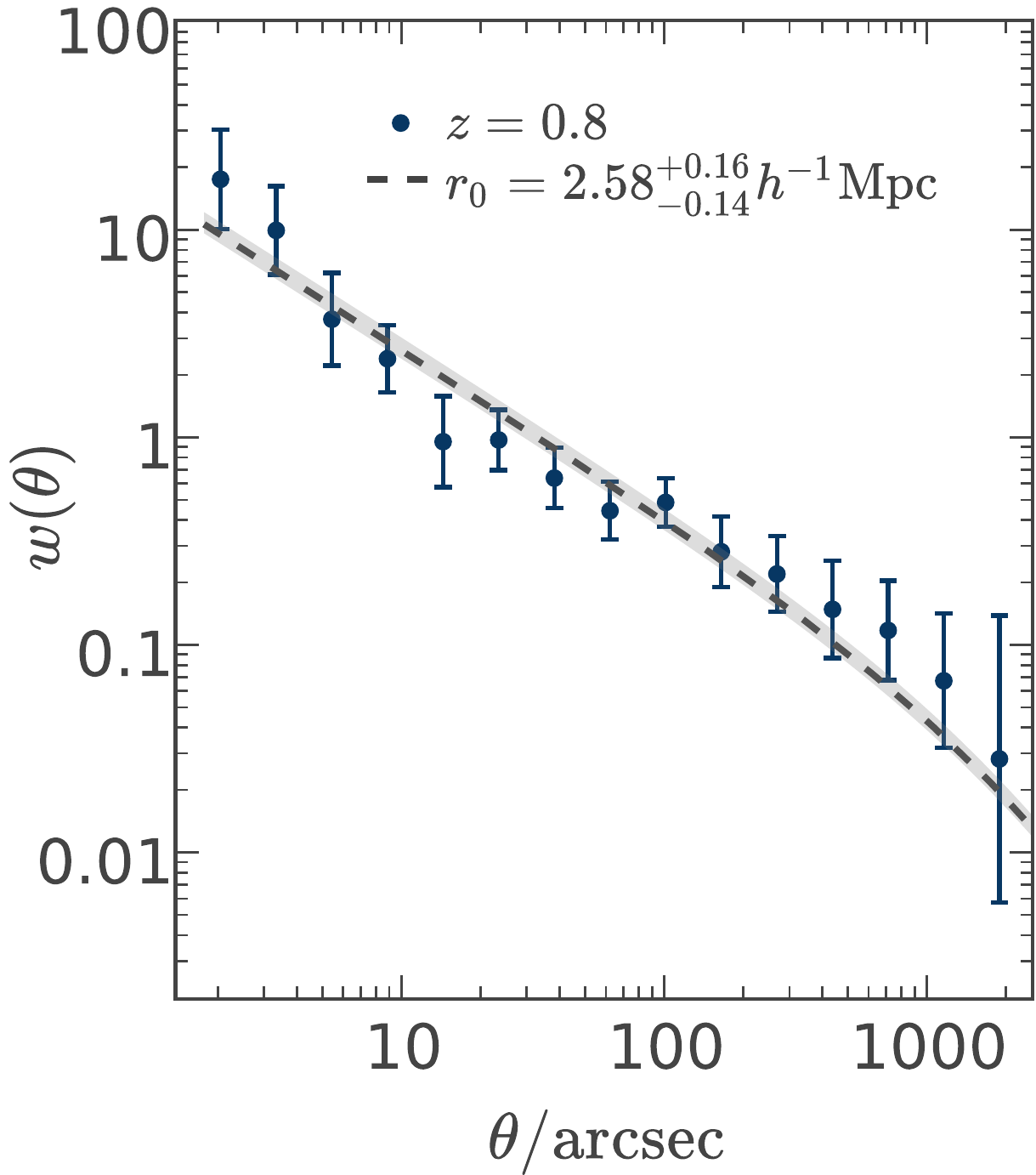}		
	\includegraphics[scale=0.5]{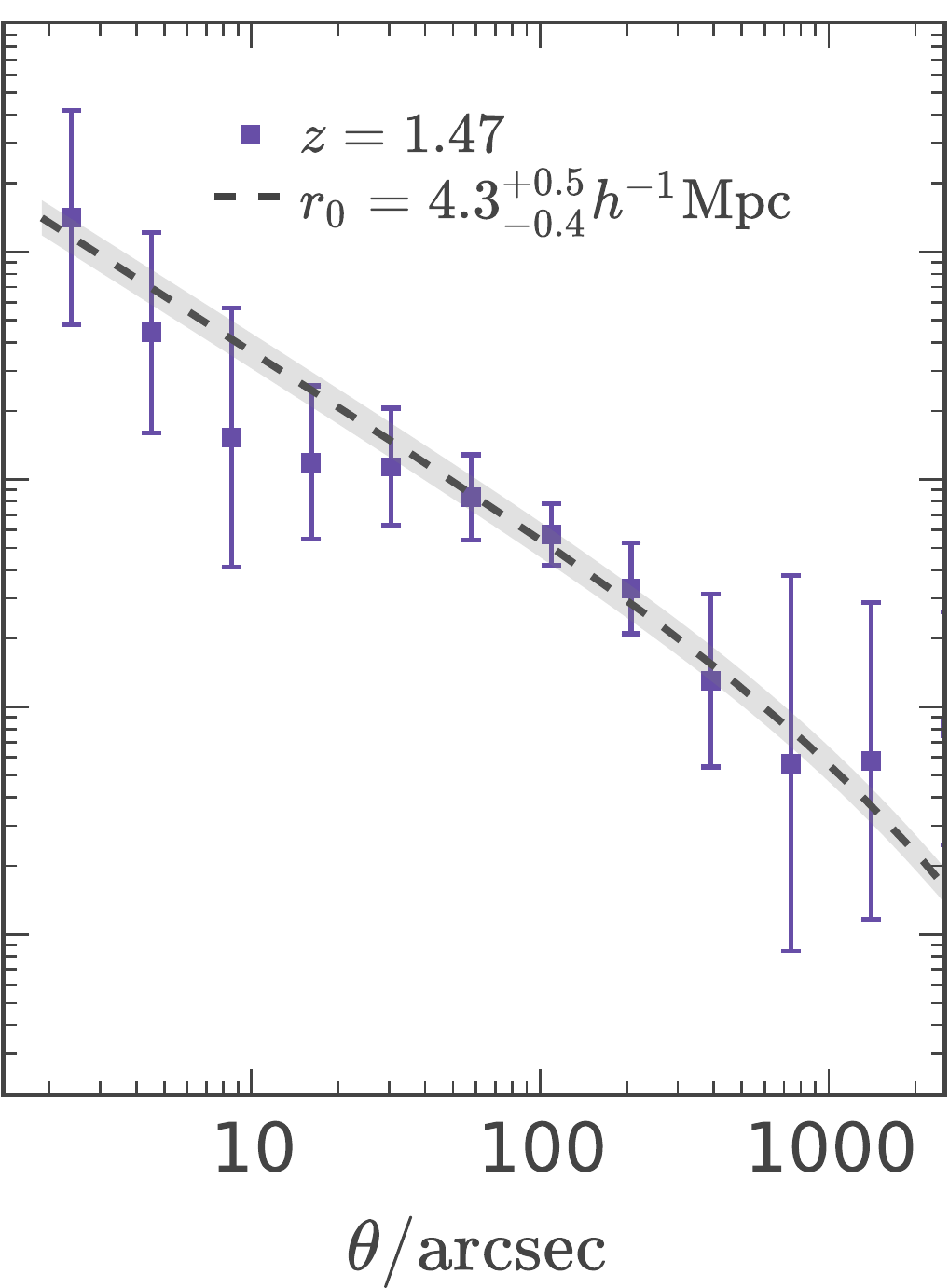}	
	\includegraphics[scale=0.5]{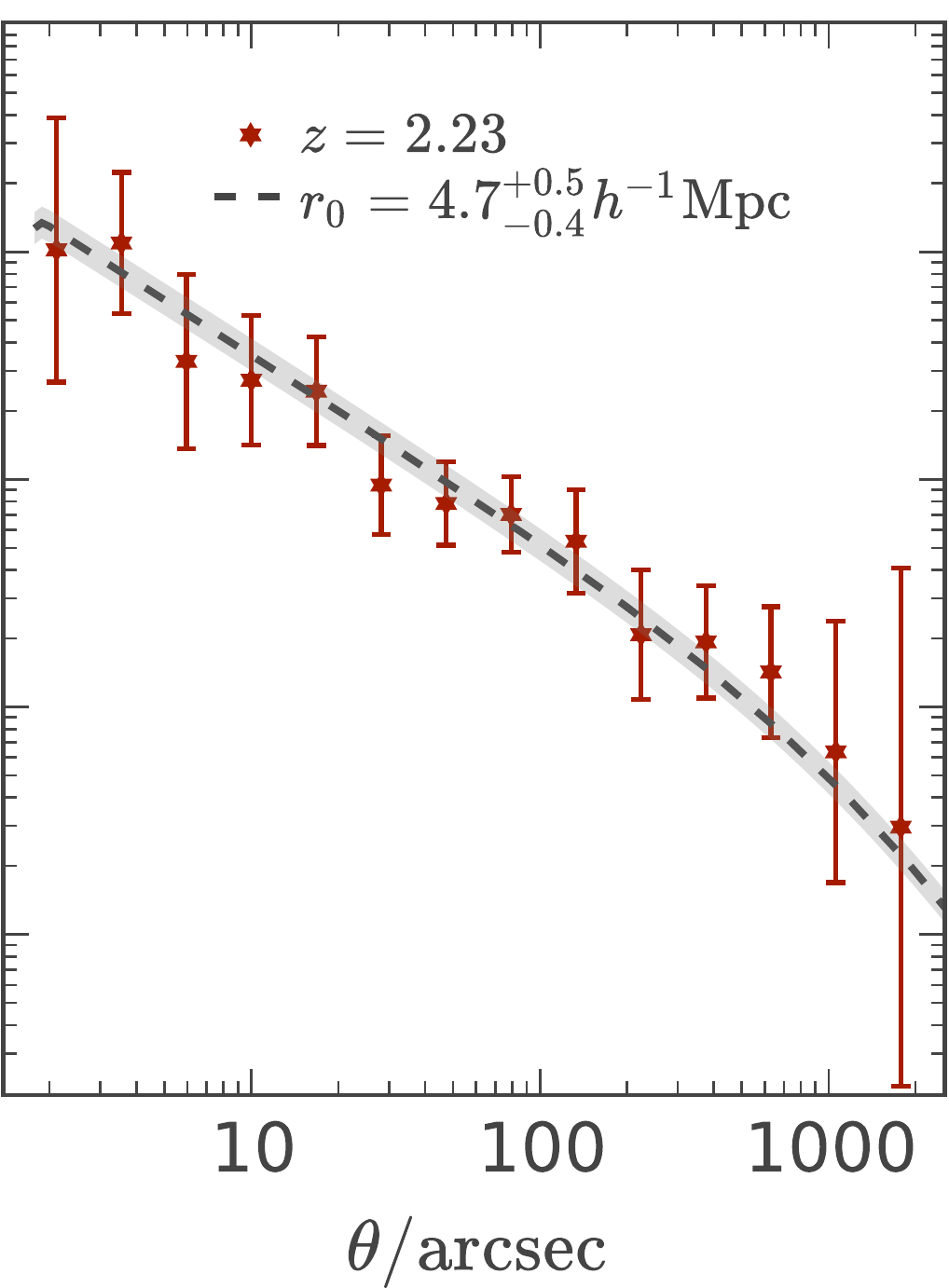}
	\includegraphics[scale=0.5]{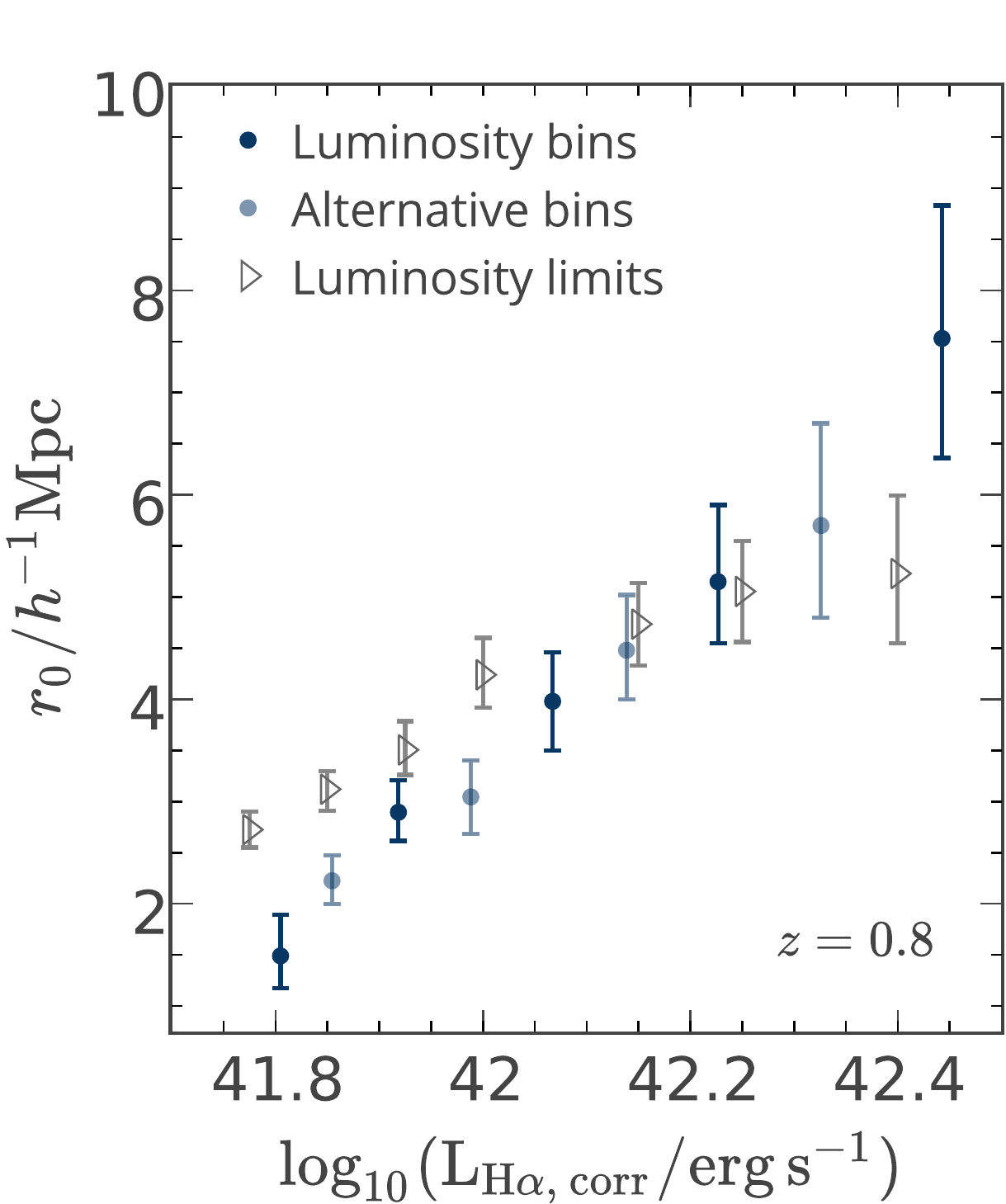}
	\includegraphics[scale=0.5]{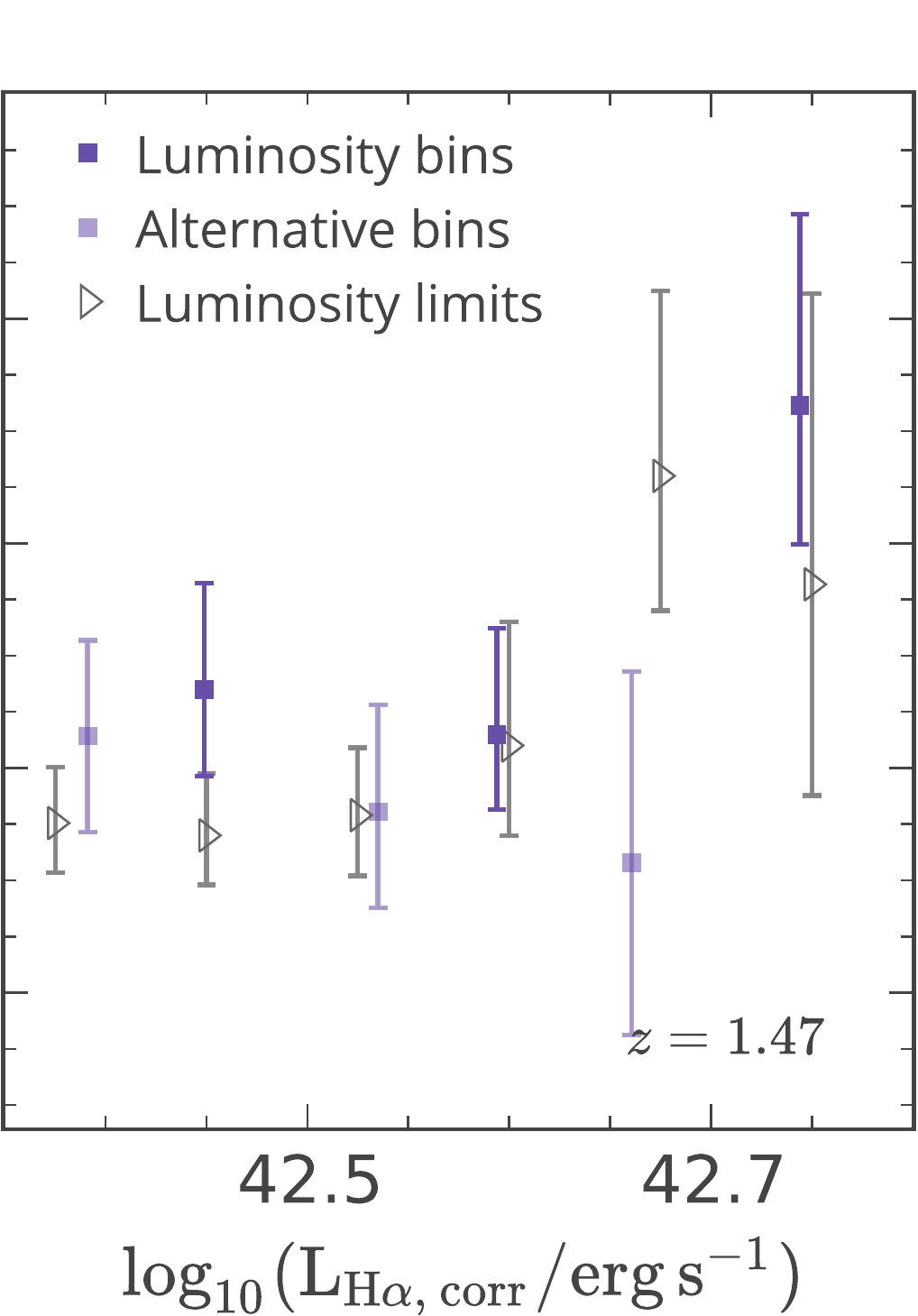}
	\includegraphics[scale=0.5]{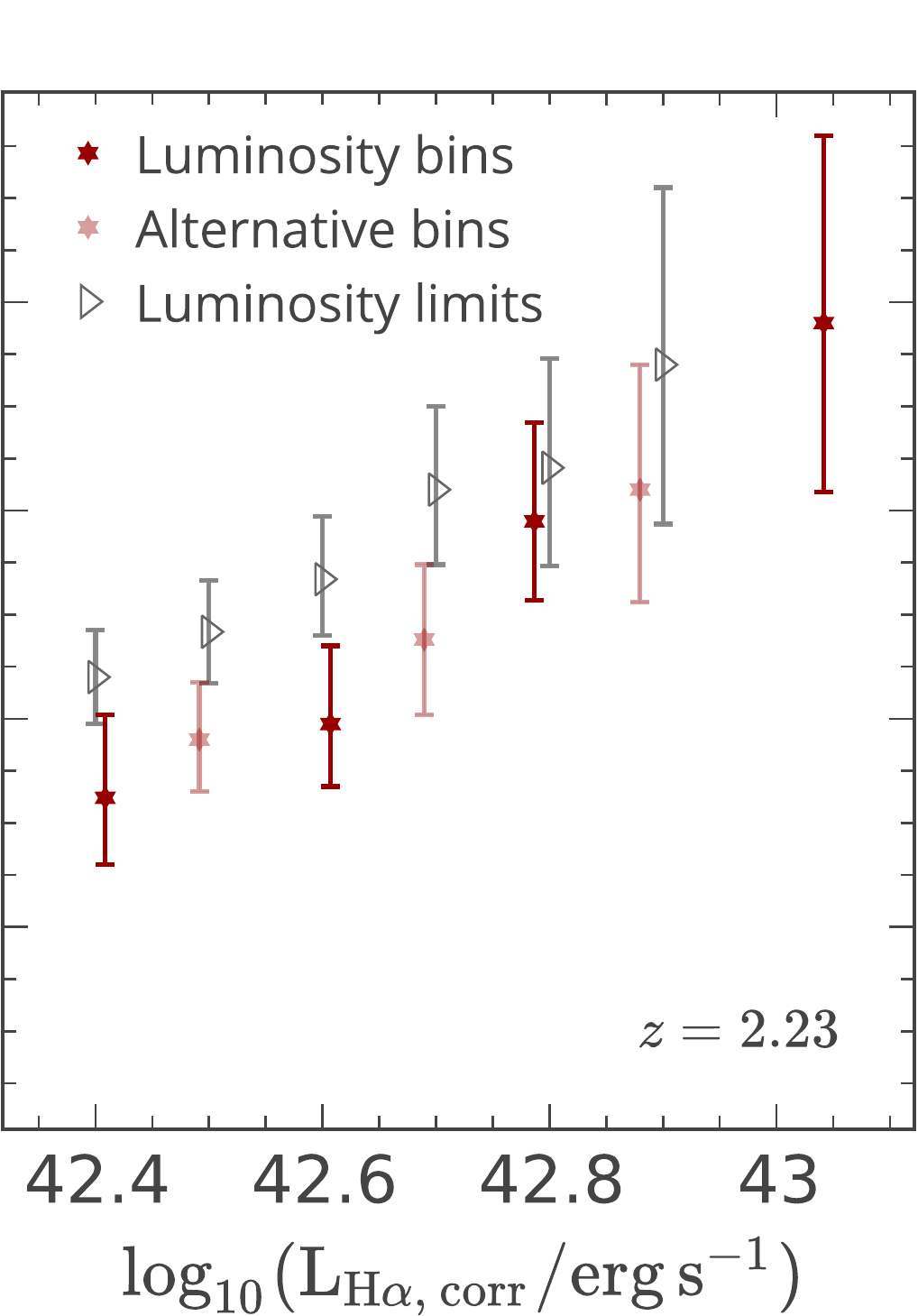}

	\caption{Top: power law fits (with the correction to Limber's approximation at large scales) to the measured angular correlation functions at three redshifts, each over the same range in $L_{\rm{H}\alpha}/L^{*}_{\rm{H}\alpha}$. Bottom: derived clustering strength, $r_{0}$, for $\rm{H}\alpha$ luminosity-binned and luminosity-limited samples. We also show alternative binning. The plotted luminosity value is the mean value of $\log_{10}(L_{\rm{H}\alpha})$ for the luminosity-binned samples, and the lower limit for the luminosity-limited samples. The clustering strength increases with $\log_{10}L_{\rm{H}\alpha}$ for all three redshifts surveyed in a broadly linear manner.}
    \label{fig:corr_z}
\end{figure*}
\subsection{Whole samples at well-defined redshifts}\label{sec:whole_samples_z}
We have derived angular correlation functions for large samples of $\rm{H}\alpha$  emitters at each redshift and fitted these with power-law models (see Figure \ref{fig:corr_z}). The exact luminosity ranges of these samples, given in Table \ref{table:r0_table}, are chosen to compare similar samples at each redshift, and span the same range in $L_{\rm{H}\alpha}/L_{\rm{H}\alpha}^{*}$: $-0.4<\log_{10}(L_{\rm{H}\alpha}/L_{\rm{H}\alpha}^{*})<0.3$ \textcolor{black}{(albeit with non-matched distributions within this range)}. The fits shown are those described in Section \ref{sec:correlation_length}, with a power-law of fixed gradient $-1.8$ for the spatial correlation function, leading to a slope of $-0.8$ in the angular correlation function on small scales and the correction to Limber's approximation at large scales where the angular separation is much greater than the separation along the line of sight. This parameterisation is sufficient to derive indicative clustering strengths. However, the correlation functions of all three samples do show clear departures from the traditional power-law relation fitted here. At angular scales of order 10s of arcseconds the power-law fit consistently overestimates the observed $w(\theta)$, indicative of a dominant contribution from a separate 1-halo term at small angular separations. We explore this further in Section \ref{sec:halo_fitting}.
\subsection{Clustering strength as a function of galaxy H$\alpha$ luminosity}\label{sec:r0_lum_dependence}
\begin{figure} 
	\centering{}
	\includegraphics[scale=0.5]{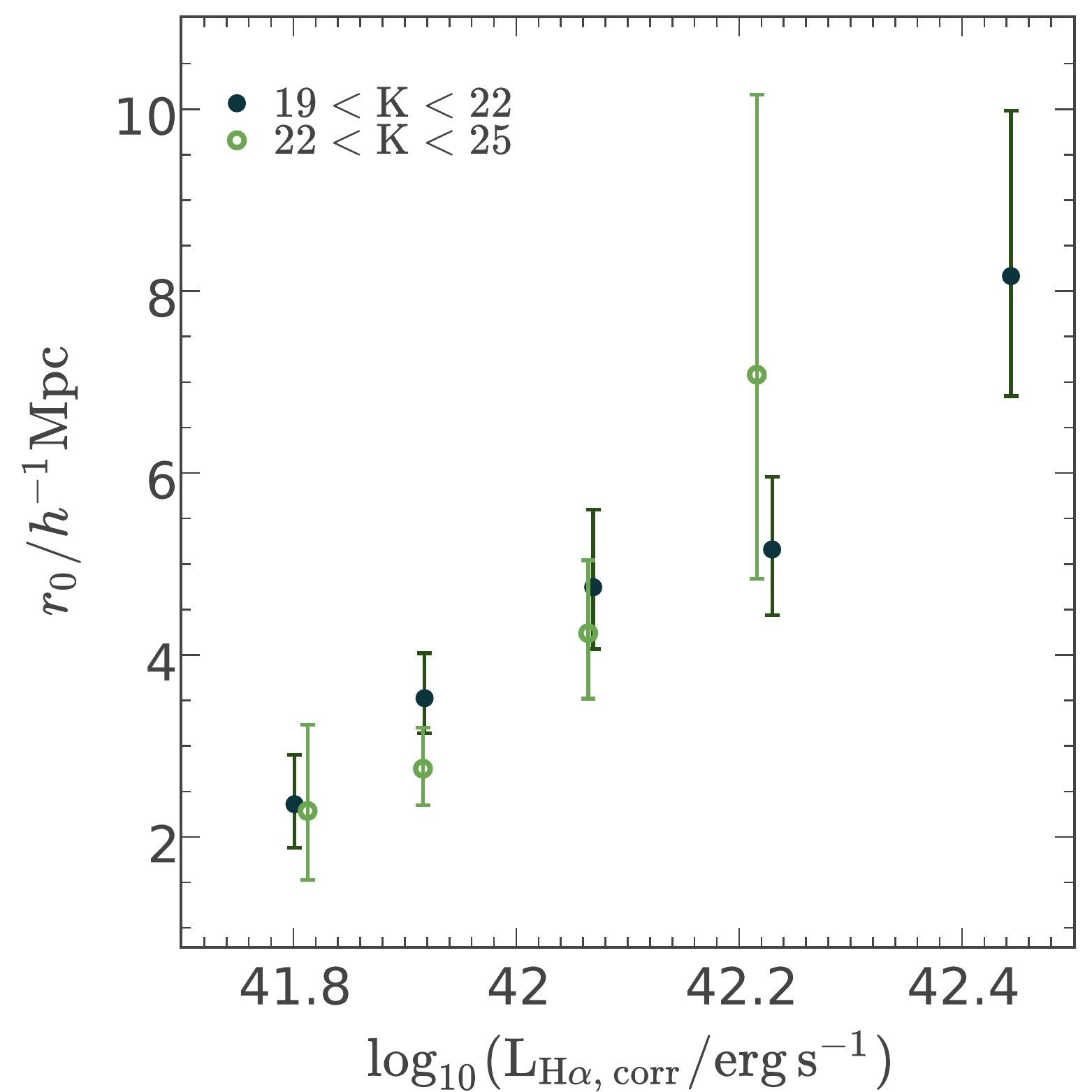}	
	\caption{\textcolor{black}{To investigate whether trends with $L_{\rm{H}\alpha}$ are driven by stellar mass, we plot $\rm{r}_{0}$ against $L_{\rm{H}\alpha}$ for observed K-band magnitude-selected subsamples of the $z=0.8$ HiZELS emitters. We find that the strong trends of clustering strength with $\rm{H}\alpha$ luminosity hold for these subsamples. This indicates that trends with $L_{\rm{H}\alpha}$ are not driven primarily by stellar mass.}}
    \label{fig:k_CORRS_lum}
\end{figure}

We have fitted both luminosity-binned data and luminosity-limited data with the same power-law models (see Table \ref{table:r0_table} \textcolor{black}{and Appendix \ref{sec:appendixA}}). As shown in the lower panels of Figure \ref{fig:corr_z}, the clustering strength, $r_{0}$, increases roughly linearly with galaxy $\rm{H}\alpha$ luminosity for the luminosity-binned samples, showing that more highly star-forming galaxies are more strongly clustered, and hence may live in more massive dark matter halo environments. The trends are similar for the luminosity-limited samples: these also show an increase in clustering strength with galaxy luminosity. The results for the two sample types do not agree exactly because luminosity-limited samples of galaxies with faint limits have their clustering increased by the inclusion of a small number of bright sources, and therefore have a greater clustering strength than that of galaxies entirely within a faint luminosity bin.\\
\indent Although the absolute values of $r_{0}$ agree (within errors) with the previous HiZELS study of a smaller sample of $\rm{H}\alpha$ emitters at $z=0.8$, the apparently linear relationship is at odds with the results of \cite{Sobral2010}, who found tentative hints of a more step-like behaviour around the characteristic luminosity. With our much larger sample of $\sim3000$ emitters, there is no longer evidence for a break in the $r_{0}$ vs $\log_{10}(L_{\rm{H}\alpha})$ relationship, and a linear relation provides a far better fit. The trends at $z=1.47$ and $z=2.23$ also show no clear departure from a simple linear trend, albeit that the $z\sim1.47$ results are noisier. \textcolor{black}{These results are also broadly consistent with previous studies. We find $r_{0} = 4.3^{+0.5}_{-0.4}h^{-1}\rm{Mpc}$
\onecolumn
\begin{table}
\begin{center}
\begin{tabular}{l|c|c|c|c|c|c|c|c}
Redshift & $\log_{10}(L_{\rm{H}\alpha}$ range/$\rm{erg\,s}^{-1})$ & Mean $\log_{10}(L_{\rm{H}\alpha})$ & $\rm{r}_{0}/h^{-1}\rm{Mpc}$ & $\rm{b}_{\rm{eff}}$ & $\log_{10}(M_{\rm{eff}}/M_{\odot})$ & $\log_{10}(M_{\rm{min}}/M_{\odot})$ & $f_{\rm{sat}}$ \\
\hline
\multicolumn{2}{l}{\bf{`Full' samples:}$-0.4<\log_{10}(L_{\rm{H}\alpha}/L_{\rm{H}\alpha}^{*})<0.3$} \\
0.8 & 41.72-42.42 & $41.96$ & $2.6^{+0.2}_{-0.1}$ & $1.12^{+0.06}_{-0.05}$ &$12.13^{+0.10}_{-0.09}$ &$11.12^{+0.11}_{-0.15}$ & $0.05^{+0.01}_{-0.01}$\\
1.47 & 42.16-42.86 & $42.52$ & $4.3^{+0.5}_{-0.4}$ & $1.78^{+0.06}_{-0.08}$ &$12.16^{+0.07}_{-0.09}$ &$11.45^{+0.06}_{-0.08}$ & $0.05^{+0.02}_{-0.02}$\\
2.23 & 42.47-43.17 & $42.71$ & $4.7^{+0.5}_{-0.4}$ & $2.52^{+0.07}_{-0.09}$ &$11.96^{+0.05}_{-0.07}$ &$11.41^{+0.06}_{-0.06}$ & $0.05^{+0.02}_{-0.02}$\\
\hline
\multicolumn{2}{l}{$\mathbf{\rm{H}\alpha}$ \bf{Luminosity-selected subsamples}} \\
& \bf{Bins} & & & & & \\
0.8 & 41.7-41.85 & $41.80$ & $1.5^{+0.4}_{-0.3}$ &$1.07^{+0.09}_{-0.03}$ &$11.92^{+0.14}_{-0.04}$ &$11.26^{+0.13}_{-0.08}$ & $0.03^{+0.01}_{-0.01}$\\
0.8 & 41.775-41.925 & $41.85$ & $2.2^{+0.3}_{-0.2}$ &$1.16^{+0.06}_{-0.03}$ &$12.01^{+0.14}_{-0.06}$ &$11.46^{+0.08}_{-0.07}$ & $0.02^{+0.01}_{-0.01}$\\
0.8 & 41.85-42.0 & $41.92$ & $2.9^{+0.3}_{-0.3}$ & $1.33^{+0.07}_{-0.06}$ &$12.31^{+0.14}_{-0.11}$ &$11.69^{+0.07}_{-0.08}$ & $0.02^{+0.01}_{-0.01}$\\
0.8 & 41.925-42.075 & $41.99$ & $3.1^{+0.4}_{-0.4}$ & $1.32^{+0.09}_{-0.07}$ &$12.30^{+0.16}_{-0.12}$ &$11.71^{+0.10}_{-0.10}$ & $0.02^{+0.01}_{-0.01}$\\
0.8 & 42.0-42.15 & $42.07$ & $4.0^{+0.5}_{-0.5}$ & $1.49^{+0.08}_{-0.09}$ &$12.55^{+0.11}_{-0.14}$ &$11.91^{+0.08}_{-0.12}$ & $0.02^{+0.01}_{-0.01}$\\
0.8 & 42.075-42.25 & $42.14$ & $4.5^{+0.5}_{-0.5}$ & $1.56^{+0.07}_{-0.09}$ &$12.63^{+0.08}_{-0.12}$ &$12.01^{+0.07}_{-0.09}$ & $0.02^{+0.01}_{-0.01}$\\
0.8 & 42.15-42.35 & $42.23$ & $5.2^{+0.8}_{-0.6}$ & $1.63^{+0.08}_{-0.11}$ &$12.71^{+0.09}_{-0.13}$ &$12.09^{+0.07}_{-0.12}$ & $0.03^{+0.02}_{-0.01}$\\
0.8 & 42.25-42.475 & $42.33$ & $5.7^{+1.0}_{-0.9}$ & $1.79^{+0.08}_{-0.12}$ &$12.86^{+0.07}_{-0.12}$ &$12.28^{+0.07}_{-0.11}$ & $0.03^{+0.02}_{-0.01}$\\
0.8 & 42.35-42.6 & $42.44$ & $7.5^{+1.3}_{-1.2}$ & $2.02^{+0.08}_{-0.13}$ &$13.05^{+0.06}_{-0.10}$ &$12.50^{+0.06}_{-0.09}$ & $0.05^{+0.03}_{-0.02}$\\
& \bf{Limits} & & & & &\\
0.8 & >41.775 & $41.99$ & $2.7^{+0.2}_{-0.2}$ & $1.18^{+0.06}_{-0.06}$ &$12.18^{+0.10}_{-0.11}$ &$11.24^{+0.10}_{-0.14}$ & $0.04^{+0.01}_{-0.01}$\\
0.8 & >41.85 & $42.03$ & $3.1^{+0.2}_{-0.2}$ & $1.26^{+0.05}_{-0.06}$ &$12.28^{+0.08}_{-0.11}$ &$11.40^{+0.08}_{-0.11}$ & $0.04^{+0.01}_{-0.01}$\\
0.8 & >41.925 & $42.10$ & $3.5^{+0.3}_{-0.3}$ & $1.32^{+0.05}_{-0.07}$  &$12.38^{+0.08}_{-0.11}$ &$11.54^{+0.07}_{-0.10}$ & $0.04^{+0.01}_{-0.01}$\\
0.8 & >42.0 & $42.18$ & $4.2^{+0.4}_{-0.3}$ & $1.41^{+0.05}_{-0.07}$ &$12.52^{+0.07}_{-0.10}$ &$11.69^{+0.07}_{-0.09}$& $0.05^{+0.02}_{-0.02}$\\
0.8 & >42.075 & $42.25$ & $4.7^{+0.4}_{-0.4}$ & $1.48^{+0.05}_{-0.07}$  &$12.57^{+0.06}_{-0.09}$ &$11.84^{+0.07}_{-0.09}$ & $0.03^{+0.02}_{-0.01}$\\
0.8 & >42.15 & $42.33$ & $5.1^{+0.5}_{-0.5}$ & $1.57^{+0.06}_{-0.08}$ &$12.66^{+0.07}_{-0.10}$ &$11.98^{+0.07}_{-0.09}$ & $0.03^{+0.02}_{-0.01}$\\
0.8 & >42.25 & $42.44$ & $5.2^{+0.8}_{-0.7}$ & $1.71^{+0.07}_{-0.11}$  &$12.79^{+0.07}_{-0.12}$ &$12.18^{+0.07}_{-0.10}$ & $0.03^{+0.02}_{-0.01}$\\
0.8 & >42.4 & $42.57$ & $5.8^{+1.5}_{-1.3}$ & $2.00^{+0.11}_{-0.17}$  &$13.03^{+0.09}_{-0.14}$ &$12.53^{+0.07}_{-0.13}$ & $0.03^{+0.02}_{-0.01}$\\
& & & & & &\\
& \bf{Bins} & & & & &\\
1.47 & 42.3-42.45 & $42.39$ & $4.3^{+0.9}_{-0.8}$ & $2.12^{+0.09}_{-0.13}$  &$12.4^{+0.07}_{-0.12}$ &$11.88^{+0.06}_{-0.10}$ & $0.02^{+0.01}_{-0.01}$\\
1.47 & 42.375-42.525 & $42.45$ & $4.7^{+1.0}_{-0.9}$ & $2.21^{+0.09}_{-0.13}$ &$12.48^{+0.06}_{-0.11}$ &$11.97^{+0.06}_{-0.09}$ & $0.02^{+0.01}_{-0.01}$\\
1.47 & 42.45-42.6 & $42.53$ & $3.6^{+1.0}_{-0.7}$ & $2.12^{+0.16}_{-0.15}$  &$12.39^{+0.14}_{-0.14}$ &$11.97^{+0.09}_{-0.11}$ & $0.01^{+0.01}_{-0.01}$\\
1.47 & 42.525-42.675 & $42.59$ & $4.3^{+1.0}_{-0.9}$ & $2.21^{+0.20}_{-0.18}$ &$12.46^{+0.15}_{-0.15}$ &$12.04^{+0.10}_{-0.16}$ & $0.03^{+0.01}_{-0.01}$\\
1.47 & 42.6-42.75 & $42.66$ & $3.2^{+1.7}_{-1.2}$ & $2.28^{+0.22}_{-0.18}$  &$12.50^{+0.16}_{-0.15}$ &$12.12^{+0.11}_{-0.13}$ & $0.02^{+0.01}_{-0.01}$\\
1.47 & 42.675-42.85 & $42.74$ & $7.2^{+1.7}_{-1.5}$ & $2.72^{+0.13}_{-0.20}$ &$12.79^{+0.07}_{-0.11}$ &$12.34^{+0.07}_{-0.12}$ & $0.07^{+0.06}_{-0.04}$\\
1.47 & 42.75-43.3 & $42.87$ & $6.8^{+2.6}_{-2.2}$ & $2.67^{+0.16}_{-0.26}$ &$12.76^{+0.09}_{-0.16}$ &$12.33^{+0.07}_{-0.15}$ & $0.04^{+0.03}_{-0.02}$\\
& \bf{Limits} & & & & &\\
1.47 & >42.2 & $42.55$ & $3.7^{+0.5}_{-0.4}$ & $1.85^{+0.07}_{-0.11}$  &$12.18^{+0.08}_{-0.12}$ &$11.58^{+0.06}_{-0.10}$ & $0.02^{+0.01}_{-0.01}$\\
1.47 & >42.375 & $42.59$ & $3.5^{+0.5}_{-0.4}$ & $1.90^{+0.09}_{-0.12}$  &$12.22^{+0.09}_{-0.13}$ &$11.67^{+0.07}_{-0.10}$ & $0.02^{+0.01}_{-0.01}$\\
1.47 & >42.45 & $42.63$ & $3.4^{+0.6}_{-0.4}$ & $1.90^{+0.16}_{-0.13}$  &$12.21^{+0.16}_{-0.15}$ &$11.71^{+0.09}_{-0.13}$ & $0.02^{+0.01}_{-0.01}$\\
1.47 & >42.525 & $42.66$ & $3.6^{+0.6}_{-0.5}$ & $2.07^{+0.12}_{-0.14}$  &$12.36^{+0.11}_{-0.13}$ &$11.88^{+0.08}_{-0.10}$ & $0.02^{+0.01}_{-0.01}$\\
1.47 & >42.6 & $42.73$ & $4.2^{+1.1}_{-0.8}$ & $2.22^{+0.14}_{-0.17}$  &$12.47^{+0.11}_{-0.14}$ &$12.02^{+0.08}_{-0.12}$ & $0.02^{+0.01}_{-0.01}$\\
1.47 & >42.675 & $42.80$ & $6.6^{+1.7}_{-1.2}$ & $2.56^{+0.12}_{-0.18}$  &$12.71^{+0.07}_{-0.11}$ &$12.22^{+0.07}_{-0.11}$ & $0.07^{+0.06}_{-0.04}$\\
1.47 & >42.75 & $42.87$ & $5.6^{+2.6}_{-1.9}$ & $2.69^{+0.14}_{-0.25}$  &$12.77^{+0.08}_{-0.15}$ &$12.34^{+0.07}_{-0.14}$ & $0.03^{+0.03}_{-0.02}$\\
& & & & & &\\
& \bf{Bins} & & & & &\\
2.23 & 42.2-42.5 & $42.41$ & $3.2^{+0.8}_{-0.6}$ & $2.30^{+0.11}_{-0.17}$ &$11.79^{+0.09}_{-0.15}$ &$11.25^{+0.07}_{-0.13}$ & $0.03^{+0.02}_{-0.01}$\\
2.23 & 42.35-42.6 & $42.49$ & $3.8^{+0.6}_{-0.5}$ & $2.50^{+0.10}_{-0.15}$ &$11.93^{+0.07}_{-0.12}$ &$11.43^{+0.06}_{-0.09}$ & $0.02^{+0.01}_{-0.01}$\\
2.23 & 42.5-42.7 & $42.61$ & $4.0^{+0.8}_{-0.6}$ & $2.67^{+0.13}_{-0.20}$ &$12.03^{+0.09}_{-0.14}$ &$11.58^{+0.07}_{-0.12}$ & $0.02^{+0.01}_{-0.01}$\\
2.23 & 42.6-42.8 & $42.69$ & $4.8^{+0.7}_{-0.7}$ & $2.87^{+0.09}_{-0.16}$ &$12.14^{+0.06}_{-0.10}$ &$11.70^{+0.05}_{-0.09}$ & $0.02^{+0.01}_{-0.01}$\\
2.23 & 42.7-42.9 & $42.79$ & $5.9^{+1.0}_{-0.8}$ & $3.05^{+0.09}_{-0.14}$ &$12.24^{+0.05}_{-0.08}$ &$11.81^{+0.05}_{-0.07}$ & $0.02^{+0.01}_{-0.01}$\\
2.23 & 42.8-43.0 & $42.88$ & $6.2^{+1.2}_{-1.1}$ & $3.23^{+0.10}_{-0.16}$ &$12.33^{+0.05}_{-0.08}$ &$11.92^{+0.05}_{-0.07}$ & $0.03^{+0.02}_{-0.01}$\\
2.23 & 42.9-43.6  &$43.04$ &  $7.8^{+1.8}_{-1.6}$ & $3.23^{+0.11}_{-0.19}$  &$12.35^{+0.05}_{-0.09}$ &$11.93^{+0.05}_{-0.08}$ & $0.05^{+0.03}_{-0.03}$\\
& \bf{Limits} & & & & &\\
2.23 & >42.2 & $42.66$ & $4.3^{+0.5}_{-0.4}$ & $2.23^{+0.07}_{-0.09}$ &$11.77^{+0.06}_{-0.08}$ &$11.12^{+0.06}_{-0.07}$ & $0.07^{+0.02}_{-0.02}$\\
2.23 & >42.3 & $42.66$ & $4.3^{+0.5}_{-0.4}$ & $2.29^{+0.07}_{-0.09}$ &$11.80^{+0.05}_{-0.08}$ &$11.21^{+0.06}_{-0.07}$ & $0.04^{+0.02}_{-0.01}$\\
2.23 & >42.4 & $42.69$ & $4.4^{+0.5}_{-0.5}$ & $2.39^{+0.07}_{-0.11}$ &$11.86^{+0.06}_{-0.10}$ &$11.30^{+0.06}_{-0.07}$ & $0.04^{+0.01}_{-0.01}$\\
2.23 & >42.5 & $42.73$ & $4.8^{+0.5}_{-0.5}$ & $2.52^{+0.07}_{-0.10}$ &$11.95^{+0.05}_{-0.08}$ &$11.42^{+0.05}_{-0.07}$ & $0.04^{+0.02}_{-0.01}$\\
2.23 & >42.6 & $42.78$ & $5.3^{+0.6}_{-0.5}$ & $2.65^{+0.08}_{-0.12}$ &$12.03^{+0.05}_{-0.09}$ &$11.53^{+0.05}_{-0.07}$ & $0.04^{+0.01}_{-0.01}$\\
2.23 & >42.7 & $42.87$ & $6.2^{+0.8}_{-0.7}$ & $2.83^{+0.08}_{-0.12}$ &$12.14^{+0.05}_{-0.07}$ &$11.66^{+0.05}_{-0.07}$ & $0.04^{+0.02}_{-0.02}$\\
2.23 & >42.8 & $42.95$ & $6.4^{+1.1}_{-1.0}$ & $3.03^{+0.09}_{-0.15}$ &$12.24^{+0.05}_{-0.09}$ &$11.79^{+0.05}_{-0.07}$ & $0.04^{+0.02}_{-0.02}$\\
2.23 & >42.9 & $43.06$ & $7.4^{+1.7}_{-1.5}$ &$3.26^{+0.12}_{-0.19}$ &$12.35^{+0.06}_{-0.10}$ &$11.93^{+0.05}_{-0.08}$ & $0.05^{+0.04}_{-0.03}$\\
\end{tabular}
\caption{$r_{0}$ values and key parameters derived from HOD fitting, for samples of $\rm{H}\alpha$ emitters at different redshifts and luminosities. We find a clear trend towards increasing $r_{0}$, $b_{\rm{eff}}$, $M_{\rm{min}}$ and $M_{\rm{eff}}$ for samples of galaxies with higher $\rm{H}\alpha$ luminosities at all redshifts, but little evidence for changing satellite fractions for these SFR-selected samples.}
\label{table:r0_table}
\end{center}
\end{table}
\twocolumn
\noindent for our sample at $z=1.47$, while \cite{Kashino2017} obtain $r_{0} = 5.2\pm0.7h^{-1}\rm{Mpc}$ for $\rm{H}\alpha$ emitters at $1.43\le z \le1.74$.} We find $r_{0}=4.7^{+0.5}_{-0.4}h^{-1}\rm{Mpc}$ for the full sample at $z=2.23$, which is slightly higher than \cite{Geach2012} found using a smaller sample at the same redshift ($r_{0}=3.7\pm0.3$), but this depends critically on the luminosity range studied. \\
\indent \textcolor{black}{In Figure \ref{fig:k_CORRS_lum}, we show the $\rm{H}\alpha$ luminosity-dependent clustering of $z=0.8$ HiZELS emitters split into two observed K-band magnitude bins. Observed K-band magnitude is believed to be a rough proxy for galaxy stellar mass. We find that the clustering strength increases broadly linearly with $\log_{10}(L_{\rm{H}\alpha})$ within each of the broad K-band magnitude bins, and that this trend is much larger than any differences between the two K-band magnitude bins. We will explore the stellar mass-dependence of the clustering of star-forming galaxies more thoroughly in a subsequent paper, but we stress here that the strong trends of clustering strength with $\rm{H}\alpha$ luminosity presented in this paper are not driven primarily by galaxy stellar mass.}
\begin{figure*}
	\centering
	\includegraphics[scale=0.55]{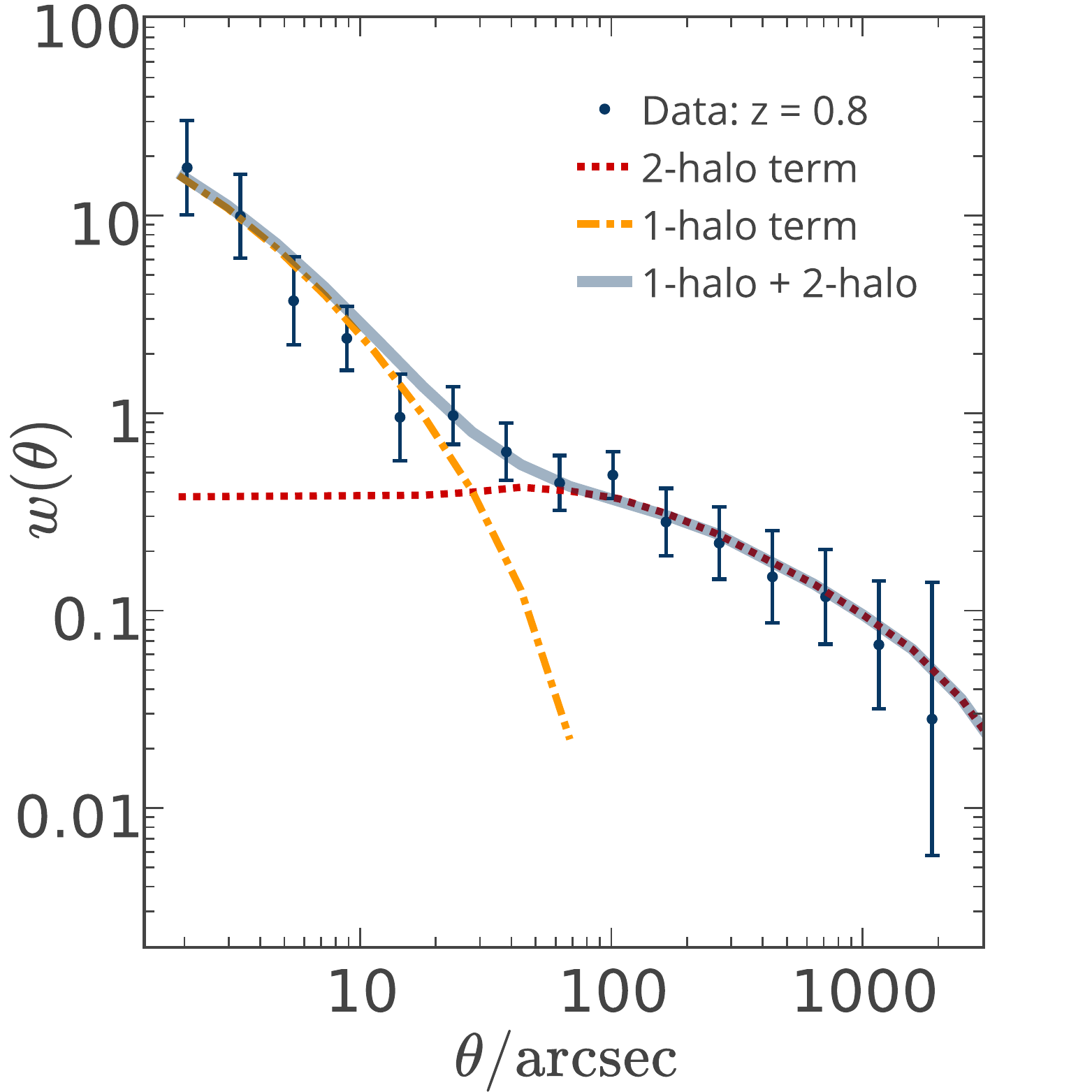}	
	\includegraphics[scale=0.55]{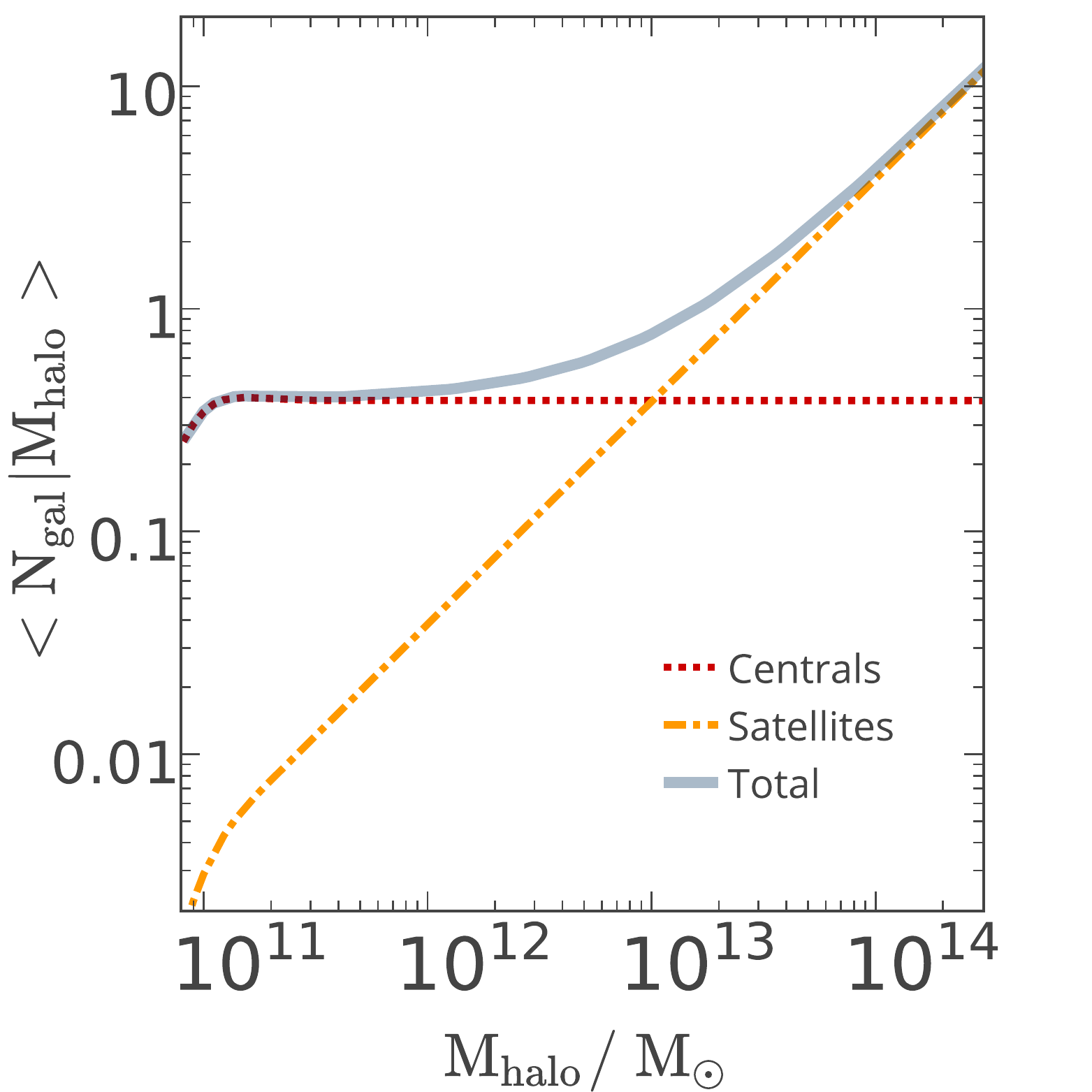}	
	\caption{Left: \textcolor{black}{Halo Occupation Model} fit to the correlation function of the whole $z=0.8$ sample \textcolor{black}{using HALOMOD.} This multi-parameter model provides a better fit to data than the single power law model and shows the separate contributions of satellite and central galaxies. Right: the best-fitting halo occupation distribution model. \textcolor{black}{The contribution from satellite galaxies becomes significant only in halos more massive than $\sim10^{13}M_{\odot}$.}}
	\label{fig:hod_whole_J}
\end{figure*}
\section{Modelling galaxy populations via Halo Occupation Distribution fitting}\label{sec:halo_fitting}
The Halo Occupation Distribution (HOD) formalism extends dark matter halo models to galaxy populations: given a set of input parameters, we can predict the average number of galaxies of a certain type as a function of dark matter halo mass, $\langle N|M\rangle$. A combination of a cosmological model and an HOD enables us to predict any clustering statistic on any scale; usually observations of galaxy clustering (or weak lensing) are then used to constrain cosmological or galaxy evolution models. Here, HOD modelling enables us to estimate typical host halo masses for HiZELS galaxies. We can also do better than the straight-line $r_{0}$ fit; HOD fitting takes into account the small dip observed on angular scales of order 10s of arcseconds, below which the clustering is dominated by correlations between galaxies within a single dark matter halo. We can now include the effects of the satellite galaxy population on the observed clustering, no longer assuming that a power-law relationship holds on the smallest scales. \\
\indent A number of different halo occupation parameterisations have been used to fit 2-point galaxy correlation functions. Typically, 3 or 5-parameter fits of \cite{Zehavi2005} and \cite{Zheng2005} are used. While these do well for stellar mass-selected samples (e.g. \citealt{Wake2011,Hatfield2015}), they \textcolor{black}{may not be} suitable for our sample. As noted by \cite{Contreras2013}, HODs for stellar-mass selected samples are very different to the HODs of SFR or cold gas mass selected samples. In particular, HODs for mass-selected samples sensibly assume that above a given halo mass, all halos contain a central galaxy. However, in not all cases does this central galaxy fall within a star-formation rate or cold gas selected sample (e.g. due to the suppression of gas cooling in high mass halos via AGN feedback), so for star-formation rate limited samples the HOD for central galaxies may be peaked rather than a step function \citep{Contreras2013}.\\
\subsection{An 8-parameter HOD model}\label{sec:full_9}
Studying the clustering of $\rm{H}\alpha$ emitters at $z=2.23$, \cite{Geach2012} developed an 8-parameter model suitable for star-formation selected samples via comparison to the predictions of the semi-analytic model GALFORM \citep{Cole2000a,Bower2006}. In this parameterisation, the mean numbers of central\footnote{In \cite{Geach2012}, the factor of $\frac{1}{2}$ in the second term of the central galaxy parameterisation was excluded. We include it here, so that a halo can host a maximum of one (rather than two) central galaxies.} and satellite galaxies in a halo of mass M are given by:
\begin{equation}
\begin{split}
\langle N_{\rm{cen}}|M\rangle = F_{c}^{B}(1-F_{c}^{A}){\rm{exp}}\Bigg[-\frac{\log(M/M_{c})^{2}}{2(\sigma_{\log M})^{2}}\Bigg] \\
+ \frac{1}{2}F_{c}^{A}\Bigg[1+{\rm{erf}}\Bigg(\frac{\log(M/M_{c})}{\sigma_{\log M}}\Bigg)\Bigg],
\end{split}
\end{equation}
\begin{equation}
\langle N_{\rm{sat}}|M\rangle = F_{s}\Bigg[1+{\rm{erf}}\Bigg(\frac{\log(M/M_{\rm{min}})}{\delta_{\log M}}\Bigg)\Bigg]\Bigg(\frac{M}{M_{\rm{min}}}\Bigg)^{\alpha}.
\end{equation}
The key parameters are:
\begin{itemize}[leftmargin=0.3cm]
  \item[--] $M_{c}$: the halo mass at which the probability of hosting a central galaxy peaks.
  \item[--] $\sigma_{\log M}$: the width of the Gaussian distribution of centrals around its peak, $M_{c}$.
  \item[--] $M_{\rm{min}}$: the threshold halo mass for satellite galaxies, above which the distribution follows a power law $\langle N_{\rm{sat}}|M\rangle \approx F_{s} \Big(\frac{M}{M_{\rm{min}}}\Big)^{\alpha}$.
  \item[--] $\delta_{\log M}$: characterises the width of the transition to  $\langle N_{\rm{sat}}|M\rangle = F_{s} \Big(\frac{M}{M_{\rm{min}}}\Big)^{\alpha}$ around $M_{\rm{min}}$.
  \item[--] $\alpha$: the slope of the power law for $\langle N_{\rm{sat}}|M\rangle$ in halos with $M>M_{\rm{min}}$.
  \item[--] $F_{c}^{A,B}$: normalisation factors, in range [0,1].
  \item[--] $F_{s}$: the mean number of satellite galaxies per halo, at $M = M_{\rm{min}}$.
  \end{itemize}
  \indent \cite{Geach2012} did not have large enough samples to fit all 8 parameters simultaneously, so fixed the following parameters:
\begin{itemize}[leftmargin=0.3cm]
  \item[--] $M_{c} = M_{\rm{min}}$. The minimum mass halo hosting a satellite galaxy is the mass at which the central HOD peaks.
  \item[--] $\sigma_{\log M} = \delta_{\log M}$. The smoothing of the low-mass cut-off for satellite galaxies is not critical, as satellites in low mass halos contribute little to the overall HOD.
  \item[--] $\alpha = 1$. This is consistent with the literature for mass-selected samples.
  \end{itemize}
The total number of galaxies is given by:
\begin{equation}
\langle N|M\rangle = \langle N_{\rm{cen}}|M\rangle + \langle N_{\rm{sat}}|M\rangle.
\end{equation}
Some implementations use $\langle N|M\rangle = \langle N_{\rm{cen}}|M\rangle [1 + \langle N_{\rm{sat}}|M\rangle]$, requiring a central for every satellite galaxy. Given that our sample is essentially star-formation rate limited, some of our galaxies could be star-forming satellites around less highly star-forming centrals which are not included in our sample. Therefore we do not impose this condition. \\
\indent  We have performed a number of tests with different HOD parameterisations (e.g. allowing $\alpha$ to vary, fitting a full 8-parameter model), and confirm that neither the reproduction of the correlation function nor the values of the derived parameters are dependent on our choice (see Appendix \ref{sec:choose_parameterisation}). We base our parameterisation on the 5-parameter model of \cite{Geach2012}, but truncate the halo occupation sharply at $M_{\rm{min}}$, allowing only halos more massive than this to host galaxies. As detailed in Appendix \ref{sec:lower_limit_hod}, we have found that allowing the HOD to reach lower halo masses results in values of $M_{\rm{min}}$ which are strongly dependent on the lower limit of the HOD integral, and which are poorly constrained. $M_{\rm{min}}$ is now the minimum mass of halo hosting central galaxies, and, due to the shape of the halo mass function, also the most common host halo mass. Reassuringly, all other derived parameters are robust against the choice of lower limit.

\begin{table*}
\begin{center}
\begin{tabular}{l|c|c|c|c|c|c|c}
Redshift & $\log_{10}(M_{\rm{min}}/M_{\odot})$ & $F_{s}$ & $F_{c}^{A}$ & $F_{c}^{B}$ & $\sigma$\\
\hline
& [``unif", 10, 13.0, 11.5] & [``unif", 0.001, 1.0, 0.01] & [``unif", 0.001, 1.0, 0.9] & [``unif", 0.001, 1.0, 0.4] & [``log", 0.05, 1.0, 0.5] 
 \\
\hline
0.8 & $11.08^{+0.12}_{-0.15}$ & $0.002^{+0.001}_{-0.001}$ & $0.3^{+0.2}_{-0.3}$ & $0.6^{+0.3}_{-0.3}$ & $0.5^{+0.3}_{-0.2}$ \\
1.47 & $11.45^{+0.06}_{-0.08}$ & $0.005^{+0.003}_{-0.002}$ & $0.7^{+0.2}_{-0.4}$ & $0.7^{+0.3}_{-0.4}$ & $0.6^{+0.3}_{-0.3}$ \\
2.23 & $11.40^{+0.06}_{-0.07}$ & $0.007^{+0.003}_{-0.003}$ & $0.7^{+0.2}_{-0.4}$ & $0.6^{+0.3}_{-0.4}$ & $0.6^{+0.3}_{-0.4}$ \\
\hline
\end{tabular}
\caption{Fitted HOD parameters, with MCMC priors used (form, minimum, maximum, starting point). We show here the derived parameters for the large samples of galaxies within a fixed $L_{\rm{H}\alpha}/L_{\rm{H}\alpha}^{*}$ range at each redshift. $M_{\rm{min}}$ is the minimum mass halo hosting a galaxy, $F_{s}$ determines the number of satellite galaxies per halo, $F_{c}^{A,B}$ are normalisation factors, and $\sigma$ is the width of the Gaussian distribution of centrals around its peak, $M_{\rm{min}}$.}
\label{table:5_params_table}
\end{center}
\end{table*}

\subsection{Physical parameters from HOD models}
When fitting the models to data, we use the observed number density of galaxies as a constraint. For a given $\langle N|M\rangle$ output from the halo model, the predicted number density of galaxies is:
\begin{equation}
n_{g} =\int dM n(M) \langle N|M\rangle
\end{equation}
where n(M) is the halo mass function. Here we use that of \cite{Tinker2010}. The observed number density of galaxies used is the integral of the luminosity function between the same limits used to select the real and random galaxy sample (using the luminosity functions derived by \cite{Sobral2012,Sobral2015} for the same data). We assume a $10\%$ error on the number density in the fitting. \\\\
For each set of HOD parameters, we may derive a number of parameters of interest for galaxy evolution. The satellite fraction is:
\begin{equation}
f_{\rm{sat}} = \frac{1}{n_{g}}\int dM n(M) \langle N_{\rm{sat}}|M\rangle,
\end{equation}
with the corresponding central fraction $f_{\rm{cen}} = 1 - f_{\rm{sat}}$.\\
The effective halo mass, the typical mass of galaxy host halo is:
\begin{equation}
M_{\rm{eff}} = \frac{1}{n_{g}}\int dM M n(M) \langle N|M\rangle.
\end{equation}
The average effective bias factor, which characterises the clustering of galaxies relative to dark matter, is:
\begin{equation}
b_{\rm{eff}} = \frac{1}{n_{g}} \int dM n(M) b(M) \langle N|M\rangle,
\end{equation}
where $b(M)$ is the halo bias, a function of halo mass M. We use $b(M)$ from \cite{Tinker2010}.\\

\subsection{Fitting HOD models to HiZELS $\rm{H}\alpha$-emitting galaxies}\label{sec:fitting_halomod}
\begin{figure}
	\centering
	\includegraphics[scale=0.55]{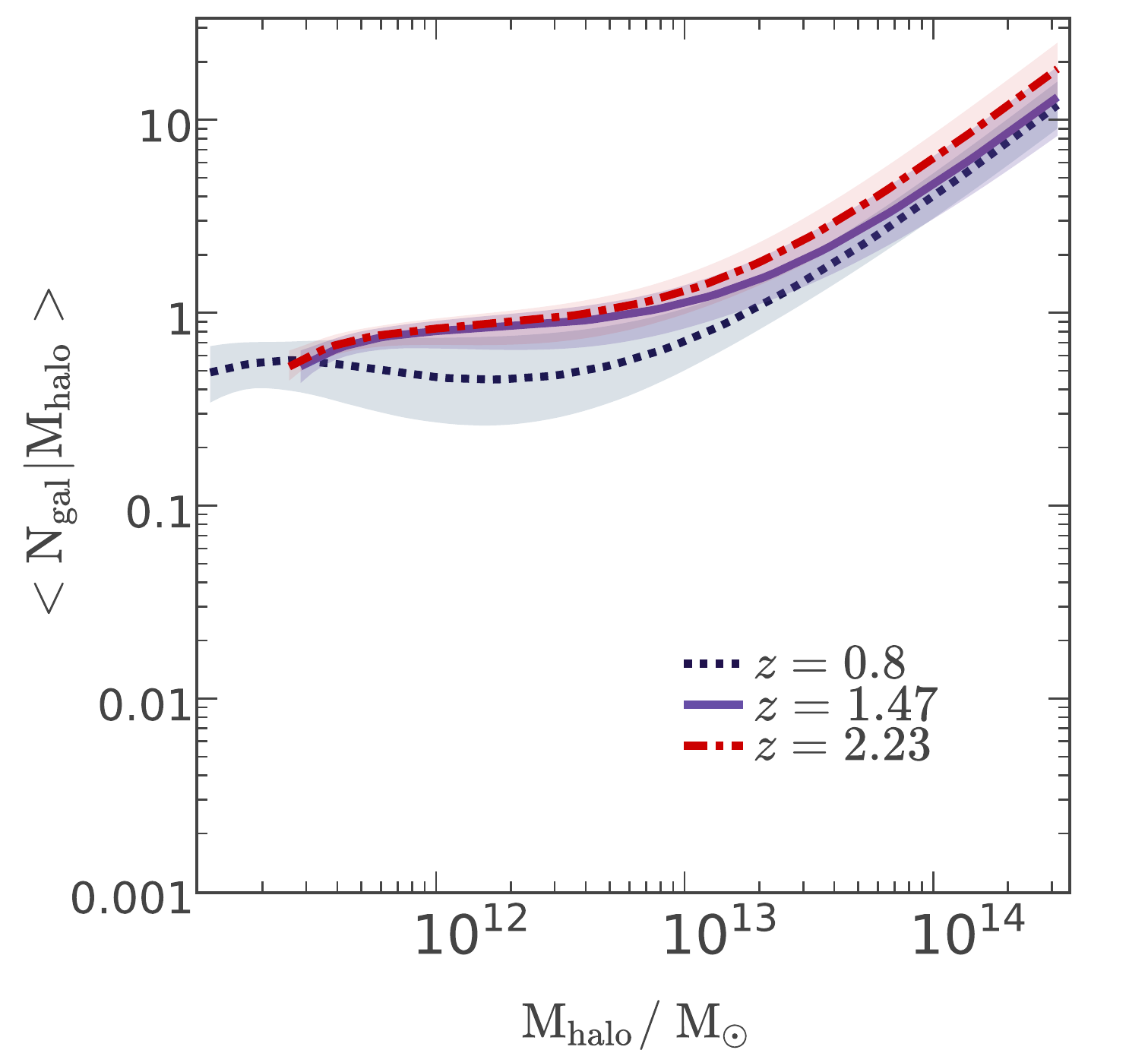}	
	\caption{HOD parameterisations of samples of galaxies at $z=0.8$, $z=1.47$ and $z=2.23$, within fixed ranges of $L_{\rm{H}\alpha}/L_{\rm{H}\alpha}^{*}$ line up closely.  \textcolor{black}{Although the $L_{\rm{H}\alpha}/L_{\rm{H}\alpha}^{*}$ distributions are not exactly the same across the different redshift ranges, galaxies selected at similar $L_{\rm{H}\alpha}/L_{\rm{H}\alpha}^{*}$ seem to trace similar dark matter halos across redshift.}}
	\label{fig:hod_all_z}
\end{figure}

We use the HMF \citep{Murray2013} and HALOMOD codes (Murray, in prep.) to fit HOD models to the correlation functions. These take an HOD parameterisation and construct real-space correlation functions for a range of parameter inputs. For each set of parameter inputs, we compare the projection of the modelled real-space correlation function with that observed, and calculate the log likelihood. We use {\it{emcee}} \citep{Foreman-Mackey2013}, a fast python implementation of an affine-invariant Markov Chain Monte Carlo (MCMC) ensemble sampler, to sample the parameter space of our 5 fitted parameters and optimise the fit to the correlation function. As discussed, we fit the number density of galaxies in the log-likelihood fitting as a further constraint. We use 500 walkers, each with 1000 steps. \\
\indent We present examples of the best-fit modelled correlation function and its HOD occupation, decomposed into the central and satellite galaxy terms, in Figure \ref{fig:hod_whole_J}. The parameterisation, shown here for a correlation function constructed using the full sample of galaxies at $z=0.8$, provides a good fit to the data, and clearly shows the separate contributions of the clustering within a single halo and between dark matter halos. \\
\indent For each correlation function to which an HOD model is fitted, we estimate the following parameters: $f_{\rm{sat}}$, $M_{\rm{eff}}$, $b_{\rm{eff}}$, $M_{\rm{min}}$. We take the 50th, 16th and 84th percentiles of the posterior distribution of each of these derived parameters, to obtain an estimate of the median and associated $1\sigma$ errors. The individual HOD input parameters $\sigma_{\log M}$, $F_{c}^{A,B}$ \textcolor{black}{and} $F_{s}$, tend to be individually less well constrained due to correlations between them. In Table \ref{table:5_params_table}, we present the five HOD parameters fitted to the correlation functions of large samples of galaxies within a fixed $L_{\rm{H}\alpha}/L_{\rm{H}\alpha}^{*}$ range at each redshift. In Appendix \ref{sec:mcmc_plot}, we show an example of the MCMC output for one of our HOD fits. \\
\indent \textcolor{black}{The selection of galaxies within a fixed $L_{\rm{H}\alpha}/L_{\rm{H}\alpha}^{*}$ range, as in Section \ref{sec:whole_samples_z} (see Table \ref{table:r0_table}), allows the comparison of similar galaxies across cosmic time. Interestingly, the derived galaxy occupations as a function of halo mass are similar, consistent within their errors (see Figure \ref{fig:hod_all_z}). Although the $L_{\rm{H}\alpha}/L_{\rm{H}\alpha}^{*}$ distributions are not exactly the same across the different redshift ranges, we deduce from this that samples of galaxies selected from HiZELS at similar $L_{\rm{H}\alpha}/L_{\rm{H}\alpha}^{*}$ trace similar dark matter halos across redshift. Intrigued by this, we compare galaxies within narrower $L_{\rm{H}\alpha}/L_{\rm{H}\alpha}^{*}$ bins in Section \ref{sec:lum_hod}.}

\begin{figure*}
	\centering
	\includegraphics[scale=0.5]{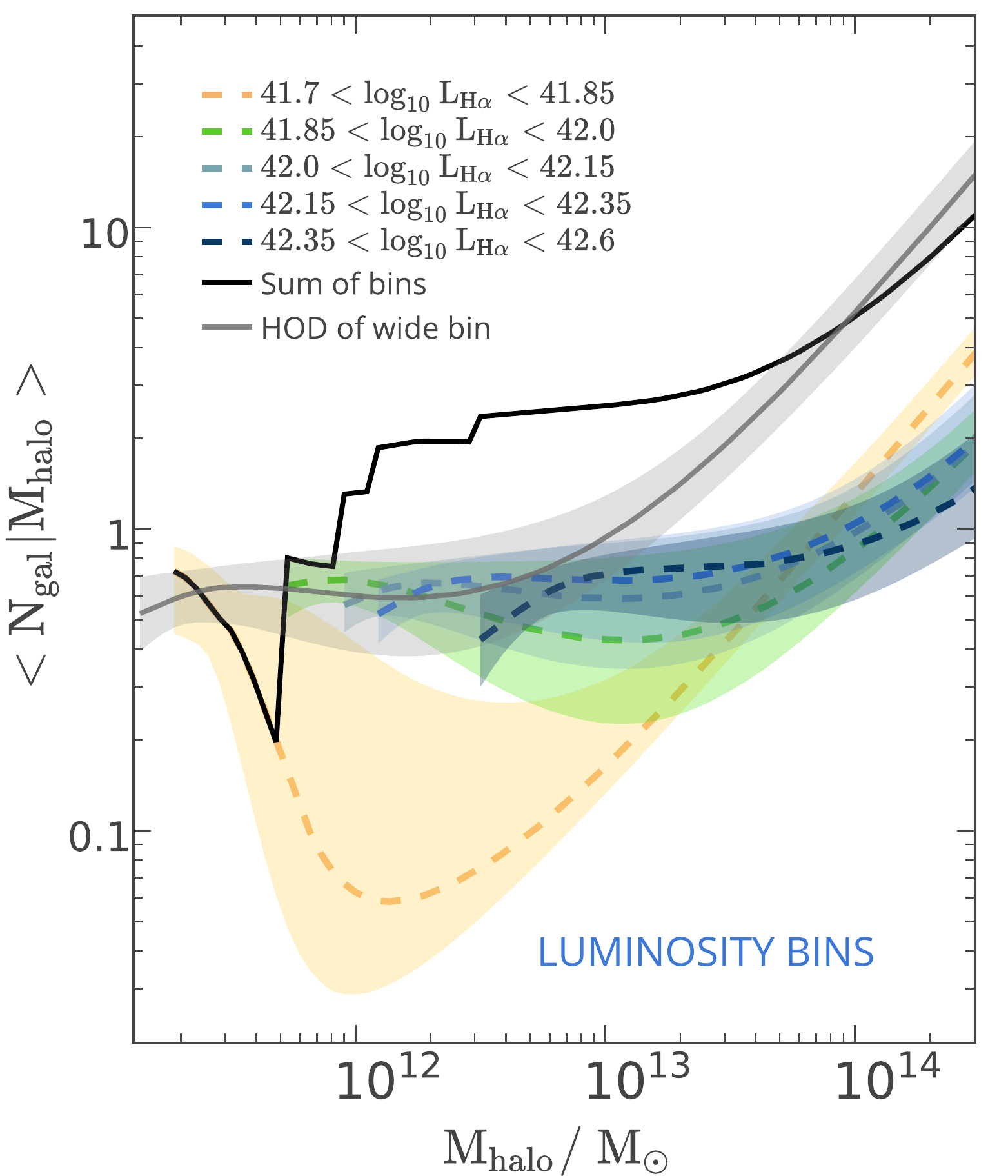}	
	\includegraphics[scale=0.5]{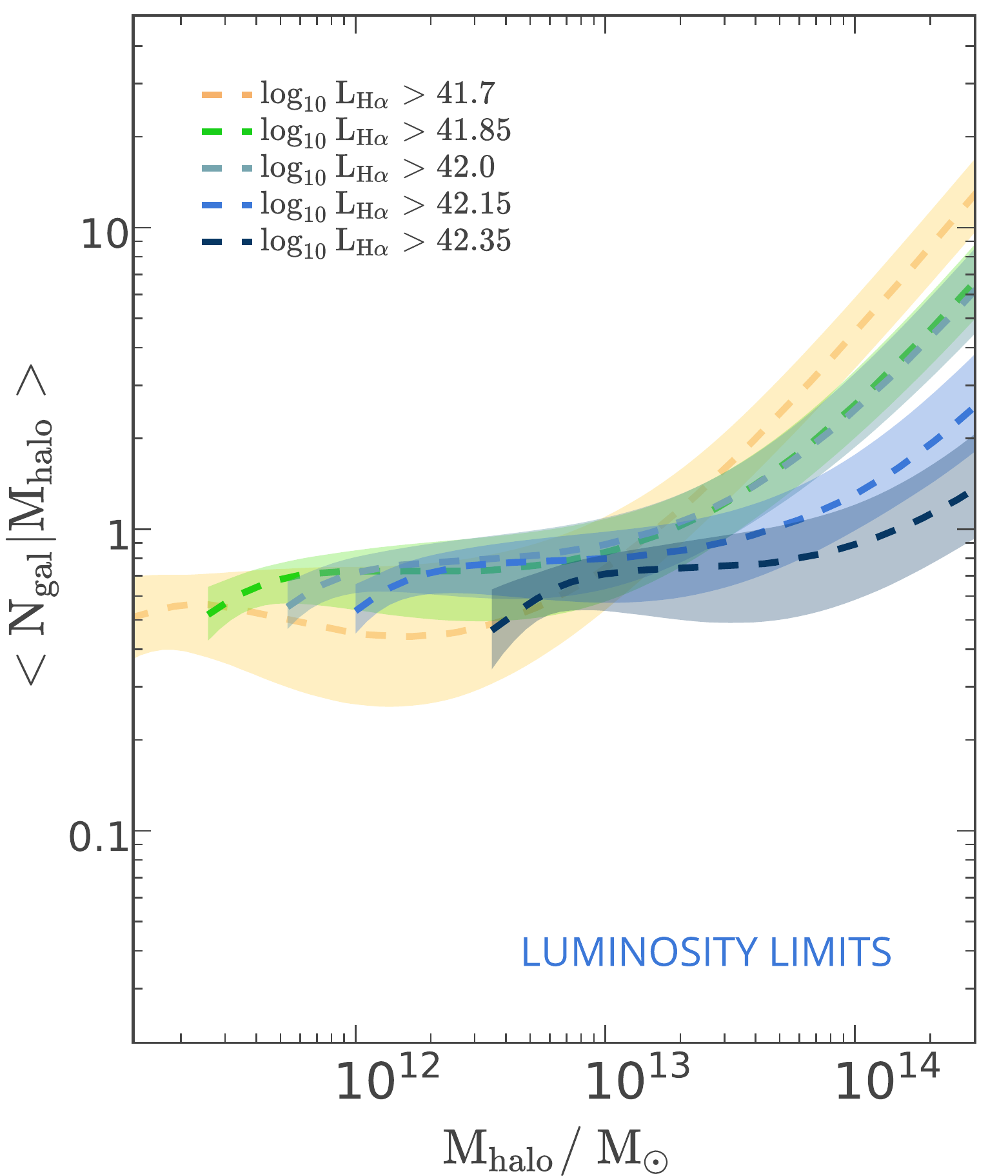}	
	\caption{Fitted halo occupation distributions for $\rm{H}\alpha$ luminosity-binned (left) and luminosity-limited (right) samples at $z\sim0.8$. Higher luminosity $\rm{H}\alpha$ emitters occupy higher mass dark matter halos. \textcolor{black}{Our results are qualitatively consistent between the luminosity-binned and luminosity-limited samples, but} trends are cleaner for the luminosity-limited samples, which are larger.}
    \label{fig:hod_lum_bl}
\end{figure*}

\begin{figure*} 
	\centering
	\includegraphics[scale=0.5]{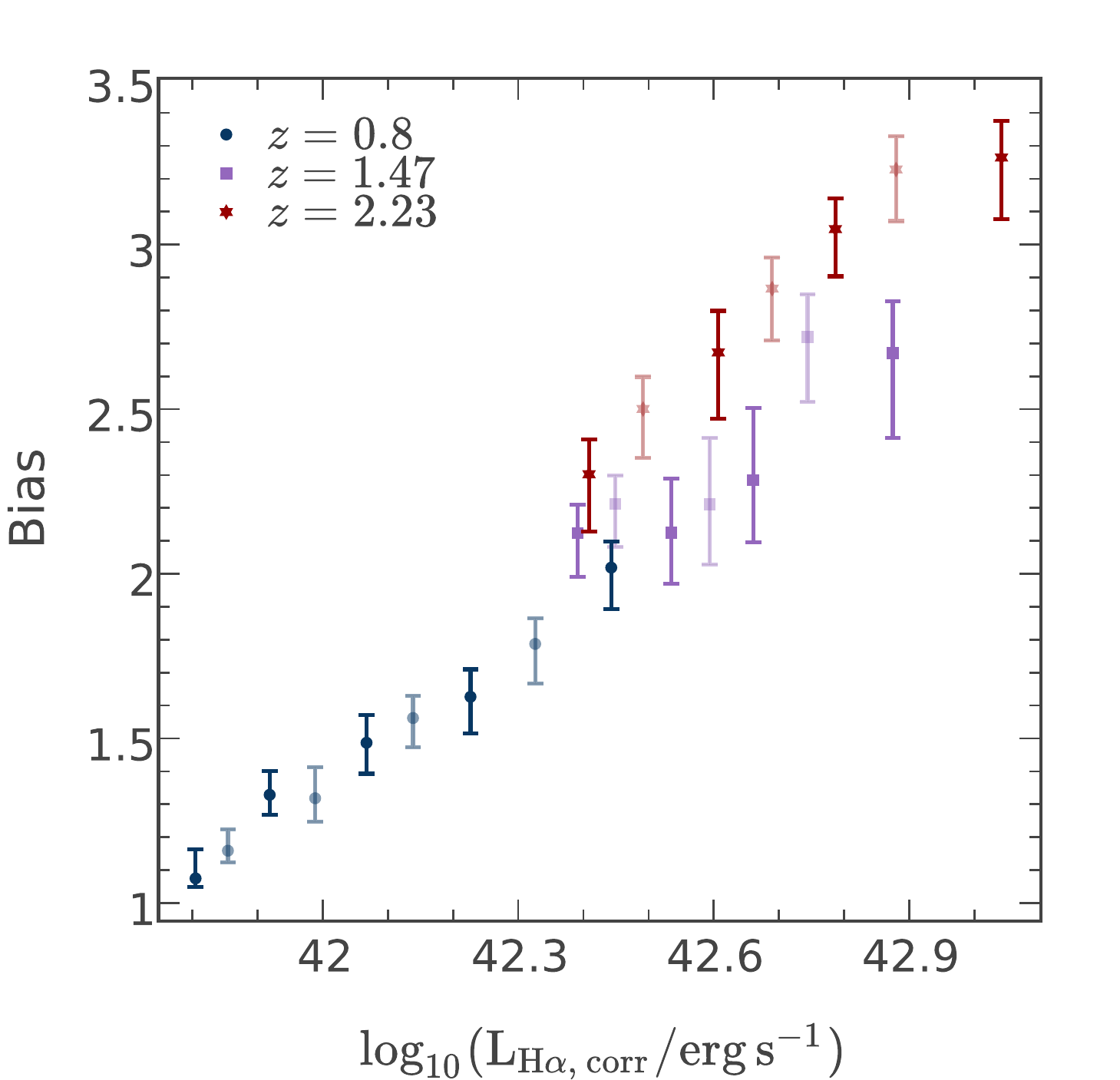}	
	\includegraphics[scale=0.5]{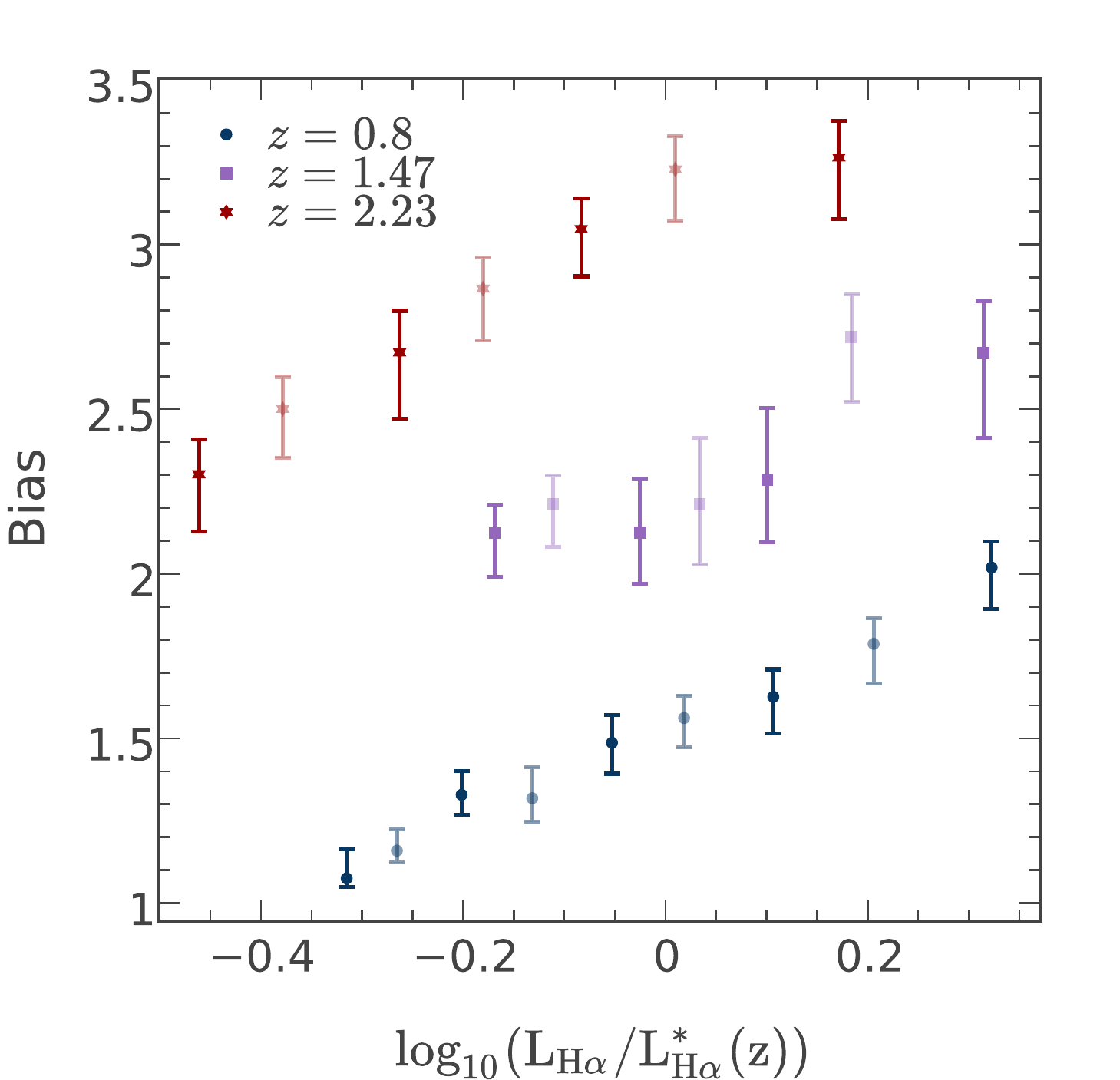}	
	\includegraphics[scale=0.5]{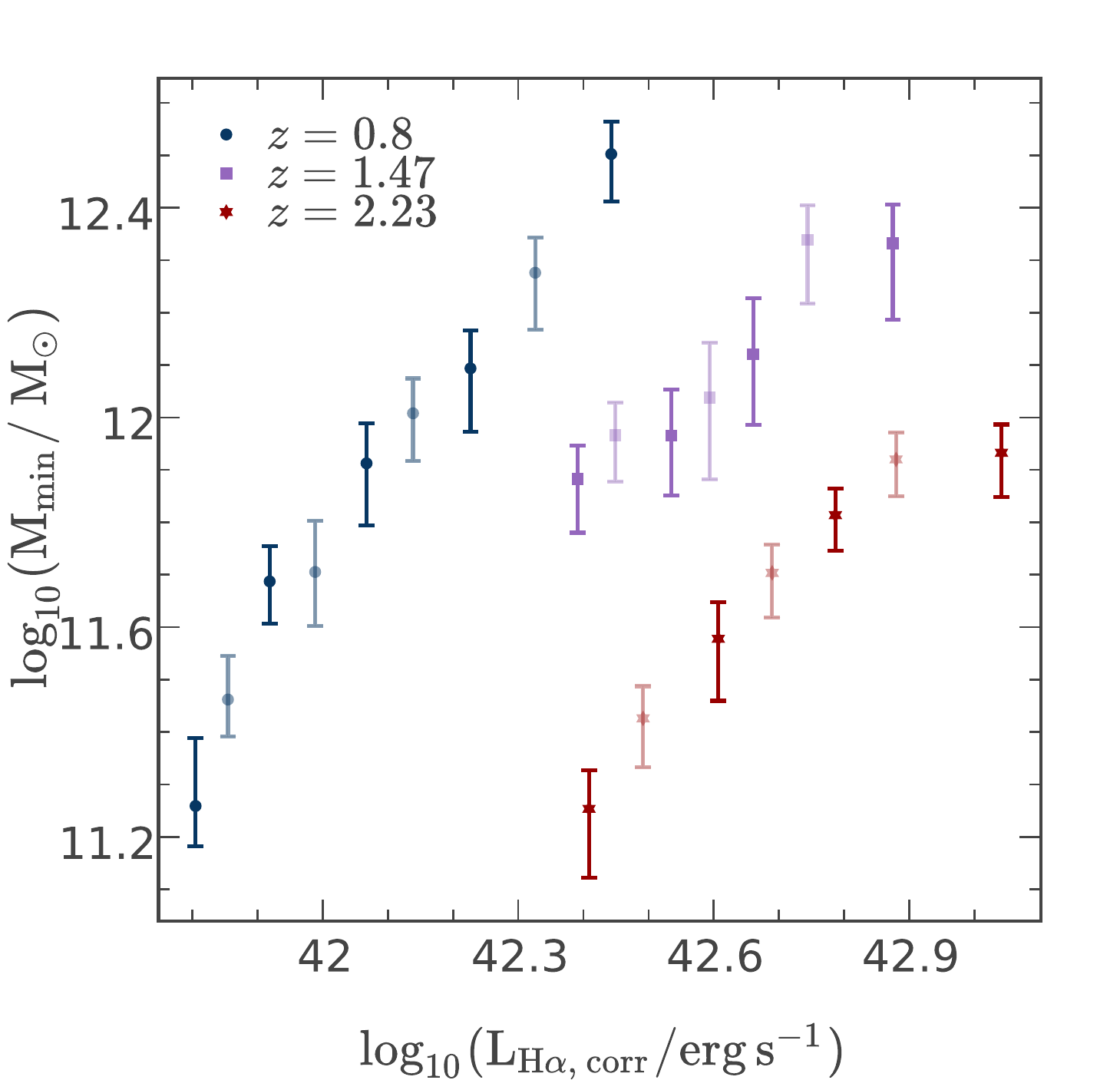}	
	\includegraphics[scale=0.5]{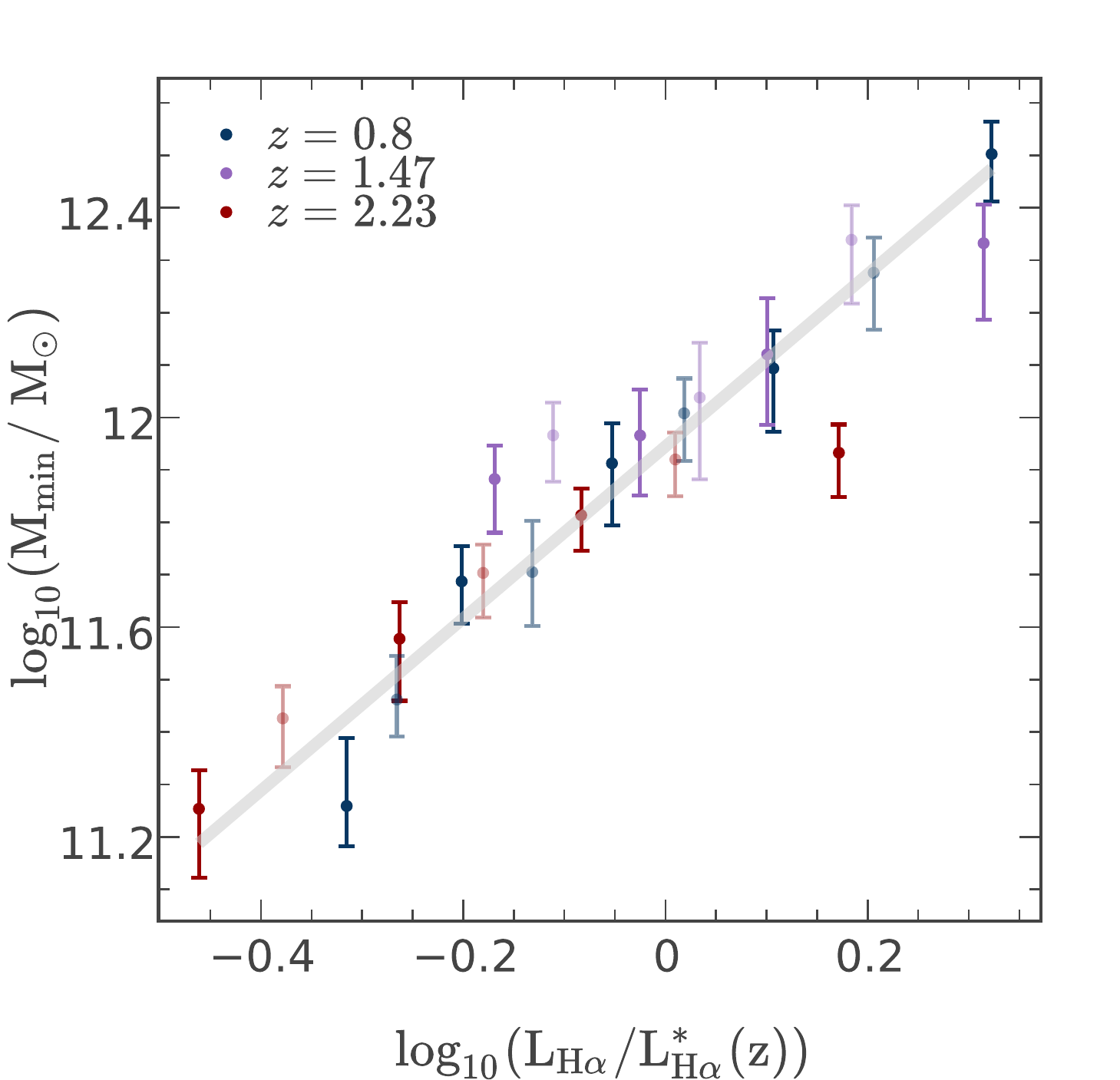}	
	\includegraphics[scale=0.5]{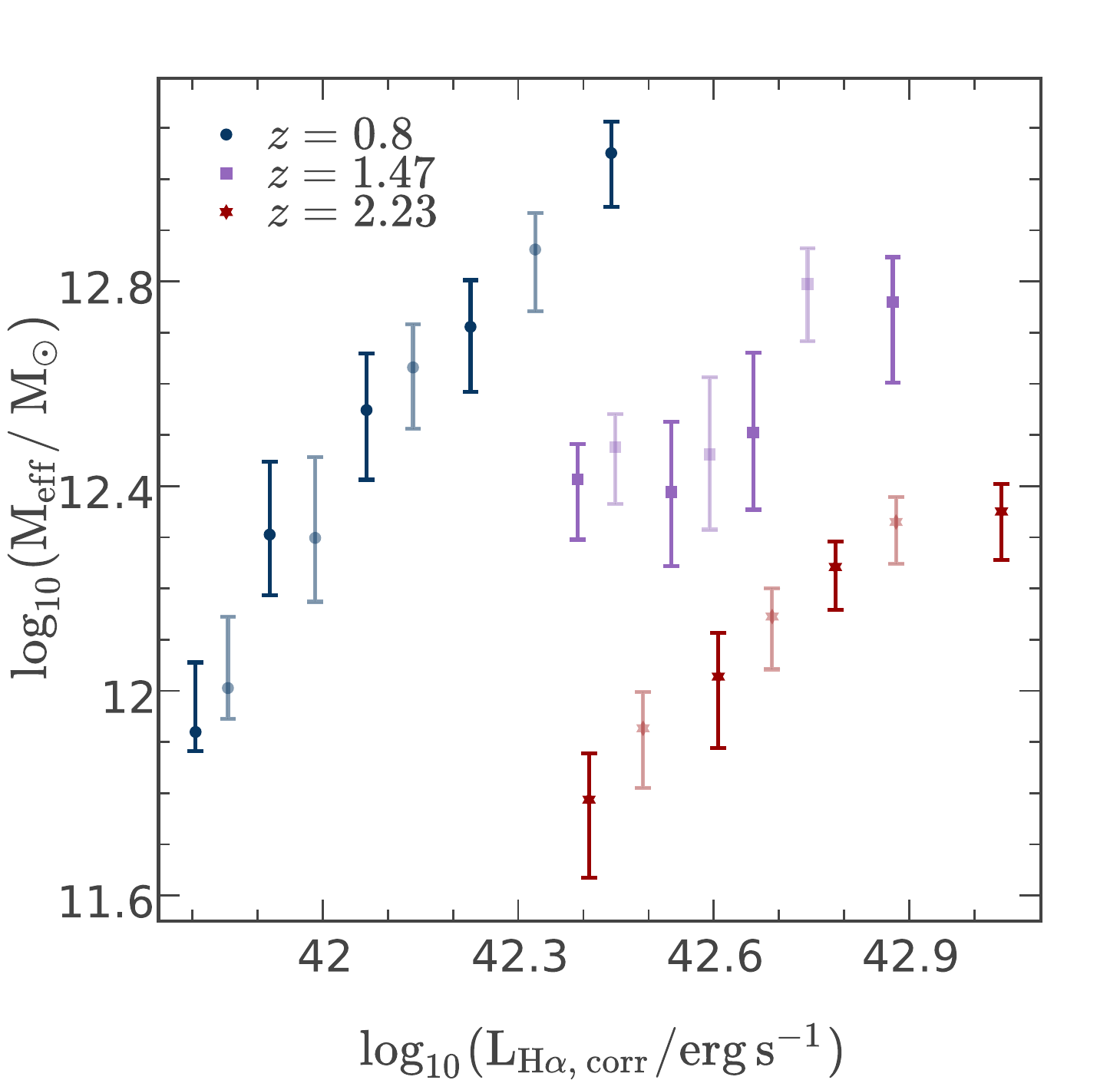}	
	\includegraphics[scale=0.5]{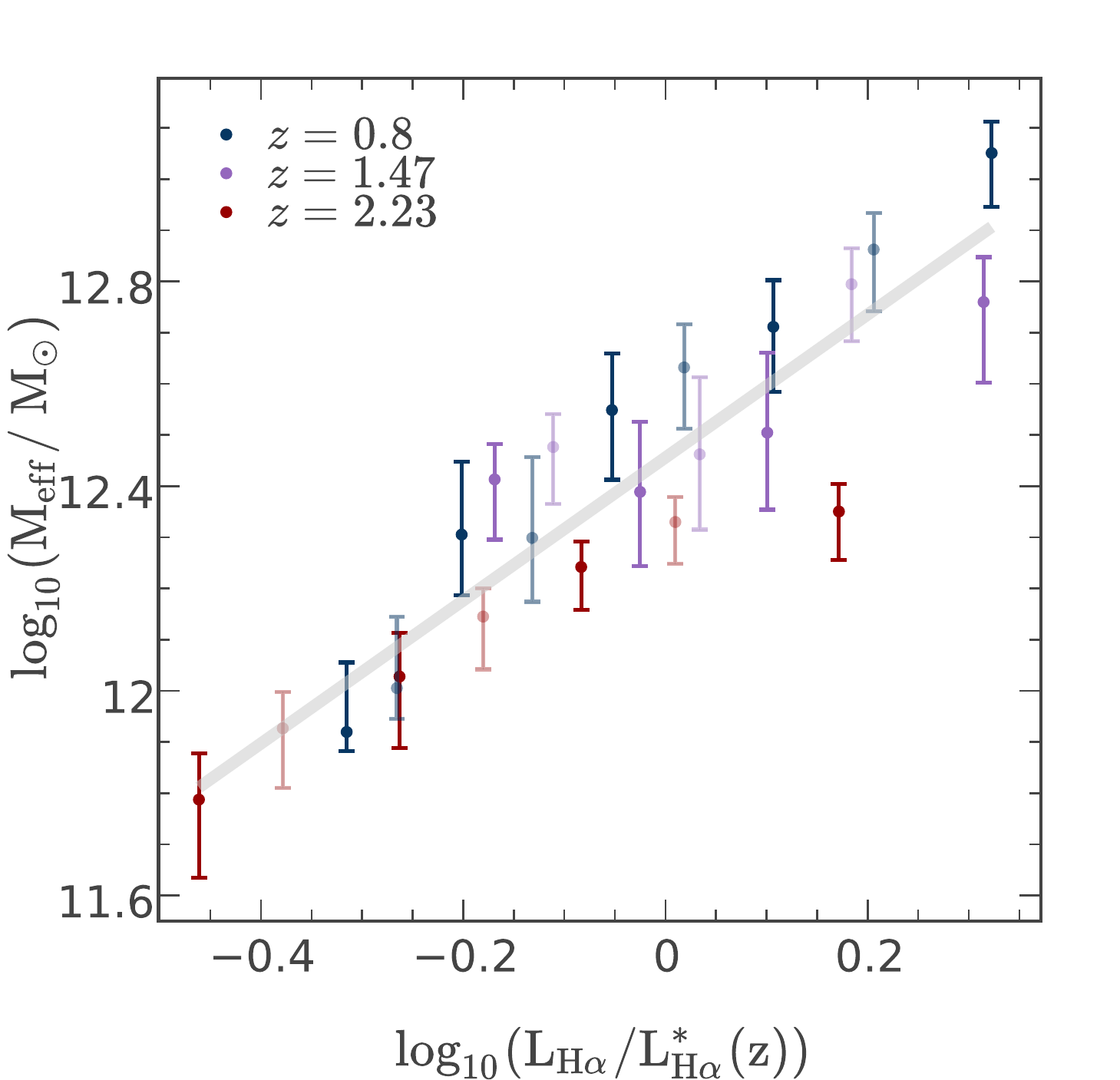}	
	\caption{Derived properties of galaxy populations of HiZELS galaxies binned by $\rm{H}\alpha$ luminosity. We find a linear, broadly redshift-independent relationship between halo mass and $\rm{H}\alpha$ luminosity. As in Figure \ref{fig:corr_z}, the paler colours denote alternative binning. The lines of best fit derived in Section \ref{sec:lum_hod} are overplotted: $\log_{10}(M_{\rm{min}}/ M_{\odot}) = (1.64\pm0.11) \log_{10}(L_{\rm{H}\alpha}/L_{\rm{H}\alpha}^{*})+(11.94\pm0.02)$, $\log_{10}(M_{\rm{eff}}/ M_{\odot}) = (1.40\pm0.12)\,\log_{10}(L_{\rm{H}\alpha}/L_{\rm{H}\alpha}^{*})+(12.46\pm0.02)$.}
	\label{fig:all_l_params}
\end{figure*}
\subsection{Luminosity dependence of HOD models}\label{sec:lum_hod}
\begin{figure} 
	\includegraphics[scale=0.58]{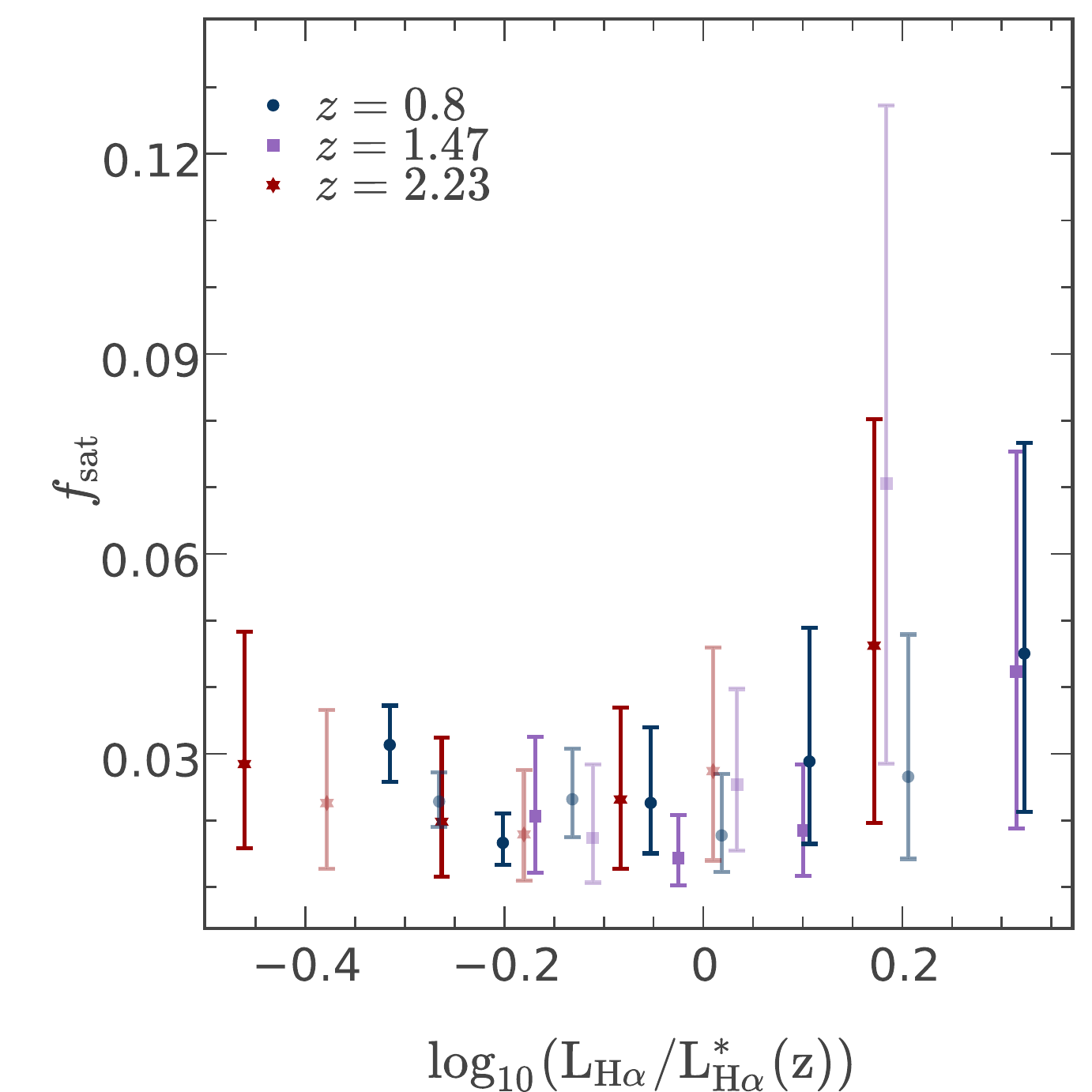}
	\caption{The derived satellite fraction is low for all redshifts and luminosity bins\textcolor{black}{, indicating that HiZELS galaxies are primarily centrals. However, the satellite fraction is the least well-constrained of the HOD output parameters.} Again, the paler colours denote alternative binning.}
	\label{fig:fsat_fig}
\end{figure}

Before extending the HOD analysis to $\rm{H}\alpha$ luminosity-binned data at all three redshifts, we show fits to luminosity-binned and luminosity-limited data at a single epoch, $z=0.8$, where we have the largest and most robust sample (Figure \ref{fig:hod_lum_bl}). For the highest luminosity (SFR) bins (e.g. dark blue line), there is a clear shift towards the right, indicating that galaxies typically occupy higher mass dark matter halos with increasing $\rm{H}\alpha$ luminosity. The lowest luminosity bin (yellow line) is also interesting: the central galaxy distribution is strongly peaked around $M_{\rm{halo}}\sim10^{11}M_{\odot}$. Therefore high mass halos do not tend to host central galaxies with these low star-formation rates. \\
\indent The luminosity-binned and luminosity-limited results are largely self-consistent, though there is some discrepancy between the sum of the HODs of independent luminosity bins (black line) and the HOD of the sum of the luminosity bins (grey). This is particularly evident at halo masses in the range $10^{12}M_{\odot}<M_{\rm{halo}}<10^{13}M_{\odot}$, where the bins sum to more than one central galaxy per halo. We attribute this to the limitations of our parameterisation, and to the uncertainties inherent in fitting HODs to correlation functions constructed using limited numbers of galaxies. \\
\indent The luminosity-limited HODs broadly agree with the halo occupation of simulated $\rm{H}\alpha$ emitters from the semi-analytic model GALFORM. \cite{Geach2012} show the HOD of GALFORM emitters with $L_{\rm{H}\alpha}>10^{42} \rm{erg\,s^{-1}}$, which is in excellent agreement with our derived HOD (Figure \ref{fig:hod_lum_bl}, right, green line). Both HODs show the occupation of central galaxies peaking at $M_{\rm{halo}}\sim10^{12}M_{\odot}$, with satellites becoming dominant at $M_{\rm{halo}}\sim10^{13}M_{\odot}$. HODs derived from the highest luminosity GALFORM sources display a dip in the occupation of halos around $10^{13}M_{\odot}$, with high mass halos in GALFORM preferentially hosting low luminosity galaxies. We see no evidence for this, but do not reach the high luminosities of $L_{\rm{H}\alpha}>10^{43} \rm{erg\,s^{-1}}$ where this is most clear in the simulated galaxies. We now explore these trends in greater detail using binned samples at all three redshifts. \\
\indent At all three redshifts, we observe strong trends in the derived HOD parameters with galaxy $\rm{H}\alpha$ luminosity (left-hand panels of Figure \ref{fig:all_l_params}, see also Table \ref{table:r0_table}). The effective bias, \textcolor{black}{which characterises the increased clustering of galaxies compared to dark matter,} increases roughly linearly with $\rm{H}\alpha$ luminosity: more highly star-forming galaxies are therefore more strongly clustered with respect to the underlying dark matter distribution. \textcolor{black}{The effective bias also increases towards higher redshifts. This reflects the growth of the dark matter correlation function with time \citep{Weinberg2004}. The first galaxies to form - those at high redshift - are more biased relative to the underlying mass distribution which itself is less strongly clustered.}\\
\indent \textcolor{black}{The effective mass ($M_{\rm{eff}}$) is the average mass of the dark matter halo inhabited by the star-forming galaxies in our samples.} The relationship between effective mass of the host dark matter halos and $\rm{H}\alpha$ luminosity is similar \textcolor{black}{to that of the bias}: galaxies with higher star-formation rates lie, on average, in more massive dark matter halos. At fixed $\rm{H}\alpha$ luminosity, the dark matter halo mass increases steeply towards low redshifts. The minimum mass of dark matter halo that hosts star-forming galaxies scales with $\rm{H}\alpha$ luminosity in a similar way: more luminous satellite galaxies are hosted by more massive dark matter halos. \\
\indent To compare similar populations of galaxies at the three different redshifts, we scale by the characteristic luminosity once again (see right-hand panels of Figure \ref{fig:all_l_params}). Values of $M_{\rm{min}}$ from samples at all 3 redshifts form a tight sequence when plotted against $\log_{10}(L_{\rm{H}\alpha}/L_{\rm{H}\alpha}^{*})$. This is key: if we select galaxies at a given luminosity relative to the characteristic luminosity at any redshift, they reside in dark matter halos of the same minimum mass. $M_{\rm{eff}}$ shows a similar, broadly redshift-independent trend, though there is tentative evidence of evolution to slightly higher masses towards lower redshifts, as the mass of typical dark matter halos grows with cosmic time. We obtain the following best-fit relations, by fitting to one set of bins at each redshift:
\begin{equation}
\log_{10}(M_{\rm{min}}/ M_{\odot}) = (1.64\pm0.11)\,\log_{10}(L_{\rm{H}\alpha}/L_{\rm{H}\alpha}^{*})+(11.94\pm0.02)
\end{equation}
\begin{equation}
\log_{10}(M_{\rm{eff}}/ M_{\odot}) = (1.40\pm0.12)\,\log_{10}(L_{\rm{H}\alpha}/L_{\rm{H}\alpha}^{*})+(12.46\pm0.02)
\end{equation}
\textcolor{black}{We test for evolution in the normalisation of these lines by fixing their gradients to those fitted above ($1.64$ and $1.40$) and fitting the intercept at each redshift individually. We find intercepts of $11.92\pm0.05$ at $z=0.8$,  $11.96\pm0.06$ at $z=1.47$, and $11.94\pm0.08$ at $z=2.23$ for the $M_{\rm{min}}$ fit. Similarly, we obtain $12.54\pm0.04$ at $z=0.8$, $12.41\pm0.06$ at $z=1.47$, and $12.36\pm0.06$ at $z=2.23$ for the $M_{\rm{eff}}$ fit. The fits are consistent to within $0.04\rm{dex}$ for $M_{\rm{min}}$ and $0.2\rm{dex}$ for $M_{\rm{eff}}$.}\\
\indent \textcolor{black}{The satellite fraction for the HiZELS samples is the least well constrained derived parameter. This is because when a halo contains only one star-forming galaxy, the two-point correlation function cannot distinguish whether this is a central galaxy or a satellite of a central quenched galaxy. Satellite galaxies are only constrained by the one-halo term in the most massive halos, and thus the determination of $f_{\rm{sat}}$ is sensitive to the form of the HOD parameterisation (which extrapolates this to lower masses). Nevertheless, we find no evidence of a change in satellite fraction with redshift with luminosity or with redshift (Figure \ref{fig:fsat_fig}).} As noted previously, this satellite fraction is only the fraction of {\it{star-forming satellites in the sample}}, and may be higher if passive populations were included. There is a slight indication of an upturn in satellite fractions at the highest luminosities, but at low significance. Figure \ref{fig:hod_lum_bl} had shown that the sum of the HODs of luminosity-binned samples clearly exceeds the HOD of the whole sample at moderate halo masses of $M_{\rm{halo}}~\sim10^{12}-10^{13}M_{\odot}$ by a factor of $\sim2$. \textcolor{black}{This suggests that the HOD fits to luminosity-binned samples are overestimating the number of central galaxies in the sample. This would decrease the satellite fraction and explain the discrepancy between the $\sim5\%$ satellite fractions derived for the whole samples (see Table \ref{table:r0_table}) compared to those of luminosity-binned samples, which stand at $\sim3\%$.} The $\sim5\%$ satellite fraction is likely to be closer to the `true' satellite fractions of our samples. Nevertheless, the main result of Figure \ref{fig:fsat_fig} is that there is no evidence that $f_{\rm{sat}}$ changes dramatically with either $L_{\rm{H}\alpha}$ or redshift.\\
\indent Finally, we note that when scaled by $L_{\rm{H}\alpha}^{*}$, the luminosity-bias relations show strong redshift dependence. This is due to the growth of the dark matter correlation function \textcolor{black}{with time.} The different redshifts align better in the $\log_{10}L_{\rm{H}\alpha}$ vs bias plot, but this is likely to be simply because at fixed $L_{\rm{H}\alpha}$, selection of brighter (relative to $L_{\rm{H}\alpha}^{*}$) galaxies at low redshift goes some way towards compensating the dark matter halo growth. \\

\section{Discussion}\label{sec:discussion}
\indent Having studied the halo environments of galaxies at three different redshifts, we draw together the main findings here. The $\rm{H}\alpha$-selected galaxies detected by the HiZELS survey are typical star-forming galaxies which reside in dark matter halos of masses $\sim10^{12}M_{\odot}$. \textcolor{black}{Our typical HiZELS limiting $\rm{H}\alpha$ fluxes correspond to star-formation rates of $\sim4M_{\odot}/\rm{yr}$ at $z=0.8$, $\sim8M_{\odot}/\rm{yr}$ at $z=1.47$ and  $\sim13M_{\odot}/\rm{yr}$ at $z=2.23$, according to the $\rm{H}\alpha$-SFR conversion of \cite{Jr1998}.} At all redshifts, in all luminosity bins, we find low satellite fractions of $\sim5\%$, with fitted HODs only reaching above one star-forming satellite per halo in halos of $\geq10^{13}M_{\odot}$. Whilst there are some uncertainties introduced by the limitations of our HOD parameterisation, the satellite fractions derived are consistently low for both luminosity-limited and luminosity-binned samples of $\rm{H}\alpha$ emitters. We conclude that the majority of the star-forming galaxies in our samples are centrals. \\
\indent The star-forming galaxies detected at lower redshifts ($z=0.8$ and $z=1.47$) have lower $\rm{H}\alpha$ luminosities than the high-redshift $z=2.23$ galaxies which reside in equally massive halos. This reflects the general trend of decreasing star-formation rates towards low redshift \citep[see][]{Daddi2007,Elbaz2007,Karim2011,Sobral2014,Lee2015}. At all three redshifts, we find an increase in estimated average host dark matter halo mass with $\rm{H}\alpha$ luminosity of galaxies studied. More highly star-forming galaxies are hosted by more massive dark matter halos. \\
\indent We emphasise here that we have performed the analysis on a sample of galaxies selected cleanly by $\rm{H}\alpha$ emission line strength. These galaxies are predominantly star-forming, with luminosities close to the characteristic luminosity at each redshift, and are therefore representative of the star-forming population \citep{Oteo2015}. If we were to probe down to much lower star-formation rates (including the passive galaxy population), trends in halo mass vs $\rm{H}\alpha$ luminosity may eventually reverse. \cite{Hartley2010}, for example, found passive galaxies to be significantly more strongly clustered than their star-forming counterparts back to $z\sim2$ \cite[see also][]{Wilkinson2016}. This fits easily into our interpretation: the passive, massive galaxies at a given redshift formed their mass early (downsizing; \citealt{Cowie1996}), and hence quickly. Indeed, we find that the most highly star-forming galaxies at all redshifts are the most strongly clustered.
\subsection{The halo mass - characteristic luminosity relation}\label{sec:gas_reg}
\indent Scaling by the characteristic luminosity at each redshift enables us to compare similar populations of galaxies. The $\log_{10}(L_{\rm{H}\alpha}/L_{\rm{H}\alpha}^{*})$ vs halo mass relations line up very tightly, \textcolor{black}{and as shown in Figure \ref{fig:k_CORRS_lum}, this is a genuine trend, not driven by stellar mass.} This indicates that the mass of the host dark matter halo is driving the typical luminosity of its star-forming galaxies. The minimum halo mass at $L_{\rm{H}\alpha}=L_{\rm{H}\alpha}^{*}$ is $\sim10^{12}M_{\odot}$ for all three redshifts. This exactly coincides with the peak of the stellar mass-halo mass relation (SHMR), the halo mass \textcolor{black}{at} which the star-formation efficiency peaks, within this redshift range \citep{Behroozi2010,Behroozi2013}. As noted by \cite{Behroozi2013}, the halo mass at which the SHMR is at its maximum is also that at which the baryon conversion efficiency (the ratio of the SFR to the halo's baryon accretion rate) is highest. Models predict that this holds across a large redshift range, until at least $z=4$. Our results support the conclusion that halos of mass $\sim10^{12}M_{\odot}$ are the most efficient at forming stars at every epoch. The SHMR decreases at higher halo masses, which are less efficient at forming stars. We obtain $M_{\rm{min}}\sim10^{12-12.4}M_{\odot}$ for our most luminous galaxies, in line with this. This is consistent with the models of \cite{Dekel2006}, which posit a roughly redshift-independent limiting halo mass of $M_{\rm{shock}}\sim10^{12}M_{\odot}$, above which efficient gas cooling is prevented by shock heating. \cite{Sobral2016} find that those HiZELS galaxies with $L_{\rm{H}\alpha}>L_{\rm{H}\alpha}^{*}$ have increasing AGN fractions, while \cite{Sobral2009} find that that these very luminous galaxies are much more likely to be mergers than their low-luminosity counterparts (the fraction of $z=0.84$ HiZELS galaxies with irregular morphologies increases from $<20\%$ below $L_{\rm{H}\alpha}=L_{\rm{H}\alpha}^{*}$ to $\sim100\%$ at $L_{\rm{H}\alpha}>L_{\rm{H}\alpha}^{*}$). This supports the argument that $L_{\rm{H}\alpha}^{*}$ is the luminosity where `normal', non-merger-driven star-formation peaks. 
\subsection{Interpretation via an equilibrium gas regulator model}\label{sec:gas_reg}
\begin{figure} 
	\includegraphics[scale=0.5]{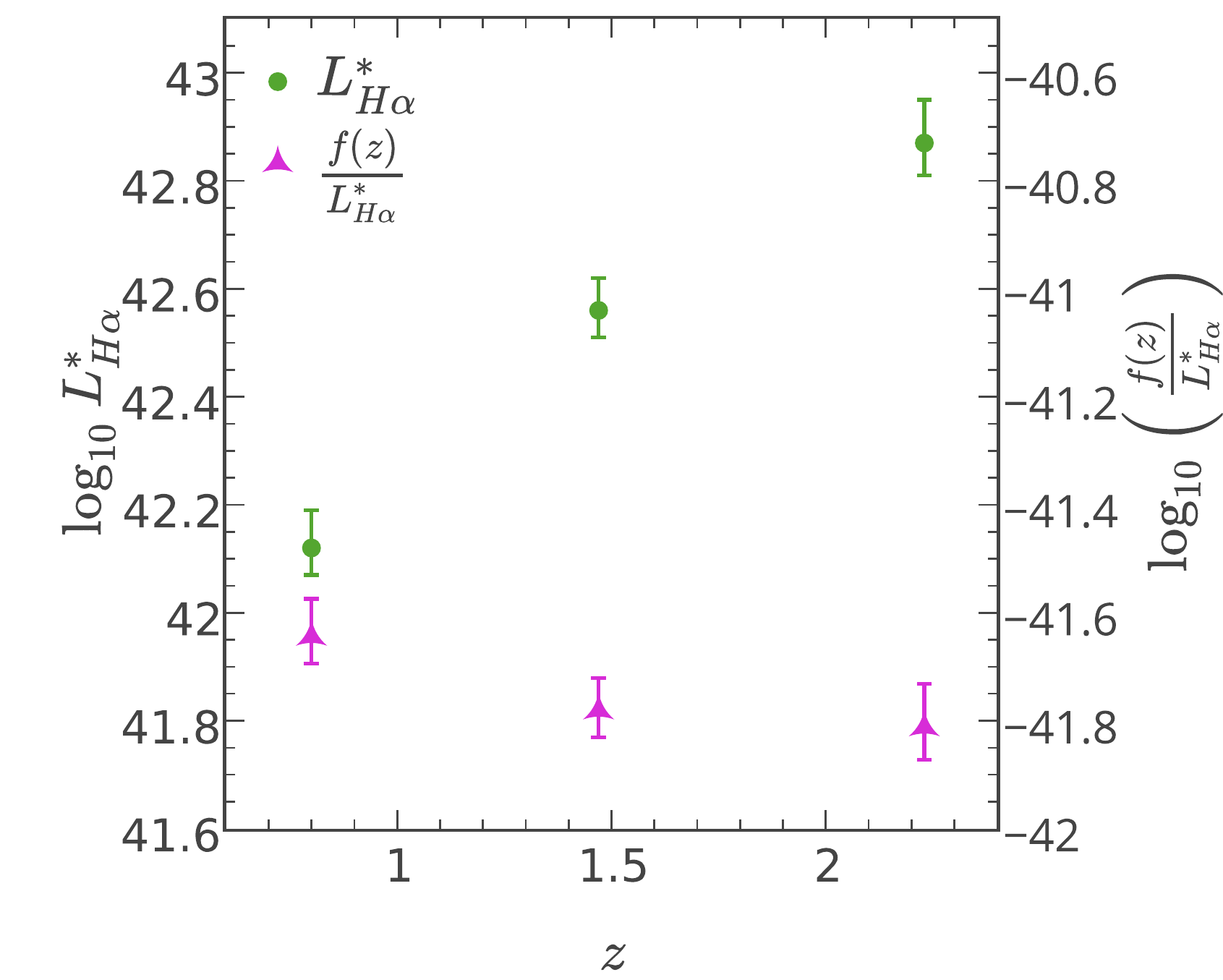}
	\caption{The characteristic $\rm{H}\alpha$ luminosity, $L_{\rm{H}\alpha}^{*}$ displays a striking increase with redshift. Once scaled by the halo mass growth factor, $f(z)$, from \protect\cite{Fakhouri2010}, we observe little evolution. This indicates that the evolution in  $L_{\rm{H}\alpha}^{*}$, and therefore in the star-formation history of the universe, is driven by dark matter halo mass accretion.}
	\label{fig:lstar_fz}
\end{figure}

In this section, we use a few simple ideas from models of the evolution of galaxies and dark matter halos to link the luminosities of the star-forming galaxies in our sample to the growth of dark matter halos over cosmic time. \\\\
\cite{Fakhouri2010} derive the mean halo mass growth as a function of mass and redshift, using the Millennium simulation:
\begin{equation}\label{eq:dmhalo}
\Bigg\langle \frac{dm_{\rm{halo}}}{dt} \Bigg\rangle = 46.1\Bigg(\frac{m_{\rm{halo}}}{10^{12}}\Bigg)^{1.1}(1+1.11z)\sqrt{\Omega_{M}(1+z)^{3}+\Omega_{\Lambda}}
\end{equation}
We gather the terms $(1+1.11z)\sqrt{\Omega_{M}(1+z)^{3}+\Omega_{\Lambda}}$ and call them $f(z)$ from here onwards. \\\\
We define the halo specific mass inflow rate, $sMIR_{\rm{DM}}$, as:
\begin{equation}\label{eq:smir}
sMIR_{\rm{DM}} = \frac{1}{m_{\rm{halo}}}\frac{dm_{\rm{halo}}}{dt}.
\end{equation}
We now attempt to relate this to star-formation in galaxies. Equilibrium models, in which star-formation in a galaxy is regulated by the instantaneous mass of gas in its reservoir and mass loss is similarly regulated by the star-formation rate, have been successful in reproducing many observed galaxy properties including gas fractions and metallicities to $z\sim2$ \citep[e.g.][]{Dav2012,Lilly2013a,Saintonge2013}. In the gas-regulated model of galaxy evolution proposed by \cite{Lilly2013a}, the specific star-formation rate of a central galaxy is related to the average specific mass accretion rate of its dark matter halo via:
\begin{equation}\label{eq:ssfr}
sSFR = \frac{1}{(1-\eta)(1-R)}sMIR_{\rm{DM}},
\end{equation}
where $\eta$ (the slope of the mass-metallicity relation) and $R$ (which determines the fraction of stars which are long-lived) are observationally-determined constants.\\\\
Substituting $sSFR = SFR/m_{\rm{star}}$, and using $SFR = 7.9\times10^{-42}L_{\rm{H}\alpha}$, from \cite{Jr1998}, then combining Equations \ref{eq:dmhalo} \& \ref{eq:smir} yields:
\begin{equation}\label{eq:lhaf}
L_{\rm{H}\alpha} = k\,m_{\rm{halo}}^{0.1}\,m_{\rm{star}}f(z),
\end{equation}
where $k$ is a numerical factor.\\\\
We found in Section \ref{sec:lum_hod} that the scaled mean $\rm{H}\alpha$ luminosity, $L_{\rm{H}\alpha}/L_{\rm{H}\alpha}^{*}$, of a sample of our star-forming galaxies is related to halo mass in a redshift-independent manner:
\begin{equation}\label{eq:my_relation}
\frac{L_{\rm{H}\alpha}}{L_{\rm{H}\alpha}^{*}(z)} \approx \Bigg(\frac{m_{\rm{halo}}}{10^{12}}\Bigg)^{1/1.6}.
\end{equation}
Dividing Equation \ref{eq:lhaf} by $L_{\rm{H}\alpha}^{*}$, we obtain: 
\begin{equation}\label{eq:lhaf2}
\frac{L_{\rm{H}\alpha}}{L_{\rm{H}\alpha}^{*}(z)} \approx k\,m_{\rm{halo}}^{0.1}\,m_{\rm{star}}\,\frac{f(z)}{L_{\rm{H}\alpha}^{*}(z)},
\end{equation}
which, from our observed relation (Equation \ref{eq:my_relation}) must remain constant with redshift for a given $m_{\rm{halo}}$. \\\\
The average galaxy stellar mass, $m_{\rm{star}}$, is also related to $m_{\rm{halo}}$ broadly independently of redshift within our range of redshifts \citep[the SHMR;][]{Behroozi2013,Birrer2014,Hatfield2015}. Therefore, to maintain Equation \ref{eq:my_relation} across cosmic time in the context of the gas regulator model,
\begin{equation}
\frac{f(z)}{L_{\rm{H}\alpha}^{*}(z)} = const.
\end{equation}
must hold.\\\\
To test this, we calculate $\frac{f(z)}{L_{\rm{H}\alpha}^{*}(z)}$ for the HiZELS samples at the three different redshifts. We find that this is, indeed, fairly constant compared to the strong evolution in $L_{\rm{H}\alpha}^{*}$ (see Figure \ref{fig:lstar_fz}). \textcolor{black}{Whereas $L_{\rm{H}\alpha}^{*}$ changes by an order of magnitude, $\frac{f(z)}{L_{\rm{H}\alpha}^{*}(z)}$ changes by less than $0.2\rm{dex}$.} Our results therefore support a model in which the evolution in $L_{\rm{H}\alpha}^{*}$ is driven solely by the halo mass growth, in line with a gas regulator model. We thus conclude that our HiZELS galaxies are dominated by typical star-forming galaxies in equilibrium, rather than extreme, merger-driven starburst systems, even at high redshifts. The halo mass accretion rate is the dominant driver of star-formation in these galaxies across the large redshift range $0.8<z<2.23$.
\subsection{Satellite fractions and environmental quenching}
\indent We have found low satellite fractions ($\sim5\%$) at all three of the redshifts studied, and in all luminosity bins, using this HOD parameterisation. The gas regulator model, shown in Section \ref{sec:gas_reg} to fit our observations well, does not include any satellite-specific mechanisms like ram pressure stripping. This supports the conclusion that the majority of HiZELS galaxies are centrals.\\
\indent As discussed earlier, the exact values we derive for the satellite fraction may be significantly dependent on the HOD parameterisation we adopted, as two-halo clustering cannot discriminate between centrals and satellites. Nevertheless, it is possible to demonstrate that the satellite fraction must be low. Many HOD models of mass-selected samples of galaxies \citep[including at these redshifts, e.g.][]{Wake2011,Hatfield2015} use a power-law satellite occupancy model with $\alpha \approx 1$, with a low-mass cut-off below halo mass $\sim 10^{13}M_{\odot}$. As shown in our HOD modelling, obtaining a good fit to our (relatively low-amplitude) correlation functions requires a substantial contribution from low mass halos, down to $<10^{12}M_{\odot}$. The scarcity of satellites in these low mass halos, coupled with the increase in the halo mass function at low halo masses, thus mandates a fairly low overall satellite fraction. To quantify this, we consider a conservative model in which the satellite occupancy of halos follows a power-law with $\alpha=1$ down to the lowest masses (i.e. no cut-off), normalised to unity at $M_{\rm{halo}}=2\times10^{13}M_{\odot}$ \citep[cf.][]{Wake2011,Hatfield2015}. Even if all potential satellite galaxies were to be star-forming, our total HOD model for the `full' sample at $z=0.8$ then permits a maximum satellite fraction of $\sim8\%$ (this increases to $\sim 14\%$ for a normalisation of $<N_{\rm{sat}}|M>=1$ at $M_{\rm{halo}}=10^{13}M_{\odot}$). We can thus safely conclude that satellite fractions must be low.\\
\indent Detailed comparison of our HOD modelling result with those of mass-selected samples at these same redshifts would require us to match the samples in stellar mass; this is beyond the scope of this paper. Nevertheless, we can gain some initial insights by comparison with the results of \cite{Hatfield2015}, who studied mass-selected samples in a similar stellar mass range as our $\rm{H}\alpha$ emitters, in overlapping redshift ranges, using the same HOD fitting code as ours (thus minimising any systematic errors). \cite{Hatfield2015} find satellite fractions of $\sim13\pm1\%$ at $z\sim0.8$ and $\sim6\pm1\%$ at $z\sim1.5$, integrating down to the lowest galaxy stellar masses within their samples. Our
redshift-independent satellite fraction of star-forming galaxies, when compared to the increasing satellite fraction amongst mass-selected galaxies towards low redshift, indicates that a significantly larger portion of satellites are star-forming at higher redshifts. These results are consistent with those of \cite{Tal2014}, who find that the quiescent fraction for satellites increases towards low redshift, from $\sim10\%$ at $z\sim1.5$ to $\sim30\%$ at $z\sim0.8$, with onset of satellite quenching taking place several Gyr after the first centrals reach quiescence. \\
\indent Our results may also provide insights into the quenching mechanisms acting at high redshifts. A number of studies find a strong excess of starbursting sub-mm galaxies in high-redshift cluster environments \citep{Elbaz2007,Smail2014,Ma2015}. In some cases these starbursting galaxies reside in the cluster core \citep[e.g.][]{Ma2015}, and in others they lie towards the cluster's outskirts, with passive galaxies dominating the central regions \citep[e.g.][]{Smail2014}. If this intense star-formation were driven by an enhanced intracluster gas supply, we would expect to see enhanced star-formation throughout these high mass halos, reflected in high satellite fractions and increased effective halo masses for our HiZELS galaxies at higher redshifts. Instead we find that both of these properties remain broadly consistent. Combined with the sub-mm view, our results support the scenario put forward by \cite{McGee2009}, in which upon infall onto a rich cluster, compression of high gas contents within galaxies may provoke intense, dust-obscured star-formation, after which quenching proceeds on fairly long timescales ($>2\rm{Gyr}$) via gas stripping or exhaustion.
\section{Conclusions}
We have used HiZELS galaxies selected by the strength of their $\rm{H}\alpha$ emission to study the clustering of star-forming galaxies at three well-defined epochs: $z=0.8$, $z=1.47$, $z=2.23$. Our samples comprise typical star-forming galaxies on and just above the `main sequence' at each redshift. We have constructed two-point correlation functions and fitted these with simple power-law fits, finding that the clustering strength, $r_{0}$, of HiZELS sources at all redshifts increases linearly with their $\rm{H}\alpha$ luminosity, from $r_{0}\sim2-3h^{-1}\rm{Mpc}$ for the lowest luminosity sources in our samples to $r_{0}\sim7-8h^{-1}\rm{Mpc}$ for the most luminous. \textcolor{black}{We have demonstrated that this is not driven by galaxy stellar mass.}\\
\indent We then used MCMC techniques to fit the same correlation functions with a more sophisticated Halo Occupation Distribution (HOD) models, deriving each galaxy population's effective bias, satellite fraction, and indicative dark matter halo masses. We summarise the key results here.
\begin{itemize}
\item Typical $\rm{H}\alpha$-emitting galaxies in the redshift range $z=0.8-2.2$ are star-forming centrals, residing in host halos of minimum mass $10^{11.2}M_{\odot}-10^{12.6}M_{\odot}$ and effective mass $10^{11.6}M_{\odot}-10^{13}M_{\odot}$. At all three redshifts, $L_{\rm{H}\alpha}^{*}$ galaxies typically reside in halos of mass $\sim10^{12}M_{\odot}$. This coincides with the halo mass predicted by theory to be maximally efficient at converting baryons into stars. \\
\item The effective bias of the galaxy populations (their clustering relative to the underlying dark matter) \textcolor{black}{decreases towards lower redshifts,} reflecting the increase of the clustering of dark matter with time. Similarly, typical masses of host halos increase with time at fixed $L_{\rm{H}\alpha}$.\\
\item  Bias increases linearly with $\rm{H}\alpha$ luminosity at all redshifts, indicating that the most highly star-forming galaxies thrive in higher dark matter overdensities, where a plentiful gas supply fuels high star-formation rates \textcolor{black}{in the central galaxies}. \\
\item Samples selected within the same $L_{\rm{H}\alpha}/L_{\rm{H}\alpha}^{*}$ range inhabit similar populations of dark matter halos. \textcolor{black}{Although the dark matter halo mass at fixed $L_{\rm{H}\alpha}$ varies by more than an order of magnitude across the three different redshifts, the relationship between scaled galaxy luminosity $L_{\rm{H}\alpha}/L_{\rm{H}\alpha}^{*}$ and dark matter halo mass is independent of redshift to within $0.04\rm{dex}$ in $M_{\rm{min}}$ and $0.2\rm{dex}$ in $M_{\rm{eff}}$.} \\
\item Comparing our results to models of galaxy evolution based on gas-regulation, we find that $L_{\rm{H}\alpha}^{*}$ evolves in line with average mass growth of the host dark matter halos. 
\end{itemize}
\noindent 
Together, these results reveal halo environment as a strong driver of galaxy star-formation rate and the evolution of the luminosity function over cosmic time. The central galaxies which dominate our samples evolve in equilibrium with their growing dark matter halos, with typical specific star-formation rate directly proportional to the specific mass accretion rate of the host dark matter halo. Satellite fractions remain low ($\sim5\%$ with the HOD parameterisation we have adopted) for all samples, regardless of redshift or luminosity. This may indicate that their star-formation is suppressed, particularly towards low redshifts and in high mass dark matter halos. This is in line with models of satellite quenching upon accretion onto a massive cluster. In a subsequent paper we will extend this study to incorporate stellar mass, exploring the clustering of HiZELS galaxies as a function of $\rm{H}\alpha$ luminosity, stellar mass and redshift. 
\section*{Acknowledgements}
This work is based on observations obtained using the Wide Field CAMera (WFCAM) on the 3.8-m United Kingdom Infrared Telescope (UKIRT), as part of the High-redshift(Z) Emission Line Survey (HiZELS; U/CMP/3 and U/10B/07). It also relies on observations conducted with HAWK-I on the ESO Very Large Telescope (VLT), programme 086.7878.A, and observations obtained with Suprime-Cam on the Subaru Telescope (S10B-144S). \\
\indent We are grateful to Steven Murray for making the HALOMOD and HMF python packages available and for extensive guidance on their use. We also thank John Peacock and Peter Hatfield for their helpful advice. \\
\indent RKC acknowledges funding from an STFC studentship. PNB is grateful for support from STFC via grant ST/M0011229/1. DS acknowledges financial support from the Netherlands Organisation for Scientific research (NWO) through a Veni fellowship and from Lancaster University through an Early Career Internal Grant A100679. IRS acknowledges support from STFC (ST/L00075X/1), the ERC Advanced Investigator programme DUSTYGAL 321334, and a Royal Society/Wolfson Merit Award. JPS gratefully acknowledges support from a Hintze Research Fellowship.

\bibliographystyle{mnras}
\bibliography{Edinburgh}

\appendix
\section{Correlation function fits}\label{sec:appendixA}

\begin{figure} 
	\includegraphics[scale=0.4]{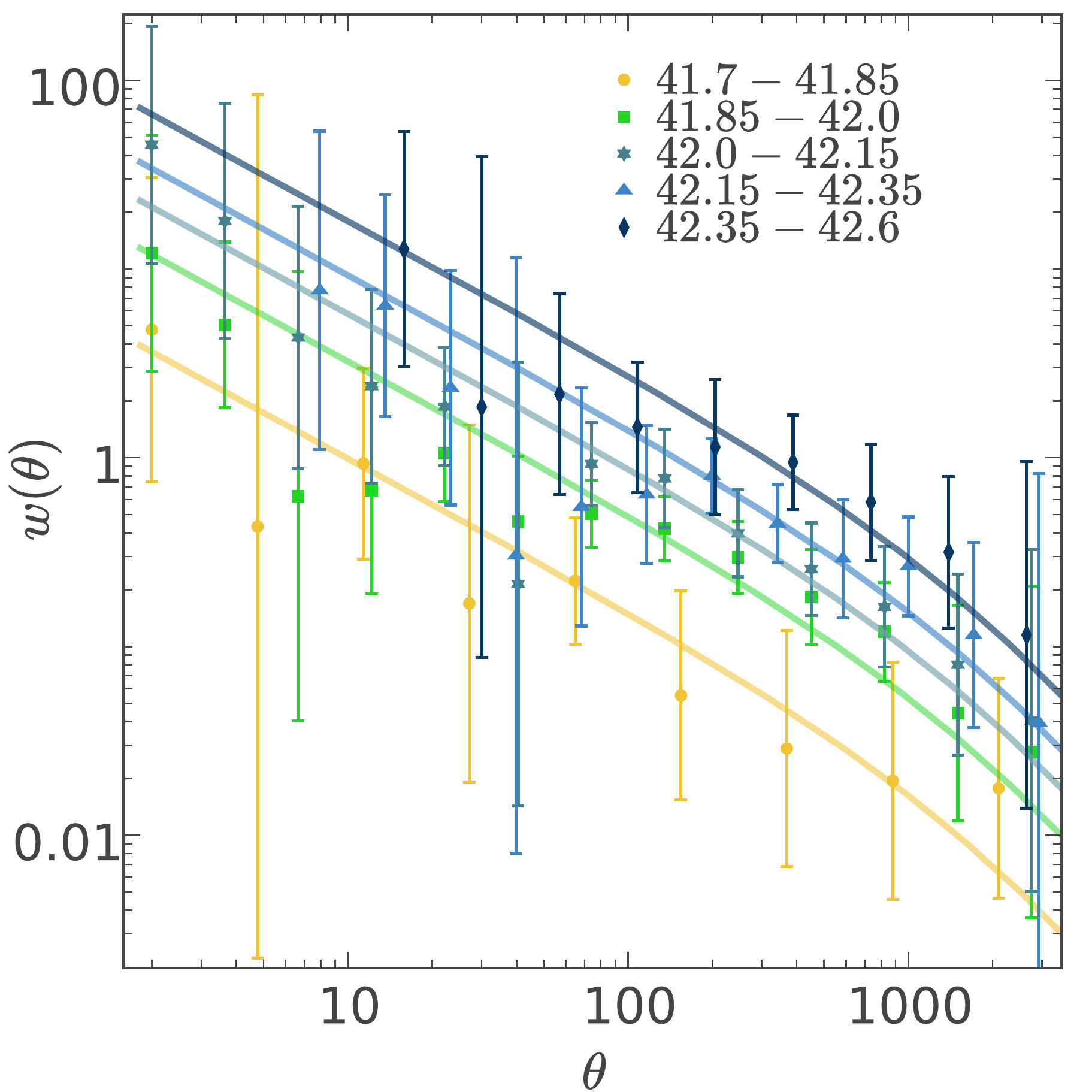}
	\caption{Examples of five correlation functions of luminosity-binned samples at $z=0.8$. Although the correlation functions are not as high quality as those of the whole samples (e.g. Figure \ref{fig:corr_z}), it is clear that the clustering strength (obtained from the amplitude of the correlation function) increases with $L_{\rm{H}\alpha}$ luminosity.}
	\label{fig:example_5corr}
\end{figure}

\textcolor{black}{In Section \ref{sec:r0_results} we calculate correlation functions in bins of $\rm{H}\alpha$ luminosity, and show the resultant $r_{0}$ values in Figure \ref{table:r0_table}. For the $z=0.8$ samples,} we show examples of the quality of the correlation functions constructed from luminosity-binned samples in Figure \ref{fig:example_5corr}.

\section{Further details on HOD parameterisations and fits}
\begin{figure} 
	\centering
	\includegraphics[scale=0.5]{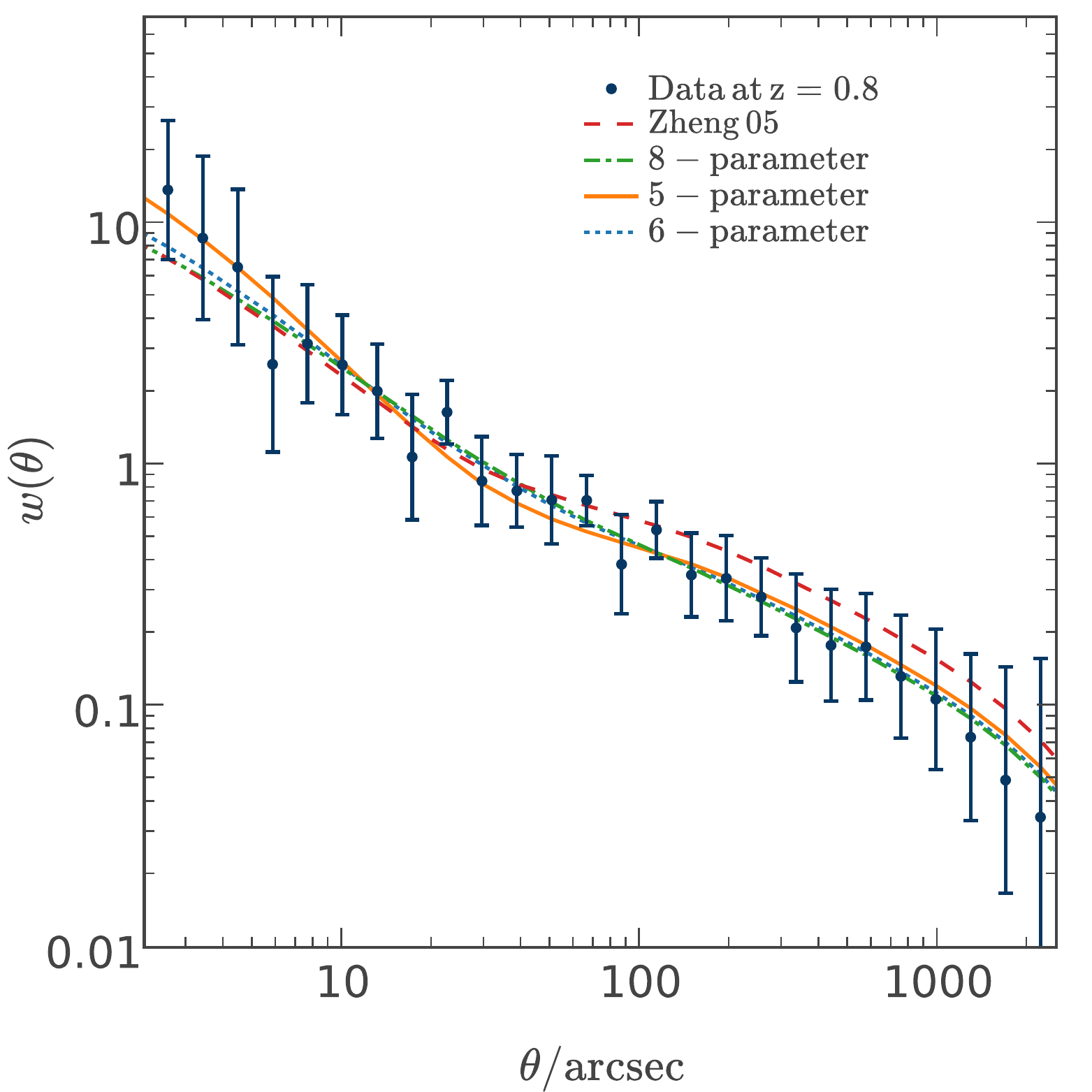}	
	\caption{Comparison of best-fit correlation functions derived using different halo occupation distribution parameterisations. We adopt the 5-parameter \citealt{Geach2012} parameterisation as it fits the correlation function well on all scales and additional parameters are not justified by any improvement to the fit. We truncate all parameterisations, using a lower integral limit of $\rm{M}_{\rm{min}}$.}
    \label{fig:hod_params}
\end{figure}
\begin{figure*} 
	\includegraphics[scale=0.55]{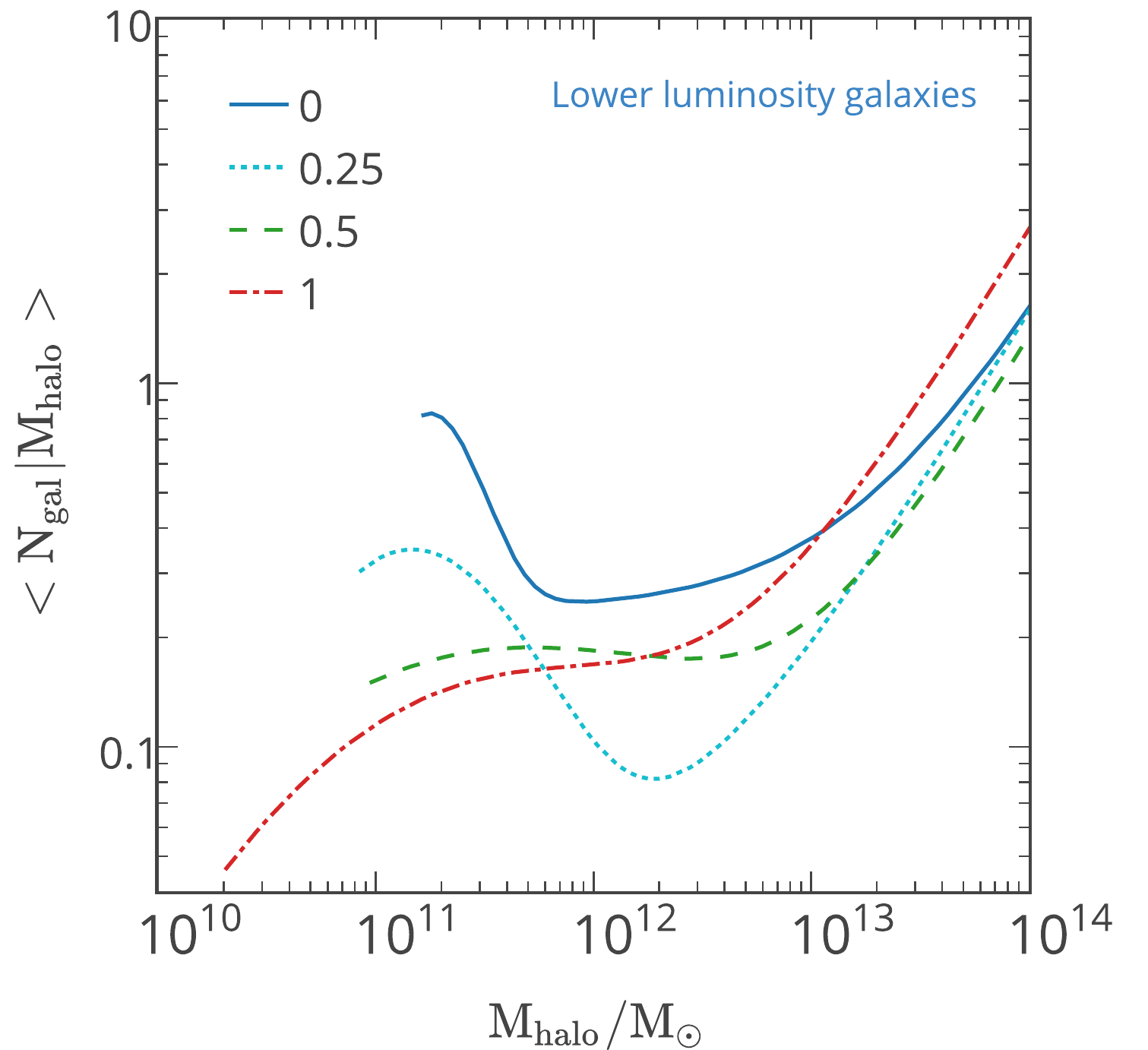}
	\includegraphics[scale=0.55]{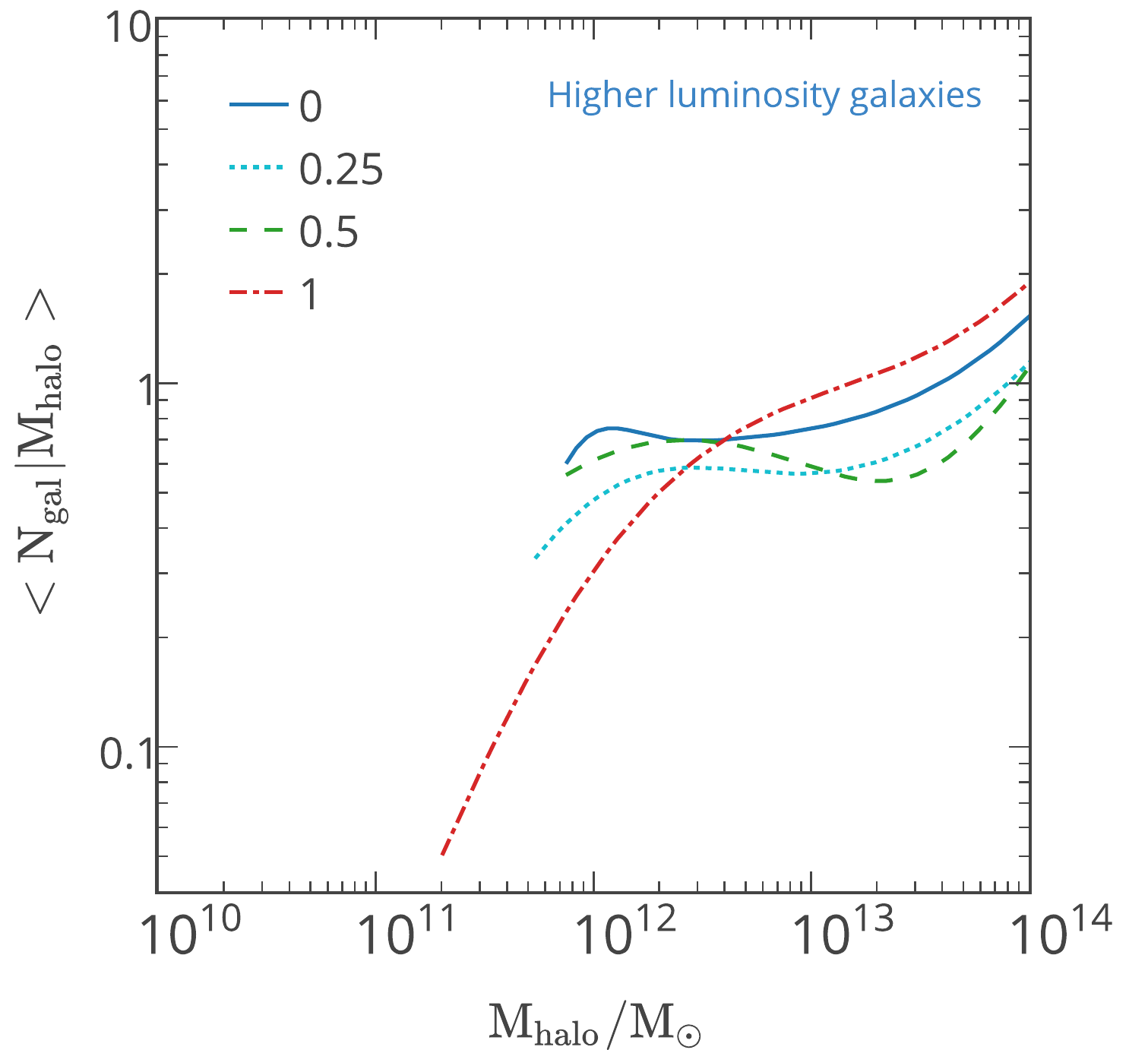}
	\caption{The best fit HODs using different lower limits. The left and right-hand panels show HODs fitted to lower and higher luminosity halves of the sample, respectively. The legends give the lower limit of the integral. For example, 0.5 indicates that we integrate to 0.5dex below $M_{\rm{min}}$. The overall shape of the HOD is strongly dependent on the lower limit. Integrating to lower limits shifts $M_{\rm{min}}$ to higher values, and leaves it less well constrained.}
	\label{fig:appendix_mmin_plot}
\end{figure*}

\begin{table*}
\vspace{0.1cm}
\begin{center}
\begin{tabular}{lc|c|c|c|c}
Parameterisation & $\rm{b}_{\rm{eff}}$ & $\rm{M}_{\rm{eff}}$ & $\rm{M}_{\rm{min}}$ & $\rm{f}_{\rm{sat}}$ \\
\hline
8-parameter (Contreras 2013) & $1.18^{+0.04}_{-0.06}$ & $12.23^{+0.13}_{-0.17}$ & $11.18^{+0.09}_{-0.18}$ & $0.04^{+0.02}_{-0.04}$ \\
6-parameter (Contreras 2013) & $1.14 \pm 0.06$ & $12.29^{+0.08}_{-0.11}$ & $11.03^{+0.17}_{-0.22}$ & $0.05^{+0.01}_{-0.02}$ \\
5-parameter (Contreras 2013/ Geach 2012) & $1.15 \pm 0.05$  & $12.20^{+0.08}_{-0.09}$ & $11.14 ^{+0.09}_{-0.12}$ & $0.06 ^{+0.02}_{-0.01}$ \\
5-parameter (Zheng 2005) & $1.24 \pm 0.02$ & $12.44^{+0.06}_{-0.05}$ & $11.21^{+0.06}_{-0.05}$ & $0.04 \pm 0.03$ \\
\end{tabular}
\caption{Derived parameters from fitting the whole sample at $z\sim0.8$ with for four different halo model parameterisations. The values of these derived parameters depend little on the choice of parameterisation.}
\label{table:different_parameterisations}
\end{center}
\end{table*}

\subsection{Choosing an HOD parameterisation}\label{sec:choose_parameterisation}
\indent We have fitted a typical HiZELS correlation function (generated using the whole sample at $z=0.8$) using four different HOD parameterisations that have been proposed for SFR-limited samples (see Section \ref{sec:fitting_halomod} for more details of the fitting procedure). The parameterisations are as follows: the full 8-parameter model described in Section \ref{sec:full_9}; the 5-parameter model adopted by \cite{Geach2012}; a 6-parameter model which is identical to the 5-parameter model apart from fitting $\alpha$ as a further parameter; and the 5-parameter model of \cite{Zheng2005}, frequently used for mass-selected samples. We truncate all paramaterisations at the lower limit $M_{\rm{min}}$. We show in Figure \ref{fig:hod_params} that these parameterisations all do well at fitting the data. Derived quantities are given in Table \ref{table:different_parameterisations}. It is important to note that the derived quantities, $\rm{b}_{\rm{eff}}$, $M_{\rm{eff}}$, $\rm{f}_{\rm{sat}}$ and $M_{\rm{min}}$ are fairly consistent between parameterisations, and the choice of parameterisation does not substantially alter the conclusions of this paper. \\
\indent We conclude that the truncated 5-parameter HOD of \cite{Geach2012} provides a sufficiently good reproduction of the correlation function. Higher order parameterisations are not justified by improvements in the fit to the correlation function. With the smaller sizes of luminosity-binned samples, minimising the number of free parameters is important to obtain good parameter constraints, and so we adopt the 5-parameter approach. We emphasize that we also checked our analyses with the 6-parameter model (allowing $\alpha$ to vary), and recovered consistent results, also finding $\alpha\approx1$.
\subsection{Testing the lower limit on the HOD integral}\label{sec:lower_limit_hod}
\indent If the halo occupation numbers fall steeply at low halo masses, we can safely set the lower limit of the integral at an arbitrary, low halo mass. In practice, we find that fits to our correlation functions produce poorly-defined gaussian peaks, with fairly flat occupations at low masses (see Section \ref{sec:lum_hod}). To test different lower limits here, we integrate to fixed distances ($0\,\rm{dex}$, $0.25\,\rm{dex}$, $0.5\,\rm{dex}$ and $1\,\rm{dex}$) below $M_{\rm{min}}$ during the HOD fitting process. We perform this test for two different correlation functions, constructed using $z\sim0.8$ sources in two luminosity bins. The best-fit HODs for each trial are shown in Figure \ref{fig:appendix_mmin_plot}. The fitted values of $\sigma$ tend to be large and highly correlated with $M_{\rm{min}}$, with a higher $M_{\rm{min}}$ and larger $\sigma$ producing the same correlation function as a lower $M_{\rm{min}}$ and smaller $\sigma$. Derived values of $M_{\rm{min}}$ are therefore highly dependent on the lower limit of the integral. Table \ref{table:comparing_int_lowers} shows the fitted parameters. All other derived values depend little on the choice of integration limit.\\ 
\indent For the purposes of this study, we fix the lower limit of the integral to be $M_{\rm{min}}$. $M_{\rm{min}}$ is then a more physical quantity: the minimum mass of halos hosting central galaxies. This is also consistent with parameterisations used by other authors, and enables easier comparison of minimum halo masses. \\

\begin{table*}
\vspace{0.1cm}
\begin{center}
\begin{tabular}{l|c|c|c|c|c|c|c}
\multicolumn{5}{c}{Fitted HOD parameters for lower luminosity HiZELS emitters at $z\sim0.8$ ($41.72<\log_{10}(L_{\rm{H}\alpha})<42.07$}) \\
\hline
Lower limit & $\rm{b}_{\rm{eff}}$ & $\log_{10}(M_{\rm{eff}}/M_{\odot})$ & $\log_{10}(M_{\rm{min}}/M_{\odot})$ & \textcolor{black}{$\sigma$} & $f_{\rm{sat}}$ \\
\hline \\
$\log_{10}(M_{\rm{min}}/M_{\odot})$ & $1.05^{+0.05}_{-0.05}$ & $11.92^{+0.09}_{-0.06}$ &$11.09^{+0.10}_{-0.09}$ & \textcolor{black}{$0.3^{+0.2}_{-0.1}$} & $0.03^{+0.01}_{-0.01}$\\
$\log_{10}(M_{\rm{min}}/M_{\odot})-0.25$ & $1.03^{+0.05}_{-0.03}$ & $11.88^{+0.10}_{-0.08}$ &$11.21^{+0.14}_{-0.12}$ & \textcolor{black}{$0.3^{+0.3}_{-0.1}$} & $0.03^{+0.01}_{-0.01}$\\
$\log_{10}(M_{\rm{min}}/M_{\odot})-0.5$ & $1.03^{+0.05}_{-0.03}$ & $11.87^{+0.11}_{-0.08}$ &$11.27^{+0.17}_{-0.13}$ & \textcolor{black}{$0.5^{+0.3}_{-0.3}$} & $0.03^{+0.01}_{-0.01}$\\
$\log_{10}(M_{\rm{min}}/M_{\odot})-1$ & $1.00^{+0.07}_{-0.04}$ & $11.85^{+0.13}_{-0.10}$ &$11.43^{+0.21}_{-0.19}$ & \textcolor{black}{$0.7^{+0.2}_{-0.3}$} & $0.02^{+0.01}_{-0.01}$\\
\hline
& & & & & \\
\multicolumn{5}{c}{Fitted HOD parameters for higher luminosity HiZELS emitters at $z\sim0.8$ ($42.07<\log_{10}(\rm{L}_{\rm{H}\alpha})<42.42$}) \\
\hline
Lower limit & $\rm{b}_{\rm{eff}}$ & $\log_{10}(M_{\rm{eff}}/M_{\odot})$ & $\log_{10}(M_{\rm{min}}/M_{\odot})$ & \textcolor{black}{$\sigma$} & $f_{\rm{sat}}$\\
\hline \\
$\log_{10}(M_{\rm{min}}/M_{\odot})$ & $1.47^{+0.06}_{-0.08}$ & $12.54^{+0.08}_{-0.11}$ &$11.85^{+0.07}_{-0.10}$ & \textcolor{black}{$0.5^{+0.3}_{-0.3}$} & $0.02^{+0.01}_{-0.01}$\\
$\log_{10}(M_{\rm{min}}/M_{\odot})-0.25$ & $1.46^{+0.05}_{-0.08}$ & $12.52^{+0.08}_{-0.12}$ &$12.02^{+0.08}_{-0.10}$ & \textcolor{black}{$0.6^{+0.3}_{-0.3}$} & $0.02^{+0.01}_{-0.01}$\\
$\log_{10}(M_{\rm{min}}/M_{\odot})-0.5$ & $1.45^{+0.05}_{-0.08}$ & $12.51^{+0.07}_{-0.12}$ &$12.17^{+0.13}_{-0.16}$ & \textcolor{black}{$0.7^{+0.2}_{-0.3}$} & $0.02^{+0.01}_{-0.01}$\\
$\log_{10}(M_{\rm{min}}/M_{\odot})-1$ & $1.42^{+0.05}_{-0.06}$ & $12.51^{+0.07}_{-0.09}$ &$12.45^{+0.21}_{-0.24}$ & \textcolor{black}{$0.7^{+0.2}_{-0.3}$} & $0.02^{+0.01}_{-0.01}$\\
\hline
\end{tabular}
\caption{Derived parameters from HOD fits to correlation functions at $z=0.8$, integrating to different lower limits. Most parameters are unaffected by this lower limit, and the projected two-point correlation functions are near-identical. However, $M_{\rm{min}}$ moves towards higher values and becomes less well constrained as we integrate to lower halo masses. We therefore adopt $M_{\rm{min}}$ as the lower limit to our HOD fits, essentially truncating the parameterisation described in Section \ref{sec:halo_fitting}.}
\label{table:comparing_int_lowers}
\end{center}
\end{table*}

\subsection{MCMC fits to correlation functions}\label{sec:mcmc_plot}
\begin{figure*} 
	\centering
	\includegraphics[scale=0.6]{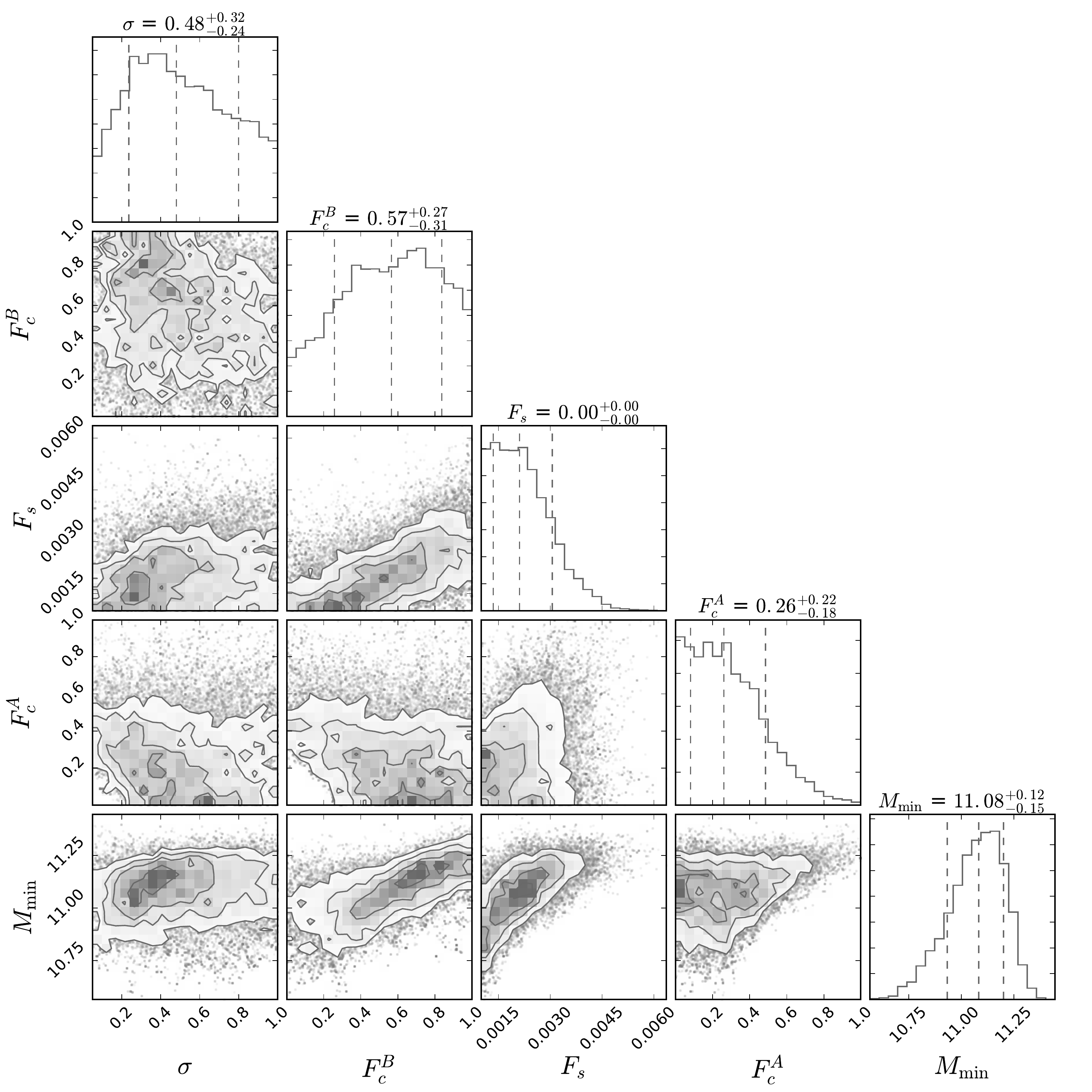}	
	\caption{An example of the output from the MCMC HOD fit to the two-point correlation function \citep{Foreman-Mackey2016}, constructed using the `full' sample of galaxies at $z=0.8$. The five fitted parameters are highly correlated, but we obtain good constraints on the derived parameters, $\rm{b}_{\rm{eff}}$, $M_{\rm{eff}}$, $M_{\rm{min}}$ and $\rm{f}_{\rm{sat}}$.}
    \label{fig:triangle_plot}
\end{figure*}

In Section \ref{sec:fitting_halomod}, we described the HOD fitting process. We primarily used the HMF \citep{Murray2013} and HALOMOD codes (Murray, in prep.), which 
make use of the MCMC fitting software {\it{emcee}} \citep{Foreman-Mackey2013} to derive HOD parameters. In Figure \ref{fig:triangle_plot} we show an example of the MCMC output. While the five individual HOD parameters are highly correlated, we can still constrain the derived parameters, $\rm{b}_{\rm{eff}}$, $M_{\rm{eff}}$, $M_{\rm{min}}$ and $\rm{f}_{\rm{sat}}$ and obtain good fits to the correlation functions.


\bsp	
\label{lastpage}
\end{document}